\documentclass[a4paper,fleqn]{cas-sc} 

\usepackage[numbers,sort&compress]{natbib} 
\usepackage{booktabs}
\usepackage{stfloats}
\usepackage{amssymb}
\usepackage{cleveref}
\usepackage{color}
\usepackage[linesnumbered,ruled,vlined]{algorithm2e}
\usepackage[justification=centering]{caption}
\def\tsc#1{\csdef{#1}{\textsc{\lowercase{#1}}\xspace}}
\tsc{WGM}
\tsc{QE}

\begin{document}
	\let\WriteBookmarks\relax
	\let\printorcid\relax
	\def\floatpagepagefraction{1}
	\def\textpagefraction{.001}
	
	\shorttitle{Socially Beneficial Metaverse: Framework, Technologies, Applications, and Challenges}    
	
	\shortauthors{Xiaolong Xu et~al.}  
	
	\title [mode = title]{Socially Beneficial Metaverse: Framework, Technologies, Applications, and Challenges}

	\tnotemark[<tnote number>] 
	
	
	%







\author[1,2]{Xiaolong Xu}
\ead{xlxu@ieee.org}

\author[1]{Xuanhong Zhou}
\ead{202083290423@nuist.edu.cn}

\author[3]{Muhammad Bilal}
\cormark[1]
\ead{m.bilal@ieee.org}

\author[4,9]{Sherali Zeadally}
\ead{szeadally@uky.edu}

\author[5]{Jon Crowcroft}
\ead{Jon.Crowcroft@cl.cam.ac.uk}

\author[6,7]{Lianyong Qi}
\ead{lianyongqi@gmail.com}

\author[8]{Shengjun Xue}
\ead{sjxue@163.com}

\affiliation[1]{organization={School of Software, Nanjing University of Information Science and Technology},
                city={Nanjing},
                country={China}}

\affiliation[2]{organization={Jiangsu Collaborative Innovation Center of Atmospheric Environment and Equipment Technology (CICAEET), Nanjing University of Information Science and Technology},
                city={Nanjing},
                country={China}}

\affiliation[3]{organization={School of Computing and Communications, Lancaster University},
                city={Lancaster},
                country={United Kingdom}} 

\affiliation[4]{o={College of Communication and Information, University of Kentucky}, c={Lexington}, cy={USA}}

\affiliation[9]{o={Imam Abdulrahman bin Faisal University (IAU)}, c={Dammam}, cy={Saudi Arabia}}

\affiliation[5]{organization={Department of Computer Science and Technology, University of Cambridge},
                city={Cambridge},
                country={United Kingdom}}

\affiliation[6]{organization={College of Computer Science and Technology, China University of Petroleum (East China)},
                country={China}}

\affiliation[7]{organization={State Key Laboratory for Novel Software Technology, Nanjing University},
                city={Nanjing},
                country={China}}

\affiliation[8]{organization={School of Computer Science and Technology, Silicon Lake College},
                city={Suzhou},
                country={China}}             

\cortext[cor1]{Corresponding author}

	\begin{abstract}
In recent years, the maturation of emerging technologies such as Virtual Reality, Digital Twins and Blockchain has accelerated the realization of the metaverse. As a virtual world independent of the real world, the metaverse will provide users with a variety of virtual activities which bring great convenience to society. In addition, the metaverse can facilitate digital twins, which offers transformative possibilities for the industry. Thus, the metaverse has attracted the attention of the industry, and a huge amount of capital is about to be invested. However, the development of the metaverse is still in its infancy and little research has been undertaken so far. We describe the development of the metaverse. Next, we introduce the architecture of the socially beneficial metaverse (SB-Metaverse) and we focus on the technologies that support the operation of SB-Metaverse. In addition, we also present the applications of SB-Metaverse. Finally, we discuss several challenges faced by SB-Metaverse which must be addressed in the future.
	\end{abstract}
	
	
%
\maketitle
	
	\section{Introduction}\label{}
While Neal Stephenson and William Gibson are often credited with conceptualizing ``cyberspace’’, which predates the introduction of the term ``metaverse’’ in Stephenson’s novel ``Snow Crash’’ \cite{chung2014empirical}, the origins of ``cyberspace’’ can actually be traced back to the late 1960s. Danish artist Susanne Ussing and architect Carsten Hoff first coined the term when they named their workspace Atelier Cyberspace. Although their enthusiasm for incorporating computers into their visual art was clear, they lacked access to the necessary technology at the time. Decades later, with the advent of the Internet, the concept of the metaverse has become a reality. This new digital realm represents an advanced form of Internet applications built upon emerging technologies such as Blockchain, Edge Computing \cite{xiang2023cloud}, Cloud Computing, and Artificial Intelligence.

The metaverse is a real-time multiverse where all the users are connected to a single online framework that enables real-time content changes, interactions, and monetization. The metaverse is characterized by sensory immersion, decentralization, open economic systems, and borderless interoperability. Thus, it will bring many positive effects to society \cite{duan2021metaverse}. Due to the isolation of geographic distance and the impact of the COVID-19 pandemic, many activities have been reduced or stopped to enable epidemic prevention. However, the metaverse can provide great accessibility to meet different social needs such as remote medical diagnosis, remote epidemic control, and real-time online teaching. In addition, the metaverse provides infinite space for expansion and seamless scene changes, which can also achieve diversity. Thus, the metaverse has great potential and broad development prospects. With the metaverse concept, the Roblox stock listed in the United States on March 10, 2021, the metaverse has gradually become known and is being applied to various fields in recent years. Many corporate giants are also deploying the metaverse \cite{lee2021all}. In May 2021, Microsoft CEO Satya Nadella said the company was building a corporate metaverse. In October 2021, Facebook founder Mark Zuckerberg announced that the company would be renamed as Meta \cite{kraus2022facebook}.

Currently, the metaverse paradigm is being leveraged by a wide range of application scenarios which are beneficial to the society. In the field of digital goods, users can select and try on goods with a highly immersive experience through the metaverse. In March 2021, Gucci launches augmented reality (AR) virtual sneakers exclusively sold in the Roblox and VRChat virtual communities \cite{kim2021advertising}. As people interact more with digital goods and their values continue to rise, digital goods can serve as a mapping of physical reality in the virtual field. In the field of telecommuting, traditional technologies such as video calls and online meetings have the shortcomings of inefficient communication and lack of real-time interaction \cite{LiaqatBilal}. The emergence of the metaverse solved these problems and it has been initially applied to telecommuting. Facebook has launched Horizon Workrooms, a remote collaboration tool that allows colleagues wearing virtual reality (VR) devices to be in the same virtual room and communicate face-to-face as much as possible. Teleworking and collaboration in the metaverse will become more commonplace, and the way we live, connect, and work is changing dramatically. In addition, the metaverse is being used in various application areas such as virtual assets, online education, and immersive travel, all of which bring various benefits to society. 

Although metaverse is becoming increasingly popular and socially beneficial, there is still a lack of sufficient research on the structure and technology of the metaverse. Moreover, the social challenges and recommended solutions of the metaverse have not been fully discussed yet. In this context, we summarize the main contributions of this paper as follows:

	\begin{itemize}
		
		\item We provide a detailed definition of the metaverse and the definition of SB-Metaverse.
		
		\item We describe the development process of the metaverse and we present. Then list several typical metaverse examples at different stages of development.
		
		\item We propose the SB-Metaverse framework and describe the specific content of each layer. 
		
		\item We present various applications of SB-Metaverse. Finally, we summarize the social and non-social challenges in SB-Metaverse, and provide corresponding solutions that solve these challenges.
		
	\end{itemize}
	
	We organize the rest of the paper as follows. In Section II, we introduce the definition and development process of the metaverse. In Section III, we propose the SB-Metaverse framework. Next, we discuss key technologies such as edge computing and blockchain of SB-Metaverse in Section IV. Section V presents applications of socially beneficial metaverse. Section VI discusses social and non-social challenges faced by the socially beneficial metaverse. Finally, Section VII concludes the paper. 

\section{What Is the Metaverse}

	\subsection{Definition of the Metaverse}
	``Metaverse’’ is composed of the prefix ``meta’’ and the root ``verse’’, in which ``meta’’ means transcending and ``verse’’ is derived from ``universe’’. The word ``metaverse’’ originated from Snow Crash, a science fiction novel written by Neal Stephenson \cite{stephenson2003snow}. In this fiction, the metaverse referred to a parallel digital world which is separated from physical world, where people can interact with each other freely. As a universe generated by computer, it has been defined in various ways since its mergence. Lee et al. \cite{lee2021all} viewed the metaverse as a virtual environment which combines the digital world with the physical world together. Similarly, Collins et al. \cite{collins2008looking} also described the metaverse as the combination of virtually enhanced physical reality and physically durable virtual space. Jaynes et al. \cite{jaynes2003metaverse} developed an unbridled, open, interactive and immersive environment which enables users to overlook the space-time barriers virtually, and they called this environment ``metaverse’’. Moreover, the metaverse has also been regarded as the next-generation Internet. 
	
	\subsection{Definition of SB-Metaverse} 
	Based on the above definitions, in this paper, we propose SB-Metaverse. SB-Metaverse could be defined as a 3D virtual space with connection awareness and sharing characteristics, which make SB-Metaverse achieve convergence and physical persistence based on future Internet and virtual enhancement. In other words, SB-Metaverse has the duality of physical reality and digital virtuality. In such a world where virtuality and reality blend, people can carry out various social activities at low cost. It also has its own ecosystem where principles such as antitrust and fairness are strictly observed so that anyone can enjoy equal treatment in the metaverse society, which is beneficial to the society.

	\subsection{Development Process of the Metaverse}
	
	Although the metaverse has been attracting significant attention recently, its concept and prototype have been developed over the past few decades, albeit with limited publicity and documentation. The development of the metaverse is closely tied to the progress of science and technology. If we view the metaverse as a spatiotemporal construct encompassing human values, ideas, technical tools, and economic models, its development can be categorized into three stages: the classical metaverse in the form of literature, art, and religion; the neoclassical metaverse in the form of video games; and the highly intelligent metaverse in the form of decentralized applications.

    While the metaverse's conceptual roots are often associated with literature like Neuromancer \cite{gibson2019neuromancer} and Snow Crash \cite{stephenson1994snow}, advancements in virtual reality (VR) technology predate these narratives. As early as the 1960s, VR prototypes such as the Sensorama provided immersive experiences. By the 1970s, text-based Multi-User Dungeons (MUDs) like LambdaMOO emerged, offering interactive environments that laid the groundwork for social virtual worlds. In the 1990s, Nottingham University pioneered mixed reality games, which further blurred the lines between the physical and virtual worlds, serving as a precursor to today's augmented and virtual reality innovations. These milestones underscore the intertwined evolution of technology and the metaverse concept. Fig. 1 illustrates the timeline of the metaverse's development. In the following sections, we detail each stage of its evolution, highlighting key milestones and their relevance to the study. Fig. 1 is adapted from \cite{duan2021metaverse}, with modifications to highlight key milestones relevant to this study.
	

	\subsubsection{Stage 1: The Classical Metaverse in the Form of Literature, Art and Religion}
	
	In this stage, since the personal computer was not widely available, people placed their illusions about the best universe and the parallel world in fictional novels or myths. For example, the Bible and the Divine Comedy belong to the metaverse. Similarly, the journey to the West showed a metaverse with oriental color. As people’s knowledge about the universe increased, science fictions such as Neuromancer \cite{gibson2019neuromancer} and True Names \cite {vinge2015true} described a more magical metaverse which showed us what the metaverse would be like. Although they did not meet people’s illusions about the metaverse, they really indicated the direction of the metaverse’s development. Due to the lack of computer science and technology, metaverse in this stage only existed in people’s imagination. As a result of the growth and popularity of computers and the birth of the Internet, the digital realization of the metaverse entered a new stage.
	
	\subsubsection{Stage 2: The Neoclassical Metaverse in the Form of Video Games}
	
	\begin{itemize}

		\item Interactive games based on text:

		Text-based interactive games are also regarded as interactive novels or text adventures. These games simulate an environment where players use natural languages to interact with the world. By reading a text description of the current state and writing text commands, users can change the environment and report subsequent changes \cite{dambekodi2020playing}. Multi-user dungeon (MUD) is a typical example. It is an instant virtual world of multiple people which is usually based on a literal description. MUD supports functions such as role-playing, interactive fiction and online chat, allowing players to read or view descriptions of rooms and items. Users can also make specific actions in the virtual world. Players in this virtual world often interact with others by inputting instructions which are like natural languages \cite {shah1995playing}. After the emergence of MUD, many versions such as DikuMUD \cite{zen2003impacts}, AberMUD \cite{bartle2003virtual} and TinyMUD \cite{ mauldin1994chatterbots} evolved from it. It is worth mentioning that TinyMUD allows players to make a novel world which is accessible to other players which is also called User generated content (UGC). In UGC, users present their original content through the Internet platform and make it available to other users, which is an important part in today’s metaverse. However, text-based virtual worlds did not give users an intuitive feel. Users want a better virtual world experience, so these games needed to be visually enhanced.

		\item Interactive virtual world based on 3D graphics:

		Advances in computer graphics led to the emergence of 3D-graphics-based interactive virtual worlds. These virtual worlds provide 3D content and real-time authoring. In this environment, people could log in as their own virtual characters and create content in the virtual worlds. Communication occurs through text and voice. Since these virtual worlds are equipped with 3D graphics, they provide the users with an immersive, first-person perspective on the virtual world \cite{livingstone2008multi}. Active World is a typical example wherein everything in this virtual world could be hyperlinked to the Internet, including audio, video, and image files. Its server could provide more than 65,000 users and unlimited virtual space at the same time \cite{tatum2000active}. Other 3D-graphics-based interactive virtual worlds such as Web World and Onlive Traveler \cite {kim2002korean}, together with Active World, could be regarded as early prototypes of the metaverse.
		
	\end{itemize}
	
	\subsubsection{Stage 3: The Highly Intelligent Metaverse in the Form of Decentralized Applications } 
	\begin{itemize}
		
		\item Massive multi-player online video games (MMOG):
		
		Unlike earlier online games, MMOG’s goal was to support thousands of concurrent users. In addition, MMOG is usually a role-playing style game in which players can participate in a story continuously \cite {chan2004strifeshadow}. In 2003, Linden Lab, an American Internet company, launched Open3D-based Second Life \cite {rymaszewski2007second}. Second Life allows users to build another identity, which is a copy of their real-life self or a totally different self. Users can send voice or text messages to others to communicate or take transportation to other places. In particular, users in Second Life have the copyright of their original content, and they could sell their content to gain virtual money called Linden Dollars. Users can also use real-world money in exchange for Linden Dollars and deposit the virtual money in the bank to earn interest, then they can redeem them into real-world currency to make up for their real-world income. Second Life enabled the Interaction and coexistence of the real world and the virtual world \cite{kaplan2009fairyland} \cite{boulos2007second}, and it was a milestone moment which demonstrated the development of the metaverse into a new stage. Then, in 2006, Roblox Corporation released Roblox, a game that was compatible with virtual worlds, casual and user-built content games \cite{jagneaux2018ultimate}. In 2009, the Swedish gaming company Mojang Studios developed Minecraft \cite{duncan2011minecraft}. Instead of trying to create a closed community where everyone must go to, Minecraft allows people to create their own world. This virtual world is not monopolized by any company or individual. Therefore, Minecraft embodies the decentralization of the metaverse.
		
		\item Decentralized virtual world:
		
		In 2008, ``Satoshi Nakamoto’’ released the Bitcoin white paper and established the first distributed network \cite{nakamoto2008bitcoin}. He found a solution for the centralized network, which is a distributed network of blockchain. The blockchain used distributed storage and computing power to guarantee that the data in the system is essentially maintained by the entire network node \cite{chen2020blockchain}. As a result, the blockchain became an important technology to enable the decentralization of the metaverse later. Facebook Horizon become the social VR world in 2019, and Decentraland, which uses Ethereum as a platform to support users in owning and operating virtual assets \cite {chaudhari2019decentraland}, developed the main historical node for this stage of the metaverse.

	\end{itemize}

    \begin{figure*}[!t]
		\centering
		\includegraphics[width=0.9\textwidth]{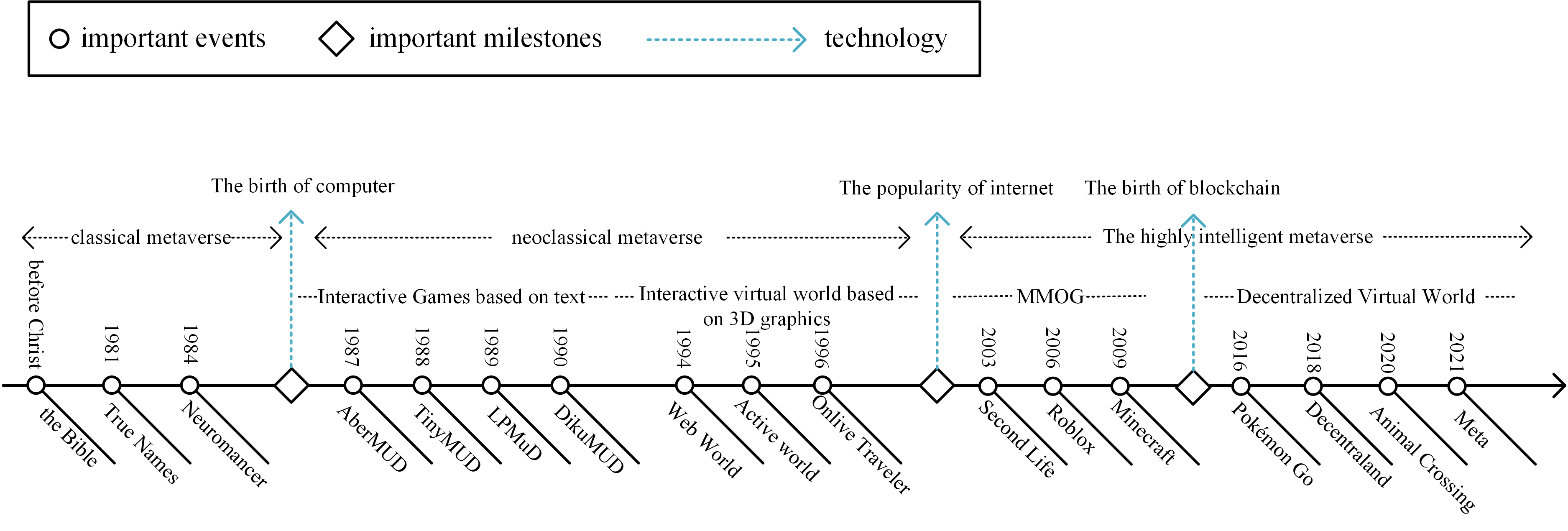}
		\caption{The timeline of the development of the metaverse.}
		\label{fig1}
	\end{figure*}

\section{Framework of Socially Beneficial Metaverse}
Metaverse is a complex ecosystem composed of multiple technological layers. One widely recognized model divides the metaverse into seven key layers: infrastructure, human interface, decentralization, spatial computing, creator economy, virtual platforms, and experience. These layers together enable the metaverse to provide immersive, decentralized, and interactive virtual environments. Kang et al. \cite{kang2021metaverse} proposed a metaverse framework consisting of eight core stacks, which are hardware, network, compute, virtual platform, interchange tools and standards, payments, content, and user behaviors. Ning et al. \cite{ning2021survey} divided the technical architecture into three aspects: network infrastructure, management technology, and basic public technology. Lee et al.\cite{lee2021all} considered the interdisciplinary nature of the metaverse and divided the structure of the metaverse into two key aspects. Then they introduced some important technologies such as edge computing and computer vision in these aspects. In addition, the six-layer structure of the metaverse has also been proposed, which are the underlying hard technology, hardware computing platform, system, software, application, and the economic system that allows the entire metaverse ecosystem to operate \cite{lee2021all}. 

\subsection{Seven Layers of the Metaverse}
The metaverse is a complex ecosystem made up of several key layers, each playing a crucial role in enabling immersive, interactive, and decentralized virtual environments.

\subsubsection{Infrastructure Layer}
The foundation of the metaverse, this layer consists of the hardware, networks, and computing resources required to support virtual environments. It includes powerful servers, data centers, cloud computing, and decentralized networks such as blockchain, which ensure low-latency processing, secure data storage, and efficient communication between users and devices in the metaverse.

\subsubsection{Human Interface Layer}
This layer facilitates the interaction between users and the metaverse. It includes all devices and technologies that enable users to access and engage with virtual worlds, such as virtual reality (VR) headsets, augmented reality (AR) glasses, haptic feedback devices, and motion tracking tools. The human interface is crucial for providing an immersive and intuitive experience for users.

\subsubsection{Decentralization Layer}
Decentralization is a core principle of the metaverse, ensuring that control and ownership are distributed rather than centralized. This layer utilizes technologies like blockchain to provide transparency, secure transactions, and decentralized governance. It ensures that users retain control over their data and digital assets while promoting a fair and open virtual economy.

\subsubsection{Spatial Computing Layer}
This layer focuses on the integration of physical and virtual spaces through augmented reality (AR), virtual reality (VR), and other spatial computing technologies. It enables the creation of interactive 3D environments that users can explore and manipulate in real time. Spatial computing bridges the gap between the digital and physical worlds, allowing users to interact with digital objects as though they exist in the real world.

\subsubsection{Creator Economy Layer}
The creator economy is centered on empowering users to generate, monetize, and exchange digital content within the metaverse. This includes virtual goods, digital art, and user-created environments. Platforms that support user-generated content (UGC) are crucial in this layer, enabling creators to gain ownership, recognition, and revenue from their contributions to the metaverse.

\subsubsection{Virtual Platform Layer}
Virtual platforms are the environments and applications where users can engage with each other and interact with digital content. These platforms range from virtual worlds like Roblox and Decentraland to social spaces and professional environments like Horizon Workrooms. The virtual platform layer is essential for hosting multiplayer interactions, collaborative activities, and social experiences within the metaverse.

\subsubsection{Experience Layer}
The experience layer encompasses the sensory and emotional aspects of users' engagement with the metaverse. It focuses on creating rich, immersive experiences that are personalized and interactive. This layer integrates all other components, from the physical interactions enabled by the human interface to the content and virtual worlds accessed through platforms. It strives to offer seamless and meaningful interactions, whether for entertainment, education, socializing, or other purposes.

	In this paper, we propose the SB-Metaverse framework (as Fig. 2 shows), which consists of three layers: infrastructure, interaction, and ecosystem. These three layers help to achieve the socially beneficial activities and criteria in SB-Metaverse. In addition, dozens of key technologies are incorporated into this framework. We present the details of each layer below.
	
	\subsection{Infrastructure}
	Infrastructure Layer corresponds to the infrastructure and decentralization layers of the seven-layer model and leverages advanced technologies such as 5G, edge computing, blockchain, and cloud computing. These technologies ensure high-speed data collection, low-latency processing, secure storage, and decentralized communication, creating a robust foundation for the metaverse.We describe the infrastructure layer in terms of four parts:
	
	\subsubsection{Data Collection}
	Data collection devices include VR/AR, sensors and smart devices. They collect information from users and send task requests, which are the entrance to SB-Metaverse.
	
	\subsubsection{Data Process}
	Data processing equipment includes chips, computing centers, and so on. They use deep learning, speech processing and other technologies to process the raw data and convert it into data that is easy to store and easy for machines to understand. Since SB-Metaverse will generate massive amounts of data to be processed, it has very high performance requirements on the data processing equipment.
	
	\subsubsection{Data Storage}
	There are various ways of storing data in SB-Metaverse. Personal private data can be stored locally on storage devices such as memory, and data security is ensured through blockchain nodes. For larger amounts of data, edge-cloud collaborative storage can be achieved through edge nodes.
	
	\subsubsection{Data Transmission}
	In SB-Metaverse, IoT enables data transmission and communication between various smart devices. Connection can be established between any devices, which is beneficial to the digital twin. In addition, edge-cloud collaboration provides low latency and high stability to data transmission, which help to achieve high immersion in SB-Metaverse.

	\subsection{Interaction}
	
	Interaction Layer maps to the human interface and spatial computing layers and focuses on enabling seamless and immersive user experiences. Technologies like VR/AR, digital twins, and 3D modeling play a critical role in connecting the virtual and physical worlds, enhancing user immersion and interaction in real-time.The intersection layer consists of three parts: digital twins, immersive user experience, and 3D modeling interface.
	
	\subsubsection{Digital Twins}
	
	Digital twins play a pivotal role in the metaverse by creating digital replicas of physical environments, enabling human users and their virtual avatars to interact, innovate, and simulate within these virtual worlds. SB-Metaverse must connect the real world and the virtual world, and digital twins helps achieve this \cite{fuller2020digital}. Digital twins are connected to real-time data streams collected from sensors and other devices, enabling the analysis and prediction of the behavior of real-world equivalents \cite{tao2018digital}. There are already many applications that construct virtual characters in the form of digital twins \cite{divya2024digital,jeyalakshmi2024digital}. Despite their potential, the current adoption of digital twins in the metaverse remains limited. Many existing metaverse applications have yet to incorporate digital twin technology, primarily due to technical and economic barriers. Key limitations include the high computational and data storage requirements for creating and maintaining real-time digital replicas, the need for robust and high-speed communication networks, and significant concerns surrounding data privacy and security \cite{balasubramaniam2024machine}. Epic Games, known for its virtual engine development, exemplifies the strides being made in digital twin applications \cite{lawrence2022epic}. The goal of Epic Games is to make it easy to capture the real world and then build it into virtual engine as a digital twin. Currently the developers of Epic Games are using digital twins to communicate between the physical and the virtual worlds. The developers have launched an app, MetaHuman Creator, which allows users to create lifelike avatars \cite{fang2021metahuman}. In the future, these virtual characters can even wear digital clothes designed by virtual clothing design companies \cite{de2021inspeccao}.

	\subsubsection{Immersive User Experience}
	
	From the early mouse and keyboard to the current VR/AR, these types of equipment have been continuously upgraded, and the immersion experience has continuously improved. In 2020, Valve released the VR game Half-Life: Alyx, where players can interact with almost any object in the game \cite{kwon2020study}. However, it only supports hand interaction with a certain degree of freedom. In this game, only the hands can be seen from the player’s perspective, which only achieves a preliminary and partial immersive user experience. In the movie Number One Players, players can enter the game through VR and wearable devices to experience real-time interactions. By collecting players’ information and outputting feedback in real time through multiple devices, players feel more real in the virtual space, thereby experiencing a full immersive user experience \cite{hu2019ready}. Moreover, the real and virtual worlds can be joined together by mixed-reality boundaries \cite{benford1998understanding}. In the future, through more interactive technologies, perceptual experiences such as smell and touch can be achieved along with interaction with the virtual world, and user immersion experience can be further improved.
	
	\subsubsection{3D Modeling Interface}
	In SB-Metaverse, character design and scene construction are inseparable from 3D modeling technology. Based on the requirements of SB-Metaverse, future modeling technologies should focus on immersive modeling, providing users with physically accurate and realistic visual effects. In addition, due to the huge metaverse scene, AI-based automated modeling technology is also an important research direction.
	
	\subsection{Ecosystem}
	
	The SB-Metaverse ecosystem provides users in the virtual world with a brand new experience that is different from the real world. The ecosystem layer enforces rules and operating procedures in SB-Metaverse, and it is the guarantee for maintaining the smooth progress of user activities. Ecosystem Layer aligns with the creator economy, virtual platforms, and experience layers and incorporates user-generated content (UGC), fair virtual economic systems, and personalized avatars. The ecosystem layer includes content creation, virtual economy, service provision and avatar. Next, we discuss content creation and virtual economy.
	
	\subsubsection{Content Creation}
	
	The content of SB-Metaverse is created in the form of User Generated Content \cite{wyrwoll2014user}. Everyone can become a content creator, users can create games according to their own creativity, which can be played by other users. The content created by the user has ownership and it is a digital asset owned by the user in the virtual world. Users can provide services to other users to obtain virtual currency, which can be exchanged for real currency.

	\subsubsection{Virtual Economy}
	A fair economic system must be established in SB-Metaverse, and all creators can participate in this economic system. The system must have rules to ensure that consumers are treated fairly, large-scale cheating, fraud or scams are avoided. Companies can freely publish and profit from the platform. Under the framework of virtual economy, items with ownership such as virtual currency and virtual real estate can be traded fairly.
	
	\begin{figure*}[!t]
		\centering
		\includegraphics[width=0.8\textwidth]{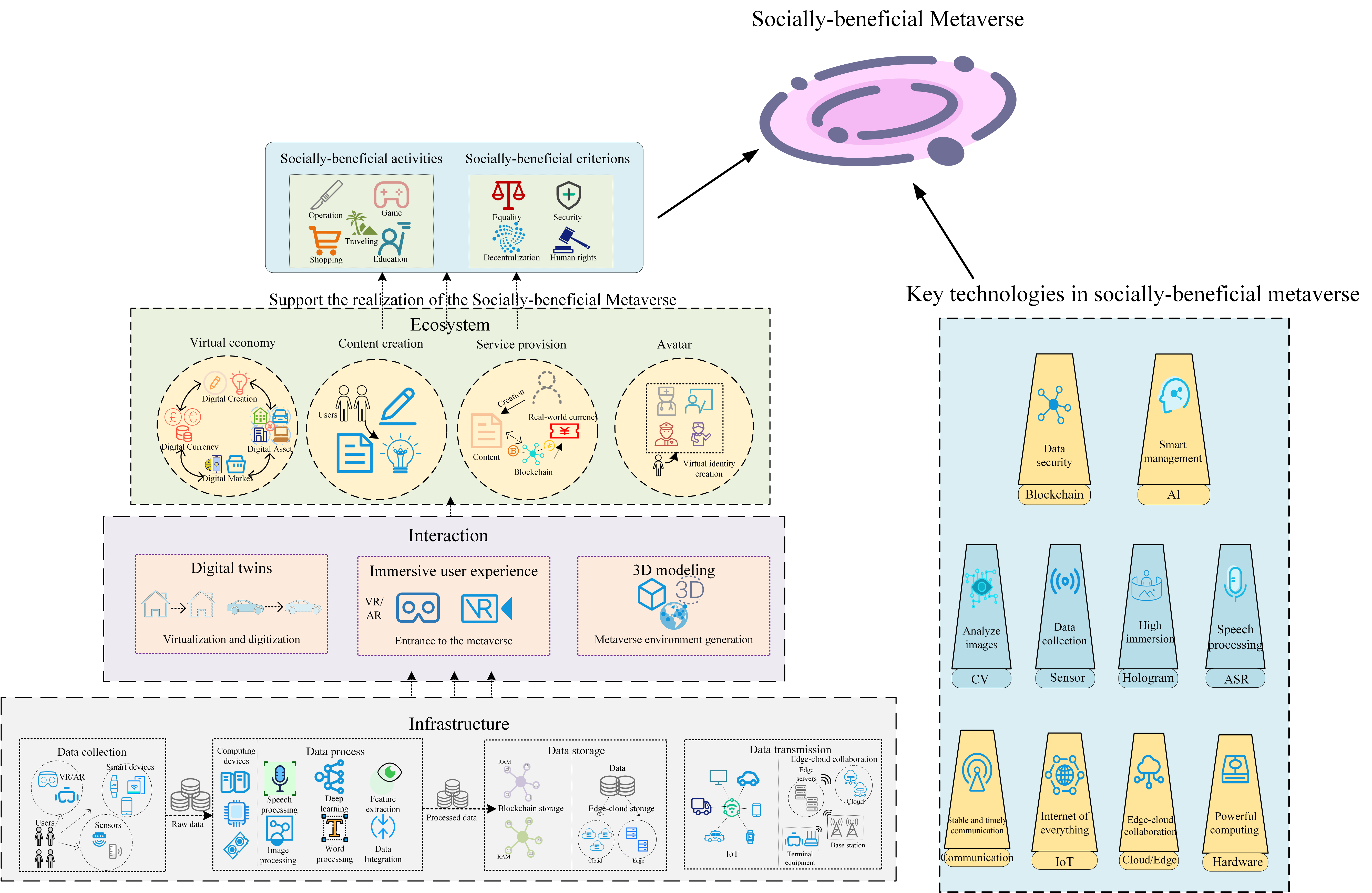}
		\caption{The framework of the socially beneficial metaverse.}
		\label{fig2}
	\end{figure*}
	
	Table I presents the characteristics of typical metaverse examples. 
	
		\begin{table*}
		\caption{ Characteristics of typical metaverse examples}\label{table 1}
		\resizebox{\textwidth}{!}{
		\begin{tabular}{|cc|lllll|ll|lll|}
			\hline
			\multicolumn{2}{|c|}{}                                                                                                                       & \multicolumn{5}{c|}{Infrastructure}                                                                                                                                                                                                                                                                                                           & \multicolumn{2}{c|}{Interaction}                                                                                  & \multicolumn{3}{c|}{Ecosystem}                                                                                                                                                                    \\ \cline{3-12} 
			\multicolumn{2}{|c|}{\multirow{-2}{*}{Metaverse examples}}                                                                                   & \multicolumn{1}{l|}{AI}                                & \multicolumn{1}{l|}{\begin{tabular}[c]{@{}l@{}}communi-\\ cation\end{tabular}} & \multicolumn{1}{l|}{\begin{tabular}[c]{@{}l@{}}Computer \\ graphics\end{tabular}} & \multicolumn{1}{l|}{3D}                                & \begin{tabular}[c]{@{}l@{}}Block-\\ chain\end{tabular} & \multicolumn{1}{l|}{AR/VR}                             & \begin{tabular}[c]{@{}l@{}}Digital \\ twins\end{tabular} & \multicolumn{1}{l|}{Avatar}                            & \multicolumn{1}{l|}{\begin{tabular}[c]{@{}l@{}}Content \\ creation\end{tabular}} & \begin{tabular}[c]{@{}l@{}}Econo-\\ mics\end{tabular} \\ \hline
			\multicolumn{1}{|c|}{}                                                                                                     & MUD             & \multicolumn{1}{l|}{{\color[HTML]{FE0000} \textbf{×}}} & \multicolumn{1}{l|}{{\color[HTML]{009901} \checkmark}}                                  & \multicolumn{1}{l|}{{\color[HTML]{FE0000} \textbf{×}}}                            & \multicolumn{1}{l|}{{\color[HTML]{FE0000} \textbf{×}}} & {\color[HTML]{FE0000} \textbf{×}}                      & \multicolumn{1}{l|}{{\color[HTML]{FE0000} \textbf{×}}} & {\color[HTML]{FE0000} \textbf{×}}                        & \multicolumn{1}{l|}{{\color[HTML]{009901} \checkmark}}          & \multicolumn{1}{l|}{{\color[HTML]{009901} \checkmark}}                                    & {\color[HTML]{FE0000} \textbf{×}}                     \\ \cline{2-12} 
			\multicolumn{1}{|c|}{}                                                                                                     & Clossal cave    & \multicolumn{1}{l|}{{\color[HTML]{FE0000} \textbf{×}}} & \multicolumn{1}{l|}{{\color[HTML]{FE0000} \textbf{×}}}                         & \multicolumn{1}{l|}{{\color[HTML]{FE0000} \textbf{×}}}                            & \multicolumn{1}{l|}{{\color[HTML]{FE0000} \textbf{×}}} & {\color[HTML]{FE0000} \textbf{×}}                      & \multicolumn{1}{l|}{{\color[HTML]{FE0000} \textbf{×}}} & {\color[HTML]{FE0000} \textbf{×}}                        & \multicolumn{1}{l|}{{\color[HTML]{FE0000} \textbf{×}}} & \multicolumn{1}{l|}{{\color[HTML]{009901} \textbf{\checkmark}}}                           & {\color[HTML]{FE0000} \textbf{×}}                     \\ \cline{2-12} 
			\multicolumn{1}{|c|}{\multirow{-3}{*}{\begin{tabular}[c]{@{}c@{}}Text-based \\ interactive games\end{tabular}}}            & Habitat         & \multicolumn{1}{l|}{{\color[HTML]{FE0000} \textbf{×}}} & \multicolumn{1}{l|}{{\color[HTML]{009901} \textbf{\checkmark}}}                         & \multicolumn{1}{l|}{{\color[HTML]{FE0000} \textbf{×}}}                            & \multicolumn{1}{l|}{{\color[HTML]{FE0000} \textbf{×}}} & {\color[HTML]{FE0000} \textbf{×}}                      & \multicolumn{1}{l|}{{\color[HTML]{FE0000} \textbf{×}}} & {\color[HTML]{FE0000} \textbf{×}}                        & \multicolumn{1}{l|}{{\color[HTML]{009901} \textbf{\checkmark}}} & \multicolumn{1}{l|}{{\color[HTML]{009901} \textbf{\checkmark}}}                           & {\color[HTML]{FE0000} \textbf{×}}                     \\ \hline
			\multicolumn{1}{|c|}{}                                                                                                     & Active world    & \multicolumn{1}{l|}{{\color[HTML]{009901} \textbf{\checkmark}}} & \multicolumn{1}{l|}{{\color[HTML]{009901} \textbf{\checkmark}}}                         & \multicolumn{1}{l|}{{\color[HTML]{009901} \textbf{\checkmark}}}                            & \multicolumn{1}{l|}{{\color[HTML]{009901} \textbf{\checkmark}}} & {\color[HTML]{FE0000} \textbf{×}}                      & \multicolumn{1}{l|}{{\color[HTML]{FE0000} \textbf{×}}} & {\color[HTML]{009901} \textbf{\checkmark}}                        & \multicolumn{1}{l|}{{\color[HTML]{009901} \textbf{\checkmark}}} & \multicolumn{1}{l|}{{\color[HTML]{009901} \textbf{\checkmark}}}                           & {\color[HTML]{009901} \textbf{\checkmark}}                     \\ \cline{2-12} 
			\multicolumn{1}{|c|}{}                                                                                                     & Onlive traveler & \multicolumn{1}{l|}{{\color[HTML]{FE0000} \textbf{×}}} & \multicolumn{1}{l|}{{\color[HTML]{009901} \textbf{\checkmark}}}                         & \multicolumn{1}{l|}{{\color[HTML]{009901} \textbf{\checkmark}}}                            & \multicolumn{1}{l|}{{\color[HTML]{009901} \textbf{\checkmark}}} & {\color[HTML]{FE0000} \textbf{×}}                      & \multicolumn{1}{l|}{{\color[HTML]{FE0000} \textbf{×}}} & {\color[HTML]{FE0000} \textbf{×}}                        & \multicolumn{1}{l|}{{\color[HTML]{009901} \textbf{\checkmark}}} & \multicolumn{1}{l|}{{\color[HTML]{009901} \textbf{\checkmark}}}                           & {\color[HTML]{009901} \textbf{\checkmark}}                     \\ \cline{2-12} 
			\multicolumn{1}{|c|}{\multirow{-3}{*}{\begin{tabular}[c]{@{}c@{}}3D-graphics-based \\ interactive games\end{tabular}}}     & Web World       & \multicolumn{1}{l|}{{\color[HTML]{FE0000} \textbf{×}}} & \multicolumn{1}{l|}{{\color[HTML]{009901} \textbf{\checkmark}}}                         & \multicolumn{1}{l|}{{\color[HTML]{009901} \textbf{\checkmark}}}                            & \multicolumn{1}{l|}{{\color[HTML]{009901} \textbf{\checkmark}}} & {\color[HTML]{FE0000} \textbf{×}}                      & \multicolumn{1}{l|}{{\color[HTML]{FE0000} \textbf{×}}} & {\color[HTML]{FE0000} \textbf{×}}                        & \multicolumn{1}{l|}{{\color[HTML]{009901} \textbf{\checkmark}}} & \multicolumn{1}{l|}{{\color[HTML]{009901} \textbf{\checkmark}}}                           & {\color[HTML]{009901} \textbf{\checkmark}}                     \\ \hline
			\multicolumn{1}{|c|}{}                                                                                                     & Second life     & \multicolumn{1}{l|}{{\color[HTML]{009901} \textbf{\checkmark}}} & \multicolumn{1}{l|}{{\color[HTML]{009901} \textbf{\checkmark}}}                         & \multicolumn{1}{l|}{{\color[HTML]{009901} \textbf{\checkmark}}}                            & \multicolumn{1}{l|}{{\color[HTML]{009901} \textbf{\checkmark}}} & {\color[HTML]{FE0000} \textbf{×}}                      & \multicolumn{1}{l|}{{\color[HTML]{FE0000} \textbf{×}}} & {\color[HTML]{FE0000} \textbf{×}}                        & \multicolumn{1}{l|}{{\color[HTML]{009901} \textbf{\checkmark}}} & \multicolumn{1}{l|}{{\color[HTML]{009901} \textbf{\checkmark}}}                           & {\color[HTML]{009901} \textbf{\checkmark}}                     \\ \cline{2-12} 
			\multicolumn{1}{|c|}{}                                                                                                     & Minecraft       & \multicolumn{1}{l|}{{\color[HTML]{009901} \textbf{\checkmark}}} & \multicolumn{1}{l|}{{\color[HTML]{009901} \textbf{\checkmark}}}                         & \multicolumn{1}{l|}{{\color[HTML]{009901} \textbf{\checkmark}}}                            & \multicolumn{1}{l|}{{\color[HTML]{009901} \textbf{\checkmark}}} & {\color[HTML]{FE0000} \textbf{×}}                      & \multicolumn{1}{l|}{{\color[HTML]{009901} \textbf{\checkmark}}} & {\color[HTML]{FE0000} \textbf{×}}                        & \multicolumn{1}{l|}{{\color[HTML]{009901} \textbf{\checkmark}}} & \multicolumn{1}{l|}{{\color[HTML]{009901} \textbf{\checkmark}}}                           & {\color[HTML]{009901} \textbf{\checkmark}}                     \\ \cline{2-12} 
			\multicolumn{1}{|c|}{}                                                                                                     & Roblox          & \multicolumn{1}{l|}{{\color[HTML]{009901} \textbf{\checkmark}}} & \multicolumn{1}{l|}{{\color[HTML]{009901} \textbf{\checkmark}}}                         & \multicolumn{1}{l|}{{\color[HTML]{009901} \textbf{\checkmark}}}                            & \multicolumn{1}{l|}{{\color[HTML]{009901} \textbf{\checkmark}}} & {\color[HTML]{FE0000} \textbf{×}}                      & \multicolumn{1}{l|}{{\color[HTML]{009901} \textbf{\checkmark}}} & {\color[HTML]{FE0000} \textbf{×}}                        & \multicolumn{1}{l|}{{\color[HTML]{009901} \textbf{\checkmark}}} & \multicolumn{1}{l|}{{\color[HTML]{009901} \textbf{\checkmark}}}                           & {\color[HTML]{009901} \textbf{\checkmark}}                     \\ \cline{2-12} 
			\multicolumn{1}{|c|}{}                                                                                                     & Pokémon go      & \multicolumn{1}{l|}{{\color[HTML]{009901} \textbf{\checkmark}}} & \multicolumn{1}{l|}{{\color[HTML]{009901} \textbf{\checkmark}}}                         & \multicolumn{1}{l|}{{\color[HTML]{009901} \textbf{\checkmark}}}                            & \multicolumn{1}{l|}{{\color[HTML]{009901} \textbf{\checkmark}}} & {\color[HTML]{FE0000} \textbf{×}}                      & \multicolumn{1}{l|}{{\color[HTML]{009901} \textbf{\checkmark}}} & {\color[HTML]{009901} \textbf{\checkmark}}                        & \multicolumn{1}{l|}{{\color[HTML]{009901} \textbf{\checkmark}}} & \multicolumn{1}{l|}{{\color[HTML]{FE0000} \textbf{×}}}                           & {\color[HTML]{009901} \textbf{\checkmark}}                     \\ \cline{2-12} 
			\multicolumn{1}{|c|}{\multirow{-5}{*}{\begin{tabular}[c]{@{}c@{}}Massive multi-player \\ online video games\end{tabular}}} & Animal crossing & \multicolumn{1}{l|}{{\color[HTML]{009901} \textbf{\checkmark}}} & \multicolumn{1}{l|}{{\color[HTML]{009901} \textbf{\checkmark}}}                         & \multicolumn{1}{l|}{{\color[HTML]{009901} \textbf{\checkmark}}}                            & \multicolumn{1}{l|}{{\color[HTML]{009901} \textbf{\checkmark}}} & {\color[HTML]{FE0000} \textbf{×}}                      & \multicolumn{1}{l|}{{\color[HTML]{009901} \textbf{\checkmark}}} & {\color[HTML]{FE0000} \textbf{×}}                        & \multicolumn{1}{l|}{{\color[HTML]{009901} \textbf{\checkmark}}} & \multicolumn{1}{l|}{{\color[HTML]{009901} \textbf{\checkmark}}}                           & {\color[HTML]{009901} \textbf{\checkmark}}                     \\ \hline
			\multicolumn{1}{|c|}{}                                                                                                     & Decentraland    & \multicolumn{1}{l|}{{\color[HTML]{009901} \textbf{\checkmark}}} & \multicolumn{1}{l|}{{\color[HTML]{009901} \textbf{\checkmark}}}                         & \multicolumn{1}{l|}{{\color[HTML]{009901} \textbf{\checkmark}}}                            & \multicolumn{1}{l|}{{\color[HTML]{009901} \textbf{\checkmark}}} & {\color[HTML]{009901} \textbf{\checkmark}}                      & \multicolumn{1}{l|}{{\color[HTML]{FE0000} \textbf{×}}} & {\color[HTML]{FE0000} \textbf{×}}                        & \multicolumn{1}{l|}{{\color[HTML]{009901} \textbf{\checkmark}}} & \multicolumn{1}{l|}{{\color[HTML]{009901} \textbf{\checkmark}}}                           & {\color[HTML]{009901} \textbf{\checkmark}}                     \\ \cline{2-12} 
			\multicolumn{1}{|c|}{}                                                                                                     & Axie infinity   & \multicolumn{1}{l|}{{\color[HTML]{009901} \textbf{\checkmark}}} & \multicolumn{1}{l|}{{\color[HTML]{009901} \textbf{\checkmark}}}                         & \multicolumn{1}{l|}{{\color[HTML]{009901} \textbf{\checkmark}}}                            & \multicolumn{1}{l|}{{\color[HTML]{009901} \textbf{\checkmark}}} & {\color[HTML]{009901} \textbf{\checkmark}}                      & \multicolumn{1}{l|}{{\color[HTML]{009901} \textbf{\checkmark}}} & {\color[HTML]{FE0000} \textbf{×}}                        & \multicolumn{1}{l|}{{\color[HTML]{009901} \textbf{\checkmark}}} & \multicolumn{1}{l|}{{\color[HTML]{009901} \textbf{\checkmark}}}                           & {\color[HTML]{009901} \textbf{\checkmark}}                     \\ \cline{2-12} 
			\multicolumn{1}{|c|}{\multirow{-3}{*}{\begin{tabular}[c]{@{}c@{}}Decentralized \\ virtual world\end{tabular}}}             & The sandbox     & \multicolumn{1}{l|}{{\color[HTML]{009901} \textbf{\checkmark}}} & \multicolumn{1}{l|}{{\color[HTML]{009901} \textbf{\checkmark}}}                         & \multicolumn{1}{l|}{{\color[HTML]{009901} \textbf{\checkmark}}}                            & \multicolumn{1}{l|}{{\color[HTML]{009901} \textbf{\checkmark}}} & {\color[HTML]{009901} \textbf{\checkmark}}                      & \multicolumn{1}{l|}{{\color[HTML]{009901} \textbf{\checkmark}}} & {\color[HTML]{009901} \textbf{\checkmark}}                        & \multicolumn{1}{l|}{{\color[HTML]{009901} \textbf{\checkmark}}} & \multicolumn{1}{l|}{{\color[HTML]{009901} \textbf{\checkmark}}}                           & {\color[HTML]{009901} \textbf{\checkmark}}                     \\ \hline
		\end{tabular}}
	\end{table*}

	\section{Key Technologies of Socially Beneficial Metaverse}
	
	In this section, we describe six technologies that support the operation and development of socially beneficial metaverse which include communication, hardware, artificial intelligence, cloud computing and edge computing, Internet of Things and Blockchain.
	
	\subsection{Communication}

	Considering that SB-Metaverse should be accessible at any time and from any place, we consider communication as an important basic technology. Since SB-Metaverse is characterized by low delay and high immersion \cite{power2013postcards}, we need low latency in network transmissions to improve service satisfaction and immediacy as well as broad coverage to ensure real-time interactions for remote users \cite{liu2018energy}. In addition, in order for users to feel truly immersed, devices need higher resolution and frame rates to transmit high-resolution content in real time. However, the throughput of multimedia applications in SB-Metaverse will increase significantly \cite{montazerolghaem2021software} and recent research showed that the tolerable delay in ordinary games will affect users’ experiences in SB-Metaverse applications \cite{ivkovic2015quantifying}. Moreover, it was shown that for devices like VR, only latency as low as milliseconds can guarantee the user’s immersive experience \cite{lincoln2016motion}. Fortunately, the development of 5G has brought possible solutions to the challenges mentioned above \cite{shafi20175g}. The 5G network has the advantages of high speed, low latency and wide outdoor coverage. It can provide an average experience rate of 1Gbps, a peak rate of 10Gbps, more than 1 million connections per square kilometer and an ultra-low air interface delay of 1ms, which are suitable for different application scenarios \cite{al2017technologies}. Therefore, 5G has become a key enabling technology of SB-Metaverse \cite{lee2021all}.

	SB-Metaverse is composed of many different smart devices, so they will form a huge heterogeneous network \cite{zhang2015cloud}. In 5G communications, by deploying small units with low power consumption to heterogeneous networks, network capacity is increased and coverage is expanded. \cite{agiwal2016next}. Several studies have been carried to improve the performance of heterogeneous networks. Lee et al. \cite{lee2014recent} solved the problem of cross-layer and co-layer interference in 5G heterogeneous networks , which can be used to improve the performance of communication between devices in SB-Metaverse. Talwar et al. \cite{talwar2014enabling} proposed an efficient allocation method to manage public resources. Since the network resources required by users are often different and dynamically changing, the implementation of this method will significantly improve the utilization of network resources, so as to meet users’ requirements.

	In addition, users can enter SB-Metaverse anytime and anywhere, which requires high quality of indoor and outdoor network communication in the real world. The indoor wireless network has the characteristics of small range and relatively stable communication quality, so it is relatively easy to achieve the entry of SB-Metaverse in the indoor wireless network environment \cite{dhawankar2021next}. The outdoor wireless network has a large range and it is easily affected by various environmental factors such as precipitation and temperature \cite{federici2016review} \cite{marfievici2013environmental}, which will cause instability in the network communication quality. So, it is important to meet the communication requirements in an outdoor environment. Fortunately, studies on improving the quality of outdoor communication networks through 5G have developed promising solutions to this problem. Wang et al. \cite{wang2020optimizing} combined Geographic Information Systems (GIS) and optimization algorithms to explicitly simulate the propagation of 5G signals, which has been proven to significantly optimize outdoor 5G service coverage. Kumari et al. \cite{kumari2020optimization} combined millimeter wave channel modeling and machine learning, which managed to improve the channel stability of 5G outdoor communications.

	However, these methods can only meet the basic requirements of SB-Metaverse’s communication, and there is still a gap between achieving the full potential of SB-Metaverse and the accessibility. Therefore, in the future, significantly improving the performance of outdoor 5G networks is an important aspect for enabling full accessibility of SB-Metaverse.

	\subsection{Hardware}
	Hardware technology is also one of the key technologies of SB-Metaverse. The hardware supports the creation and experience of the virtual content. To achieve more realistic modeling and interaction we need better hardware performance as a prerequisite. Gordon Moore et al. \cite{moore1965moore} argued that the number of components that can be accommodated on an integrated circuit doubles about every 18-24 months which will double the performance. The integrated circuit directly affects the performance of the central processing unit. Hardware performance continues to improve while hardware costs continue to decrease. However, the metaverse will continue to have the highest computing power requirements in human history, and computing resources will continue to be scarce. The performance of hardware limits and define the metaverse. Therefore, in this section, we introduce the impact of hardware on SB-Metaverse.

	The development of hardware for the SB-Metaverse must be done from the following two perspectives: the improvement of hardware performance and the reduction of hardware computing cost.

	\begin{itemize}
		\item Improvement of hardware performance:
		
		The improvement of physical hardware performance can be realized by the following methods. First, we must use new materials to improve hardware performance. Torrejon et al. \cite{torrejon2017neuromorphic} used the ability of electronic spin oscillators to interact with each other to improve the performance of parallel computing based on oscillator networks. Vandoorne et al. \cite{vandoorne2014experimental} used the photonic module to calculate the reservoir on the silicon photonics core, which significantly improves the calculation speed. Second, we must change the structural arrangement of these new devices to enhance hardware performance. Zhong et al. \cite{zhong2021dynamic} proposed an integrated storage-calculation system based on multiple memristor arrays, which significantly increases the computing power of computing equipment, and it also successfully achieved a smaller power consumption and lower hardware cost to complete complex calculations.

		\item Reduction of hardware computing cost:
		
		Microsoft proposed a model called GPT-3 in 2020, which has many human-computer interaction functions and is considered to be the most versatile AI model \cite{dale2021gpt}. But Microsoft built a 500 million dollar supercomputing center to train it, and the training cost alone was more than 4.6 million dollars. It is very costly to develop such computing power. For most companies and some countries, such economic and technical costs are too high, which violate the antitrust and decentralization requirements of SB-Metaverse. Therefore, reducing hardware costs is an important research direction. The large-scale model M6 with the world’s largest parameters developed by Alibaba Dharma Institute uses a low-carbon and high-computing algorithm model to increase the computing power and reduce carbon emissions \cite{lin2021m6}. In addition, through technological innovation on the cloud, the computing power of the data center has been more reasonably configured. The use of the database PolarDB has improved a lot the efficiency of server resource utilization and reduced hardware computing costs \cite{cao2020polardb}.

	\end{itemize}
	
	\subsection{Artificial Intelligence}
	
	Artificial intelligence (AI) is a new technical science that studies theories, methods, technologies and application systems for simulating, extending and expanding human intelligence \cite{russell2002artificial} \cite{thiebes2021trustworthy}. In recent years, the importance of artificial intelligence has accelerated, and it is currently playing an important role in the fields of autopilot \cite{rao2018deep}, face recognition \cite{li2020review}, intelligent medical diagnosis \cite{park2018methodologic} and other fields. In the architecture of SB-Metaverse, artificial intelligence technology will be equally important as it can provide functions such as intelligent system maintenance and interaction. It could also provide players with high-quality experience and services. In this section, we describe the important role of AI in SB-Metaverse from two perspectives: automatic content creation, intelligent system maintenance.
	
	\begin{itemize}
		\item Automatic content creation:
		
		In SB-Metaverse, the richness of content will be far beyond imagination. Moreover, the content will be provided to users in the form of real-time generation, real-time experience and real-time feedback. The requirements for content will far exceed the reach of manpower, and artificial intelligence technology is needed to help produce content and lower the threshold for user content creation.

		Automatic content creation by AI is achieved through the early analysis and learning of massive amounts of data, so that it can create some identical or even innovative content independently and quickly. This technique is known in the gaming world as procedural content generation \cite{shaker2016procedural}. From the simplest Tetris to open games like Minecraft, procedural content generation has been deployed in many game applications \cite{xu2016automatic}. In the space adventure survival theme game ``No Man’s Sky’’ \cite{games2016no} \cite{drago2019no}, from Non-player characters (NPCs), planetary environments, to spaceships, and even music are procedurally generated. These NPCs automatically generated by AI are different from traditional NPCs which only interact with human players based on codes determined by established conditions. Instead, they adjust their behaviors and methods by observing the actions and strategies of human players. This feature not only greatly reduces the repetitive work of human coding, but also improves the real-time nature of human-computer interaction and improves the immersive experience of users. It is worth noting that to support the increasing content needs of SB-Metaverse, the use of AI technology is vital.

		\item Intelligent system maintenance:
		
		Another important function of AI in SB-Metaverse is intelligent system maintenance. Intelligent system maintenance refers to the artificial intelligence model which not only imitate human operations, but also learn optimal strategies and automatically find bugs in the digital environment. This technology, which uses AI technology to find and fix bugs, is being developed and applied \cite{kobbacy2012application}.

		Meta has developed AI-based code-fixing tools Sapienz and SapFix \cite{marginean2019sapfix}. Sapienz is an artificial intelligence-based black-box testing tool. Its biggest advantage is that it can automatically generate test cases, conduct tests and results verification \cite{mao2016sapienz}. SapFix is an AI-based BUG fix tool \cite{marginean2019sapfix}. Metaverse requires humans to check, verify, and confirm that it works correctly. This kind of work is more suitable for artificial intelligence to complete. Due to the sheer size of the metaverse, it is difficult to check codes using traditional methods such as manual code review. At this time, it is possible to automatically detect vulnerabilities in the metaverse by using an artificial intelligence model that combines deep learning and reinforcement learning, thus reducing the workload of testing codes \cite{diaconescu2013automated}\cite{bassuday2019fault}.In the game The Witness, the developers used AI to walk around the entire game island \cite{fletcher2018witness}. Once they found some situations similar to ``passing through wall’’ that did not conform to the physical world, they would record them and give feedback to the developers \cite{bonner2016puzzle}.

		In addition to finding bugs in the game, AI in The Witness can also test and analyze the difficulties of the game in order to enhance the playability of the game. This is also a function that the AI in SB-Metaverse needs to focus on. User experience will be a key success factor for a mega-environment like the SB-Metaverse that fully simulates the real world. But The witness only implements code review, not automatic code modification. Since the amount of modification in a traditional game is relatively small, manual modification methods can be used. However, the amount of code in the metaverse is huge, and it is very likely that an error in one place must modify the code in several places at the same time. Therefore, AI in The witness only provides a good example for using AI for maintenance. Future research efforts should focus on the research of AI-based automatic review and modification technology that can handle large amounts of code, and aim to improve the accuracy and timeliness of this technology in order to satisfy the needs of SB-Metaverse.
		
	\end{itemize}
	
	\subsection{Cloud Computing and Edge Computing}
	In this section, we describe the important roles of cloud computing and edge computing for SB-Metaverse.
	
	\subsubsection{Cloud Computing}
	
	Cloud computing is a type of distributed computing, which could automatically split a huge computing processing program into countless smaller subroutines through the network. Then it submits the subroutines to a large system composed of multiple servers after searching available resources, performing calculation tasks, and analyzing the calculation results. Finally, the processing results are sent back to the users \cite{hayes2008cloud} \cite{sadiku2014cloud}. As SB-Metaverse needs computing, storage and so on, all of which are inseparable from cloud computing. In addition, SB-Metaverse will inevitably generate a huge amount of data. This data is inseparable from the underlying platform, and the cloud is the most appropriate platform for this data. Therefore, SB-Metaverse must be a field that cloud computing can empower on a large scale. Cloud computing can improve the user experience in the following ways:

	\begin{itemize}
		\item Enhanced reliability:
		
		Since SB-Metaverse needs to realize accessibility, it is necessary to set up a large number of servers to cover a wider area. However, with the increase in the number of servers, the complexity of communication between each other increases. When the amount of computation required by SB-Metaverse equipment increases, the possibility of failure or overload in a server also increases. Traditional distributed computing cannot make adjustments on time, which will affect user’s service experience. The use of cloud computing can effectively address this challenge. Cloud computing will use its elasticity to arrange new servers for computing when a server crashes, thus improving the reliability of service quality \cite{antonescu2012dynamic}.  Deng et al. \cite{deng2010fault} proposed a cloud selection strategy which decomposes the matrix multiplication problem into multiple tasks that are submitted to different clouds. In SB-Metaverse, these strategies can handle error clouds or malicious clouds, which improves the stability of the service. In addition, in terms of data transmission reliability, Liu et al. \cite{liu2017achieving} also proposed a comprehensive transmission model to achieve reliable data transmission under cloud computing by combining client/server mode and peer-to-peer mode \cite{liu2017achieving}. Therefore, cloud computing can play an important role in the improvement of service quality and reliability of data transmission in SB-Metaverse.

		\item Reduced energy consumption:
		
		Due to the huge amount of computation in SB-Metaverse, reducing server energy consumption and efficiently allocating computing resources are also major challenges that must be addressed. Cloud computing also provides possible solutions in this context \cite{hameed2016survey}. Bui et al. \cite{bui2017energy} proposed an efficient solution to coordinate computing resources based on convex optimization technology, which has achieved remarkable results in reducing energy consumption and maintaining high system performance. Since the distance between servers in SB-Metaverse may be large, data is transmitted through the network to geographically separated storage or data resources, and the servers include many data-intensive applications, the energy consumption of data transmission and processing will be high. Yang et al. \cite{yang2010sevice} proposed a cloud infrastructure service framework (CISF) to achieve guaranteed service of data-intensive applications, which significantly reduces redundant and repetitive communication calculations while optimizing the resource allocation. In addition, Banzai et al. proposed a software testing environment D-Cloud \cite{banzai2010d} which is based on cloud computing technology. D-Cloud significantly reduces the energy consumption of software testing. This technology improves the test efficiency and is suitable for SB-Metaverse with many large-scale tests. The research results dmonstrate the use of cloud computing in reducing energy consumption in SB-Metaverse.
		
	\end{itemize}
	
	In addition, cloud computing has a significant impact on improving service quality \cite{ardagna2014quality}, enhancing compatibility \cite{moqbel2014study} and facilitating human-computer interaction \cite{paul2018novel}. Therefore, cloud computing is an important underlying technology that supports the socially beneficial metaverse.

	\subsubsection{Edge computing}
	Edge computing refers to processing, analyzing and storing data closer to where it is generated. It enables fast real-time analysis and response \cite{shi2016edge}. Edge computing and cloud computing are synergistic with each other, and they are complementary to each other \cite{shi2016promise}. The combination of edge computing and cloud computing can provide low-latency, high-security and standard-open distributed cloud services for terminals \cite{peng2018survey} . Edge computing can adequately solve the problems of central traffic congestion and lack of computing resources caused by the explosive growth of terminals, and is an important path to solve future digital problems \cite{dolui2017comparison} \cite{lin2020survey}. Fig. 3 shows us the framework of edge-cloud collaboration in the socially beneficial metaverse. The massive amount of data in SB-Metaverse requires edge computing to be processed, and high-quality interactive experience requires the help of edge computing power. The traditional cloud computing model faces several challenges which include: user immersion, user privacy, real-time user interaction, excessive computing center load, and so on. We will select two of them to illustrate the benefits that edge computing can bring about.

	\begin{figure*}[!t]
		\centering
		\includegraphics[width=0.6\textwidth]{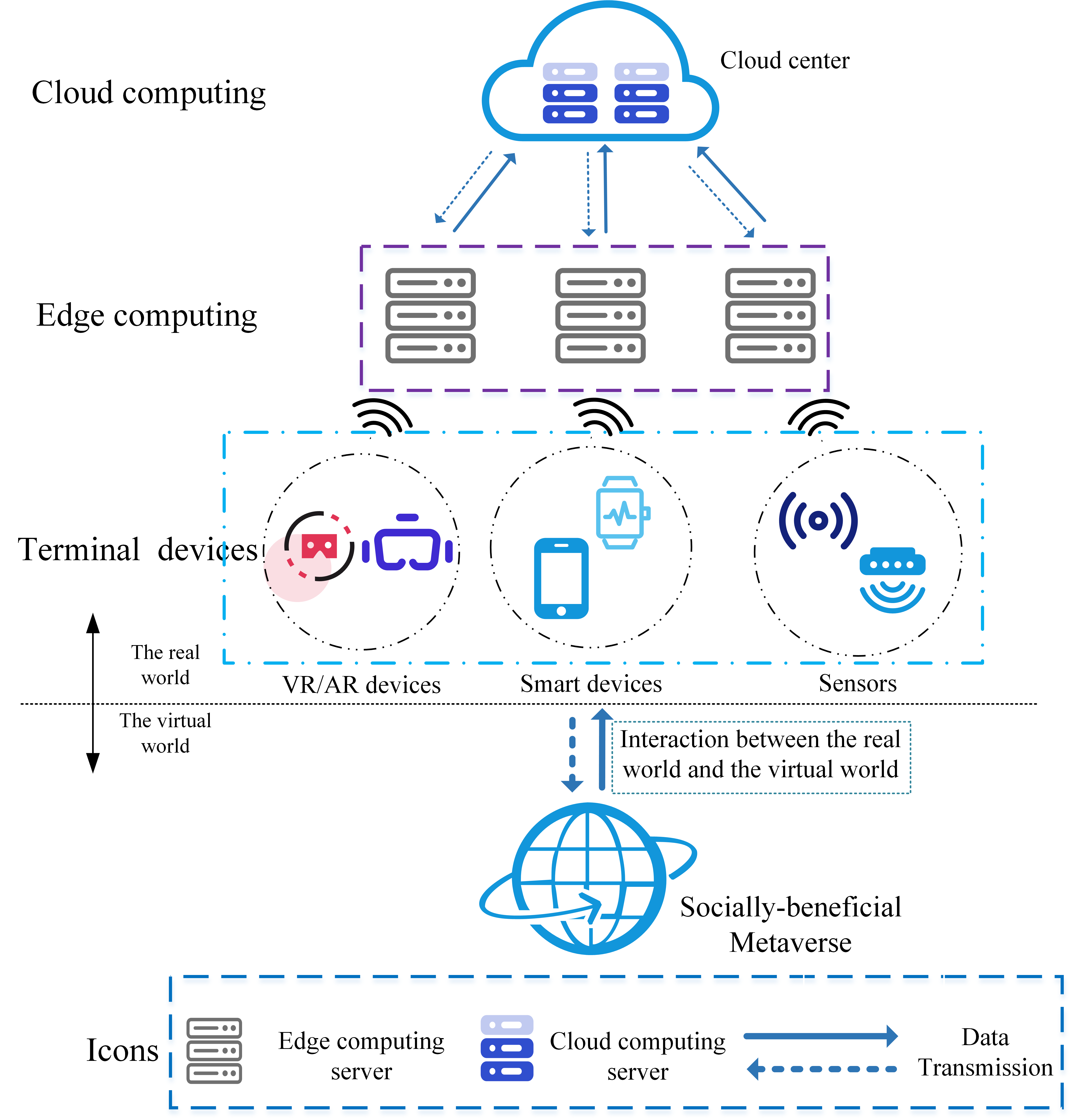}
		\caption{The framework of edge-cloud collaboration in the socially beneficial metaverse.}
		\label{fig3}
	\end{figure*}
	
	\begin{itemize}
		
		\item User Immersion:
		
		Immersion is one of the key elements of SB-Metaverse, and delay is one of the important factors that affect user’s immersion. Therefore, only by reducing the delay can users have a truly immersive experience. Ha et al. \cite{ha2014towards} found that compared with cloud computing, edge computing can reduce the delay by at least 80ms, and such optimization can bring a better sense of experience to users. Xiao et al. \cite{xiao2020system} considered multi-user and multi-server MEC systems wherein each user can choose an MEC server to perform its computing tasks. The proposed algorithm not only achieves lower delay, but also consumes energy. Niu et al. \cite{niu2019workload} proposed a workload distribution mechanism to target edge computing-based power-consuming IoT to minimize latency. The results showed that this approach can reduce latency by up to 30 per cent. Additionally, mobile edge computing can also provide low latency when the user’s geographic location changes rapidly. Zhao et al. \cite{zhao2017tasks} addressed the problem of service quality degradation when computing-intensive tasks are offloaded to the edge cloud. By coordinating heterogeneous clouds including edge cloud and remote cloud, the quality of service is enhanced and the delay is reduced. Mao et al. proposed an online computing offload and resource allocation algorithm based on Lyapunov optimization theory \cite{mao2019energy}, which reduces the communication delay of mobile edge computing and brings a stable user experience to mobile users. These past research results solves the problem of high latency caused by transmitting data to the cloud computing center in traditional cloud computing environments and reduces the latency of mobile users through mobile edge computing, thus improving service quality and stability. Therefore, the combination of edge computing and cloud computing can be applied to SB-Metaverse to reduce latency and improve service quality.

		\item User privacy:
		
		A major feature of SB-Metaverse is to provide users with an identity that is different from the real world. Therefore, we need effective user privacy data protection mechanisms. Due to frequent communications between users, a large amount of information exchange brings challenges to privacy protection \cite{leenes2007privacy}. Edge computing can offload part of the data to perform computation locally and as a result it can protect user privacy and improve data security \cite{zhang2018data}. Li et al. \cite{li2018privacy} proposed a privacy-preserving data aggregation scheme for IoT applications, which not only ensures the data privacy of terminal devices but also provides source authentication and integrity. Gheisari et al. \cite{gheisari2020edge} proposed a framework for addressing heterogeneity and privacy preservation of edge devices using a new ontology data model. Thus, edge computing has the characteristic of offloading data locally, and the optimization of edge computing privacy protection by the above technologies can improve edge computing data security in SB-Metaverse. However, due to the small scale of edge computing servers, the computing power is relatively small. Thus, it is more vulnerable to attacks, and the security of data under this condition still needs to be addressed \cite{ranaweera2021survey}.
		
	\end{itemize}
	
	In addition, Mobile Edge Computing (MEC) is a network architecture that provides services and cloud computing wirelessly. It can accelerate the rapid download of various applications in the network, allowing users to enjoy uninterrupted high-quality network experience. MEC has the characteristics of ultra-low latency, ultra-high bandwidth and strong real-time performance. It also provides convenience for real-time orchestration of multi-user interactions \cite{mehrabi2019device}. Therefore, MEC is crucial for outdoor metaverse services to understand the detailed local environment and coordinate close collaboration between nearby users or devices, enabling real-time user interactions for social AR applications in SB-Metaverse. Fig. 4 shows the framework of AR interactive devices using MEC in socially -beneficial metaverse \cite{hu2015mobile}.
	
	\begin{figure}[!t]
		\centering
		\includegraphics[width=0.8\columnwidth]{ 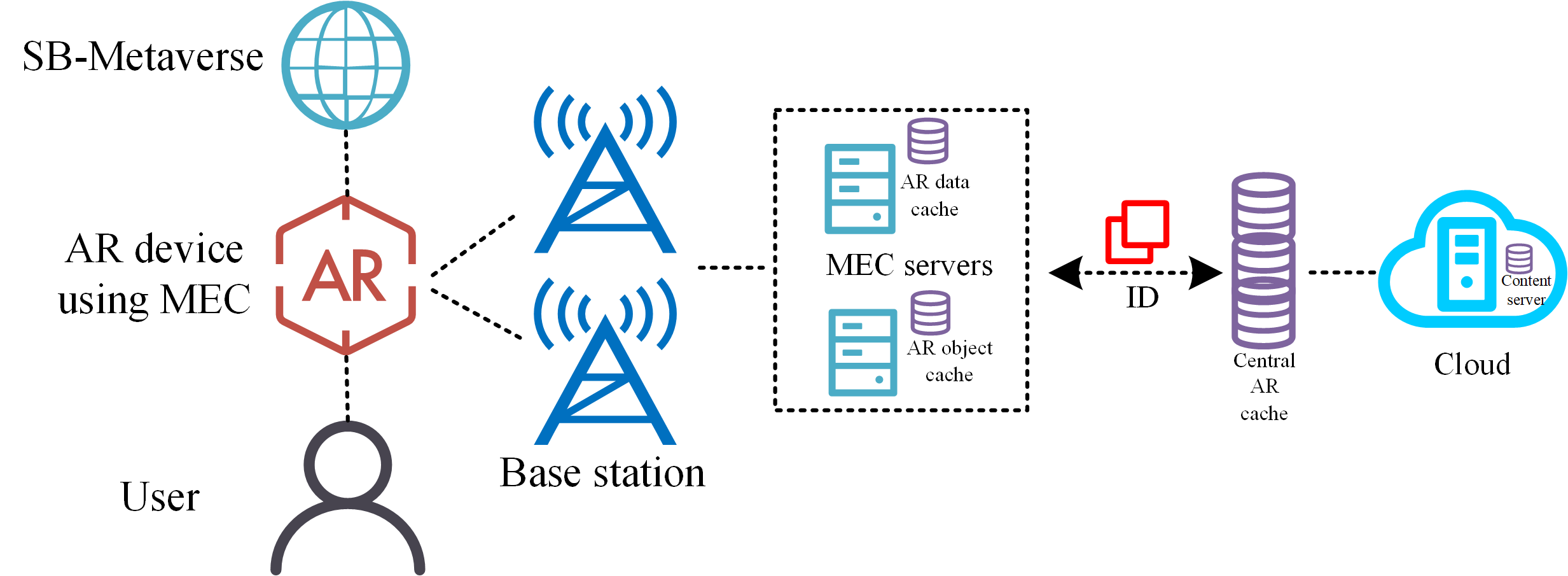}
		\caption{The framework of AR interactive devices using MEC in socially -beneficial metaverse.}
		\label{fig4}
	\end{figure}

	\subsection{Internet of Things}
	The Internet of Things (IoT) is an extension of the network based on the Internet. It is a huge network formed by integrating various information sensing devices with the network. It enables the interconnection of people, machines, and things at anytime, anywhere. IoT is the core underlying technology of SB-Metaverse, it provides reliable technical support for the connection of all things. If the intelligent wearable devices, intelligent transportation, and smart home in IoT are realized, then they will become integral components of SB-Metaverse \cite{ning2021survey}.
	IoT mainly provides the infrastructure for SB-Metaverse. Under the Internet of Things, the construction of these infrastructures must follow the following three principles:
\begin{itemize}
	\item SB-Metaverse infrastructure requires stakeholders to build together.
	\item SB-Metaverse infrastructure needs to enable two-way connectivity and closed-loop interactions between the digital and physical worlds.
	\item The builders of the SB-Metaverse infrastructure freely participate as independent digital identities and continuously form new consensus.
	\end{itemize}

	\subsection{Blockchain}
	 Since the SB-Metaverse world achieves the confirmation, pricing, transaction and empowerment of data, it can become a user-oriented, objective and open source artificial virtual parallel world. The virtual financial system is also similar to the real world, so reasonable virtual currency regulations and technology are needed to protect financial stability and security. Blockchain technology is a technical solution that does not rely on third parties and uses its own distributed characteristics to store, verify, transmit and communicate data. Blockchain plays an important role in improving security, so it is also a core underlying technology in the architecture of SB-Metaverse \cite{zheng2018blockchain}. Fig. 5 \cite{nguyen2021metachain} shows a novel blockchain-based framework for socially beneficial metaverse applications.
	
	\begin{figure}[!t]
		\centering
		\includegraphics[width=0.6\columnwidth]{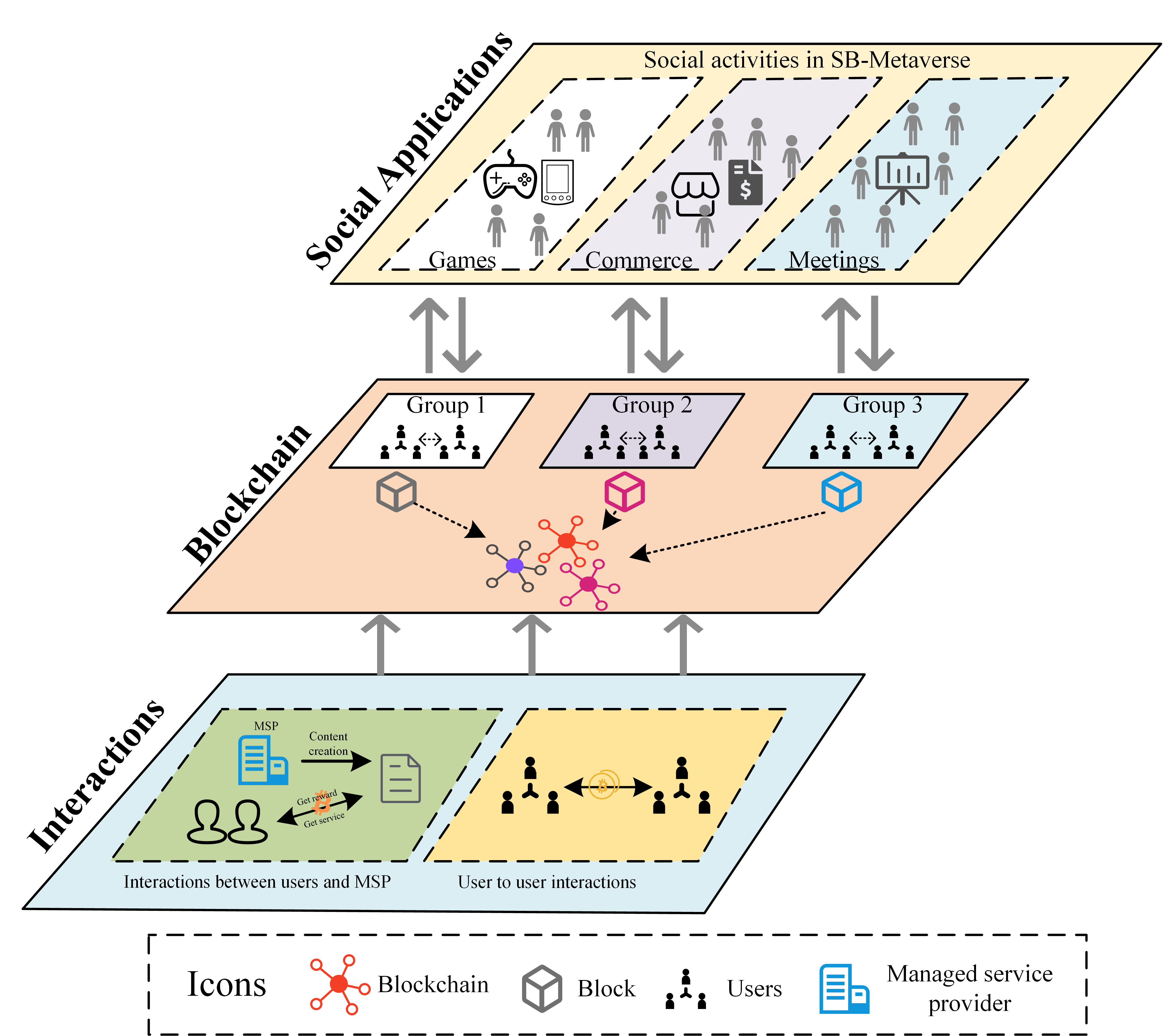}
		\caption{A novel blockchain-based framework for socially beneficial metaverse applications.}
		\label{fig5}
	\end{figure}

	\begin{itemize}
		\item Unique identifier:
		
		An important feature of SB-Metaverse is to provide each user with an identity unrelated to reality. Similar to the real world, the characters in SB-Metaverse should be unique. In the current network, a document can be easily copied, but such problems must be avoided in SB-Metaverse. With the help of blockchain, unique identifier can be achieved.
		There are already many organizations and companies implementing blockchain-based identities. Bitnation \cite{jacobovitz2016blockchain} is a governance platform powered by blockchain technology. With blockchain-based identity registration, anyone anywhere in the world can become a citizen of Bitnation by signing the constitution. With this platform, people can access social assistance and financial services \cite{jacobovitz2016blockchain}. ConsenSys is a blockchain software technology company that develops decentralized software services and applications, enabling decentralized governance \cite{mazzoni2021performance}.
		In SB-Metaverse, in addition to relying on the current traditional identity authentication system, our identity will also be connected to the blockchain identity authentication system in the future.
		Only by guaranteeing the uniqueness of the identity in the metaverse we can truly use it without worrying about our identity being copied. It is worth noting that only with this basic guarantee can SB-Metaverse truly develop.

		\item Data decentralization:
		
		To prevent the data of SB-Metaverse from being monopolized by large companies, decentralization is an important element which must be achieved. Decentralization removes power that is traditionally assigned to some central authority, and users can freely conduct peer-to-peer transactions \cite{pollitt2005decentralization}. This ensures that the data of the node is not be controlled by some central node thereby ensuring data security. Personal data is decentralized through blockchain technology, which enables data to be owned by individuals and cannot be tampered with or disposed of at will. If someone else wants to use personal data, that person needs authorization and must pay the corresponding fee. After using other blockchain encryption technologies, even obtaining information only needs to be carried out between the user and the information, and not between users, which further ensures the privacy of the information provider \cite{zhang2019security}.

		\item Asset digitization:
		
		Asset digitization is also an important aspect of the application of blockchain in SB-Metaverse. Assets in SB-Metaverse are recorded on the blockchain, enabling instant interaction while ensuring security. The establishment of the financial world in socially beneficial metaverse is different from that in the real world. Transactions in the real world can be carried out through the real world or digitization, but currency transactions in SB-Metaverse must be carried out through virtualization and asset digitization. As an immutable decentralized ledger, blockchain can digitize the assets and make instant transactions. Zhu et al. \cite{zhu2018digital} proposed a new digital asset management platform DAM-Chain which supports flexible rights management and transparent access authorization process. Harish et al. \cite{harish2021log} proposed Log-Flock, a financing platform based on blockchain technology, which supports an incentive reward mechanism to encourage users to participate. The implementation of these platforms will provide broad application prospects for the digitization of blockchain-based assets in SB-Metaverse.
		
	\end{itemize}
	
	The characteristics of decentralization and unique identification of the blockchain make it an indispensable element in the SB-Metaverse ecosystem. The blockchain enables the construction of several metaverses and achieve interconnection through cross-chains, which is conducive to the coordinated development of SB-Metaverse.

	In addition to the six basic technologies above, table II presents some other technologies that support the operation of the socially beneficial metaverse and their functions.

	\begin{table}
		\centering
		\caption{Some other technologies that support the operation of socially beneficial metaverse and their functions}\label{table 2}
		\begin{tabular}{p{60pt}p{150pt}p{40pt}}
			\hline
			Technology & Functions & References\\
			\hline
			VR/AR & Achieves the immersion required by the metaverse and improves the accessibility of the metaverse. &\cite{sherman2003understanding}\\
			Computer vision& Digitizes real-world images and provides users with a highly immersive experience.&\cite{forsyth2011computer}\\
			Speech processing & Facilitates accurate communication and eliminates the language barrier in the interaction between different system roles.& \cite{hickok2007cortical}\\
			Sensing technology & Provides a convenient and fast interaction method. &\cite{spencer2004smart}\\
			Holographic display & Allows users to interact with virtual objects in the real world and improves the immersion of users. &\cite{agocs2006large}\\
			
			\hline
		\end{tabular}
	\end{table}

\subsubsection{NFT and dNFT in the Metaverse}
Non-Fungible Tokens (NFTs) and dynamic Non-Fungible Tokens (dNFTs) are critical components in the metaverse, enabling the creation, ownership, and trade of unique digital assets. These technologies allow users to represent virtual assets such as art, collectibles, virtual real estate, and avatars, ensuring uniqueness and verifiability on the blockchain. In the context of SB-Metaverse, NFTs and dNFTs provide a foundation for the digital economy, allowing creators to monetize their content and users to gain verifiable ownership of virtual goods.

NFTs ensure the uniqueness and immutability of digital assets by utilizing blockchain’s decentralized and tamper-proof nature. For instance, virtual items like land parcels, game collectibles, and wearable items can be securely owned and transferred. dNFTs extend this functionality by allowing assets to evolve or change over time based on specific conditions or external inputs. For example, a virtual pet in a metaverse game might age or gain new abilities, with these changes recorded immutably via dNFT mechanisms.

The current blockchain infrastructure faces scalability issues, with high transaction costs and network congestion limiting the growth of NFTs and dNFTs in metaverse applications. This is compounded by the environmental impact of blockchain networks, particularly proof-of-work-based systems, raising concerns about the sustainability of NFTs. Additionally, the lack of standardized protocols hampers interoperability, restricting the seamless transfer of NFTs and dNFTs across different metaverse platforms. Legal and regulatory uncertainty further complicates the landscape, as the nascent nature of these technologies creates challenges around copyright protections, taxation, and ownership disputes. Lastly, NFTs often link real-world identity with digital assets, making user privacy and security essential to safeguard sensitive information.

In terms of opportunities, NFTs empower creators by allowing them to earn royalties from secondary sales, fostering innovation and creativity within the metaverse. Users can also create personalized, tokenized avatars and items, enhancing their engagement and self-expression in virtual spaces. Furthermore, NFTs and dNFTs have the potential to integrate with decentralized finance (DeFi) protocols, enabling users to stake, lend, or borrow against their virtual assets. As interoperability standards evolve, NFTs can facilitate seamless interactions across multiple metaverses, unlocking new collaborative possibilities. Additionally, dNFTs offer the ability to create adaptive and interactive digital content, further enhancing the immersive experience for users.

While significant progress has been made in exploring the potential of the metaverse, several technological gaps persist that hinder its full realization. Firstly, interoperability between various metaverse platforms remains limited, with the absence of standardized protocols for seamless data and asset exchange. Secondly, privacy and security concerns are magnified in decentralized virtual environments, where user data is frequently exchanged. Current privacy measures are inadequate to safeguard against potential breaches. Finally, the high cost and limited portability of VR/AR hardware continue to present a significant barrier to widespread adoption, limiting the metaverse's accessibility to a broader audience. These gaps highlight the need for an integrated approach that addresses the core challenges of interoperability, security, and hardware accessibility. The exploration of blockchain technology for enhancing cross-platform communication and ensuring data privacy through decentralized architectures can potentially offer solutions to these issues. Additionally, the development of optimization algorithms aimed at reducing the computational requirements of VR/AR devices could lead to more energy-efficient and cost-effective hardware, thus fostering greater adoption. Bridging these technological gaps would not only advance the theoretical framework of the metaverse but also enable practical applications that are accessible to a wider user base.

	\section{Applications of Socially Beneficial Metaverse}
	
	The following sections present applications of socially beneficial metaverse. Fig. 6 shows these applications.

	\begin{figure}[!t]
		\centering
		\includegraphics[width=0.55\columnwidth]{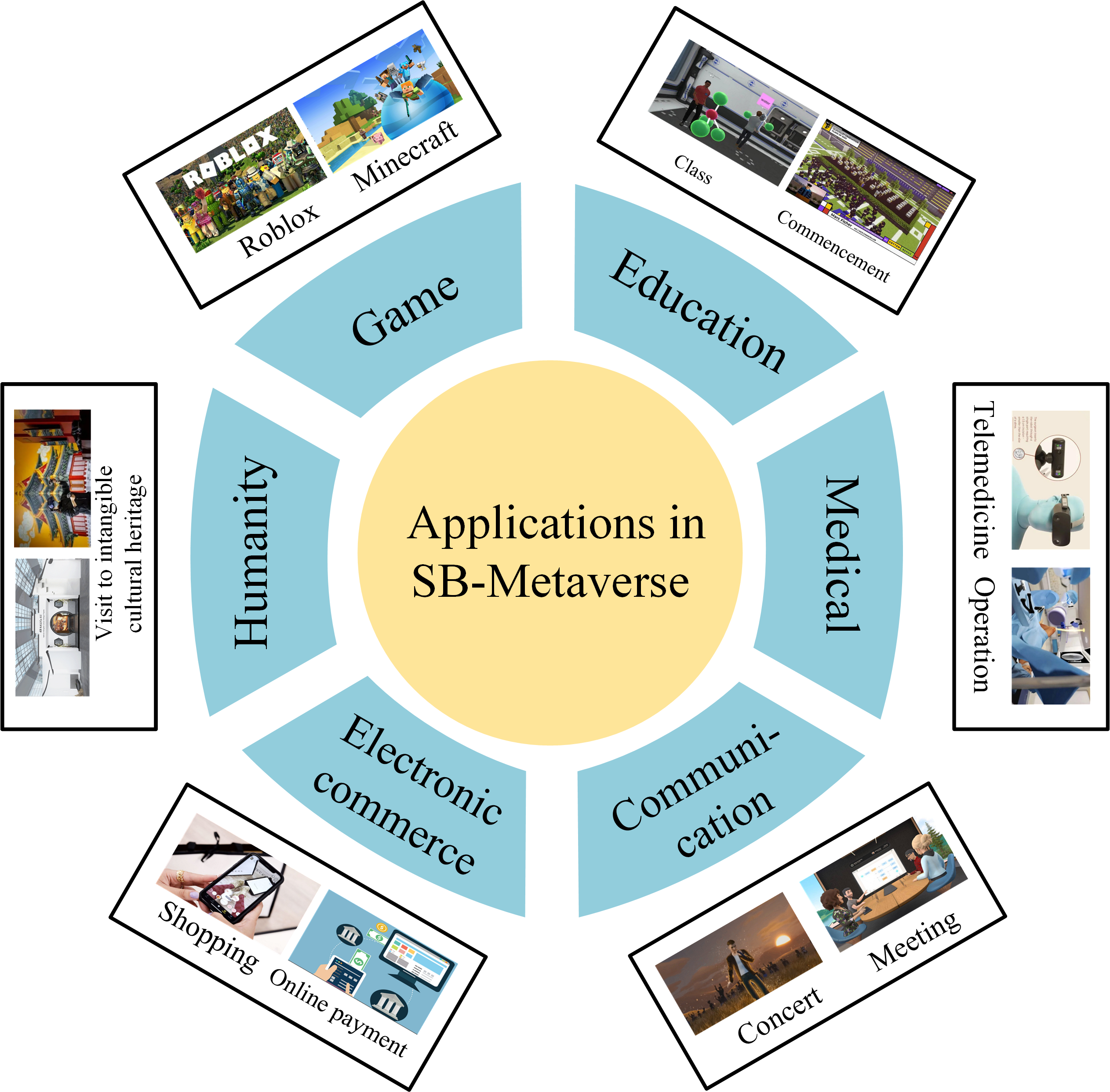}
		\caption{Applications in socially beneficial metaverse.}
		\label{fig6}
	\end{figure}

	\subsection{Game}
	Games are an indispensable part of people’s social life, and moderate play helps people improve their intelligence and relieve stress \cite{warburton2007health}. Most of the traditional games are played in a fixed scenario. Even if different players participate, the core interaction is to complete the task. However, games in SB-Metaverse pay more attention to high interactivity and high degree of freedom. This means only the basic rules of the game are designed, no tasks, no gameplay, and no social restrictions are specified. Each player can achieve a considerable degree of freedom. In addition, by sensing the player’s movements to control the activities of virtual characters in SB-Metaverse, players are encouraged to explore the metaverse world with their actions, which allows players to exercise while playing. Since Sencond Life \cite{boulos2007second}, several games such as Minecraft, Roblox, and others have emerged that contain some elements of the metaverse. Flintham et al. \cite{flintham2003line} had a particular focus on mobile mixed reality games, which use new digital content to enhance traditional physical games, and they explored the design of experiences in which mobile participants collaborate with online participants. Here we introduce several typical metaverse games, which can provide some guidance on the development of SB-Metaverse games in the future.

	Minecraft focuses on allowing players to explore, interact, and alter a dynamically generated map of one-cubic-meter-sized blocks. In addition to blocks, environmental features include plants, mobs, and items. The game also allows players to create works and make art on various multiplayer servers or in their single player mode \cite{oh2016control}. One aspect that matches Minecraft with the metaverse is that it has the characteristics of decentralization. Players can create their own metaverse world. In this world, the player is the maker of the rules and players can go to each other’s world to participate in content creation, thus avoiding the metaverse world being monopolized by big companies.

	Roblox, the world’s largest multiplayer online creation game, is a game that is compatible with virtual worlds, casual games and it relies on self-built content. Most of the works in the game are created by users \cite{long2019roblox}. Players could control avatars composed of cylinders and squares to participate in content creation. Players can share and play games created by each other. It is worth mentioning that developers in Roblox can charge the virtual currency Robux for various items and game experiences. Then they could convert these virtual currencies into real-world currencies, which will also provide ideas for the establishment of the financial system in SB-Metaverse.

	Decentraland is a virtual world platform built on Ethereum, and it is jointly owned and built by users \cite{chaudhari2019decentraland}. Each user can create a personal profile, then socialize with other users, engage in entertainment and build houses on digital land. Decentraland is built on the Ethereum network to enable ownership and an open economy. Its architecture consists of a consensus layer, a content layer, and a real-time presentation layer. The consensus layer and content layer ensure land ownership, and the real-time layer ensures that all players can be connected in the same space in real time \cite{goanta2020selling}. Decentraland provides an example for the use of blockchain to build an economic system in SB-Metaverse.

    Beyond entertainment, games in the metaverse promote social interaction, teamwork, and problem-solving skills by enabling players to collaborate in highly interactive environments. For instance, multiplayer games like Minecraft and Roblox provide platforms where users can co-create virtual worlds, fostering creativity and critical thinking. Additionally, metaverse-based games can serve as tools for physical and mental health, encouraging players to engage in active movements or simulate real-world scenarios to build resilience and manage stress. This contributes to overall well-being, particularly for younger users seeking social connections or therapeutic activities.

	Table III presents some games and their characteristics in the socially beneficial metaverse.

	\begin{table}
		\centering
		\caption{Some games and their characteristics in socially beneficial metaverse}\label{table 3}
		\begin{tabular}{p{50pt}p{170pt}p{60pt}p{30pt}}
			\hline
			Name & Characteristics &Technology &References\\ 
			\hline
			Axie Infinity &Users can make virtual asset trading and develop new characters. It also expands the way users interact with lands & Blockchain & \cite{de2022play}\\
			Decentraland & Uses blockchain technology to acquire and transfer virtual property rights, allowing users to hold these virtual property rights permanently. & Blockchain, 3D and H512& \cite{dowling2022fertile}\\
			The Sandbox & Players can create digital assets for trade. The game also has characteristics of highly immersive experience and decentralization. & Blockchain and VR& \cite{jeon2016effect}\\
			My Neighbor Alice & The game allows players to participate in the game and get rewards, distribute game assets, and decentralize the ownership of in-game assets based on NFT. &Blockchain and 3D& \cite{lee2021all}\\
			The Sims & It can achieve the interconnection between different players, and it can also generate an economic system that interacts with the real world. & 3D and VR& \cite{moore2017expression}\\
			\hline
		\end{tabular}
	\end{table}

	\subsection{Education}
	Education is an important pillar of national development, with the need to improve education quality and ensure fairness being critical social issues. The socially beneficial metaverse (SB-Metaverse) provides opportunities to address these challenges by creating digital identities for teachers, students, and other related personnel, as well as virtual teaching spaces where they can interact seamlessly \cite{collins2008looking}. This enhances accessibility and inclusivity by minimizing geographical and physical barriers. Students with disabilities or those in remote areas can participate equally, gaining access to high-quality educational resources and interactive learning experiences.

    Education in SB-Metaverse is characterized by immersive learning, improved teaching efficiency, and cost savings. Real-time simulations and virtual field trips enable students to explore complex concepts, such as scientific experiments or historical reconstructions, in a risk-free environment. For example, Kanematsu et al. \cite{kanematsu2014virtual} demonstrated the effectiveness of virtual science lectures in Second Life, significantly improving student engagement and learning outcomes. Similarly, ACAI, a leading AI academic conference, leveraged the metaverse for its 2020 seminar on Animal Crossing, providing a novel virtual space for knowledge exchange during the pandemic. As the role of SB-Metaverse in education continues to emerge, several considerations must be addressed. Developers should improve VR technology to enhance classroom immersion and interactivity, while educators must actively adopt and integrate these new technologies into their teaching. Additionally, the drawbacks of virtual classrooms, such as students becoming overly dependent on digital environments, must be mitigated. Ensuring the privacy and security of students and teachers is equally essential to foster a healthy and equitable educational ecosystem. These innovations in SB-Metaverse not only democratize access to quality education but also provide a transformative platform to enhance learning experiences globally.

	\subsection{Medicine}
	
	Medical care and people’s livelihood are closely related. Improvements in medical research is also an important social issue. Applying SB-Metaverse to medicine would address the constraints of time, distance, and technology, which will bring about numerous benefits to society. The applications of SB-Metaverse to the medical field mainly include telemedicine, surgery and psychological treatments.
	
	\begin{itemize}
		\item Telemedicine:
		
		Under the current Internet technology, patients can consult with doctors through video and voice communications. With the support of future metaverse technology, a virtual scene can be built, doctors can collect and analyze real-time physiological index data on patients through remote diagnosis equipment. At present, Meta has launched a robot skin ReSkin, which can obtain the touch of objects and collect data \cite{bhirangi2021reskin}. Telemedicine in SB-Metaverse is gaining attention and becoming possible.

		\item Surgery:
		
		Human organs can be observed through VR, which provides great help in surgery \cite{javaid2020virtual}. With more application scenarios of VR technology in the medical field, remote surgery in SB-Metaverse becomes possible. At present, there are high-precision surgical robots that can use 3D stereo high-definition images to perform remote minimally invasive surgery, effectively improving the success rate of surgery \cite{lanfranco2004robotic}.
  
		\item Psychological Treatments:
  
            Past researches have shown that controlled exposure to virtual reality systems brings benifit to the treatment of social phobia \cite{klinger2005virtual} and agoraphobia\cite{vincelli2002virtual}. M. Pavlou et al. \cite{pavlou2012effect} confirmed that virtual reality is an effective aid to vestibular rehabilitation programsd by comparing the different treatment effects of the subject patients after static and dynamic virtual reality training. The ability of VR to simulate reality enables the realization of psychotherapy, and thus the metaverse holds promise as an important adjunct to future psychological treatments \cite{freeman2017virtual}.
            
	\end{itemize}
	
	Therefore, it is not difficult to foresee that socially beneficial metaverse will create new opportunities in the field of medicine. By promoting digitalization and intelligence, and the integration of a wide range of advanced technologies (such as VR, AI, and so on) in the medical field, metaverse will bring about many opportunities.

	\subsection{Electronic Commerce}
	Electronic commerce (EC) has significantly transformed traditional lifestyles by facilitating shopping and improving the quality of life. Online shopping, as a key component of EC, has brought convenience but also faces challenges such as the inability to provide try-on services for clothes, shoes, and other products. This limitation often leads to size mismatches, returns, and exchanges, which not only frustrate users but also waste social resources \cite{kumar202120}. In this context, the socially beneficial metaverse (SB-Metaverse) introduces revolutionary changes by creating immersive and inclusive shopping experiences. Through the use of virtual reality (VR) and augmented reality (AR), the SB-Metaverse enables consumers to interact with products in real-time. For instance, VR shopping allows users to explore products in detail and understand their features more comprehensively. AR technology further enhances this experience by integrating 3D models of real scenes, enabling users to visualize how items like furniture or home decor fit into their spaces. Innovative applications, such as Gucci's AR feature for trying on sneakers and Drapr’s 3D virtual fitting room for personalized clothing recommendations, demonstrate how the SB-Metaverse enhances consumer satisfaction and confidence in online shopping \cite{ren2021cloth} \cite{kim2021advertising}.

    Moreover, the metaverse fosters equitable shopping opportunities, particularly benefiting individuals with disabilities or mobility challenges who can shop comfortably from their homes while making informed purchasing decisions. Immersive shopping experiences also reduce the need for excessive returns, minimizing environmental waste and promoting sustainability. By bridging the gap between physical and digital commerce, the SB-Metaverse supports a more responsible consumer ecosystem. In addition to these innovations, blockchain technology provides secure online payment solutions within the SB-Metaverse, ensuring that users and merchants can conduct electronic transactions safely and transparently. The integration of these advanced technologies in EC not only enhances convenience but also maximizes resource efficiency, ultimately benefiting people's daily lives and supporting sustainable development.

	\subsection{Communication}
	With the rise of globalization, exchanges and cooperation between countries are becoming more frequent. However, due to the influence of the objective factor of geographical location, the process of communication will encounter inconvenience and increase costs. For instance, due to the outbreak of COVID-19 , many activities have been suspended to ensure people’s safety, making communication at a distance difficult. But SB-Metaverse can provide high accessibility to meet different social needs. Fortnite held a concert with millions of people online, which is almost impossible to achieve in reality. In terms of immersion and visual enjoyment, such a metaverse concert brings users a new experience \cite{anderson2019getting}. Meta launches Horizon Workrooms, a holographic virtual meeting software. The user wears the headset Oculus Quest 2, enters the software working interface, makes an avatar for himself/herself, and then appears in the meeting as a three-dimensional image of a cartoon man/woman \cite{zuckerberg2021facebook}. The holographic meeting has a sense of space, and users can write and draw on the blackboard in the virtual space to make presentations just like in the real world. M. Slater et al. \cite{slater2000acting} examined the extent to which virtual reality technology could be used in the rehearsal of a play, and the study showed that actors achieved the basis for successful live performances in virtual rehearsals, an effect that could not be achieved through forms such as videoconferencing and studying scripts. Greenhalgh et al. \cite{greenhalgh1995massive} designed MASSIVE, a virtual reality teleconferencing system, which aimed to provide a flexible and natural environment for conversation. Communication in the metaverse breaks down barriers of distance, enabling people from diverse cultural and geographic backgrounds to connect seamlessly. Virtual meeting spaces reduce travel needs, thereby decreasing carbon emissions and promoting environmental sustainability. Furthermore, the metaverse supports inclusivity by offering real-time translation and accessibility features, ensuring that individuals with language or physical limitations can participate fully in global dialogues and collaborations. SB-Metaverse has become an extension of our daily lives, which helps us better communicate across geospatial constraints and meet our social needs.

	\subsection{Humanity}
    The metaverse plays a vital role in preserving and promoting cultural heritage by leveraging advanced technologies to address the challenges of accessibility and conservation. By creating virtual replicas of historical sites and artifacts, it ensures that future generations can experience and appreciate humanity's shared legacy, even if these physical assets are endangered or inaccessible. For example, the game Assassin’s Creed reconstructed the Notre Dame Cathedral as an in-game 3D model, not only aiding its real-world restoration but also allowing players to virtually visit and engage with this iconic landmark \cite{allal2022intelligent}. Such applications demonstrate how the socially beneficial metaverse (SB-Metaverse) enables cultural edification and appreciation on a global scale.

    The SB-Metaverse also provides an ecosystem for spreading intangible cultural heritage (ICH) among younger generations. As the demand for humanities grows, traditional methods of ICH transmission face significant challenges, including the overwhelming burden on inheritors and limited resources for its dissemination. Virtual characters and immersive technologies like VR and AR alleviate these pressures by offering realistic, interactive experiences that engage audiences in new and meaningful ways. Through immersive storytelling and virtual events, the SB-Metaverse fosters cultural exchange and mutual understanding among diverse communities, contributing to a more interconnected and empathetic global society. Moreover, the metaverse supports the reproduction and restoration of cultural relics, ensuring that humanity's rich heritage is preserved and revitalized for future generations. By bridging the gap between cultural tradition and technological innovation, the SB-Metaverse enhances access to and appreciation of cultural assets, benefiting individuals and communities worldwide.

    \subsection{Enabling Interoperability Across Metaverses}
    The connectivity between disparate and heterogeneous metaverses presents a significant challenge in achieving virtual interoperability. Emerging distributed technologies provide a feasible solution to bridge these distinct virtual ecosystems. At the heart of this connectivity lies blockchain technology, which offers a decentralized framework that facilitates seamless interaction between different metaverses. By utilizing smart contracts and cross-chain protocols, assets and data can be exchanged securely and transparently, thereby promoting unprecedented levels of interoperability. This technical infrastructure enables, and even encourages, the trade of digital components across metaverse boundaries. Players can tokenize their in-game assets as Non-Fungible Tokens (NFTs), which are traded on decentralized marketplaces, allowing for the buying and selling of unique digital goods. NFTs can represent a wide range of virtual items, from virtual real estate to in-game collectibles, thus laying the groundwork for a thriving digital economy both within and across metaverses. Additionally, the use of decentralized identifiers (DIDs) further enhances the ability to share services across metaverses by enabling users to maintain a consistent digital identity while navigating different virtual worlds. The convergence of these technologies fosters a more interconnected virtual landscape, where users can participate in a shared metaverse experience, overcoming the limitations inherent in isolated platforms.

	\section{Challenges in Socially Beneficial Metaverse}
	
	The development of the socially beneficial metaverse (SB-Metaverse) is still in its infancy, and there are numerous challenges that need to be addressed to realize its full potential. These challenges can be categorized into social and non-social aspects. Below, we expand on these challenges with a more technical perspective and propose potential solutions.
	
	\subsection{Social Challenges}
	
	\subsubsection{Fairness}
	
	Fairness in the SB-Metaverse is crucial to ensure that all users are treated equitably and have access to the same opportunities within virtual environments \cite{woodruff2018qualitative}. As multiple virtual worlds are established in the SB-Metaverse, each with its own set of rules and governance models, the sheer scale of user management presents significant challenges. Given the large number of users, AI-based systems are expected to play a pivotal role in achieving intelligent management. However, AI systems rely heavily on machine learning algorithms to dynamically respond to virtual objects and user behavior \cite{lee2021all}. The fairness of these algorithms is paramount because any bias introduced into the AI system—whether in data preprocessing, model training, or real-time decision-making—can lead to discrimination and negative user experiences. For instance, biased algorithms may unfairly prioritize certain user groups over others, which could result in unequal access to resources, opportunities, or services within the metaverse.

    Implementing fairness-aware machine learning techniques is essential for targeting bias elimination at various stages of the system. These techniques can be applied from data collection and processing to the design and evaluation of models \cite{mehrabi2021survey}. Preprocessing steps, such as re-weighting or re-sampling data, help balance the representation of different user groups. During model training, fairness constraints and adversarial debiasing methods adjust the model’s decision-making process to ensure equal treatment of all users, regardless of demographic attributes or other potentially biased factors \cite{wan2023processing}. Additionally, deploying real-time fairness monitoring systems allows for continuous assessment of algorithmic decisions, ensuring fairness is maintained throughout user interactions.

    Several theoretical and practical solutions have been proposed to address fairness in AI systems. Adversarial debiasing, for example, has shown promising results in reducing the impact of biased data on algorithmic outputs by adding a layer of adversarial learning that penalizes bias \cite{hong2021federated}. Moreover, reinforcement learning models with fairness constraints can help promote fairness in decision-making by adjusting rewards and penalties based on fairness criteria rather than just maximizing utility \cite{ge2022toward}. As for practical implementation, existing frameworks provide tools for continuous monitoring and correction of fairness issues, which can be integrated into the SB-Metaverse to ensure that any emerging biases are quickly detected and addressed. Further, blockchain-based decentralized governance mechanisms, such as Decentralized Autonomous Organizations (DAOs), can support fairness by enabling transparent and equitable decision-making processes, allowing users to have a voice in the development of rules and policies that affect them \cite{qin2022web3}. The incorporation of fairness-aware solutions, supported by continuous monitoring and feedback loops, ensures that the SB-Metaverse provides a fair and inclusive environment for all users. By addressing the challenges of biased algorithmic decision-making and implementing robust fairness mechanisms, we can promote a metaverse that serves the interests of a diverse user base while preventing systemic inequities.

	\subsubsection{Capital Monopoly}

    The SB-Metaverse seeks to establish a parallel world where users enjoy freedom and protection of their personal rights. However, the high capital investment required for the development of SB-Metaverse may create an environment where large corporations dominate, potentially leading to a monopoly. In such a scenario, a few powerful entities could gain control over user privacy and data, undermining the decentralization principles of the metaverse. Although blockchain technology and decentralization offer potential solutions to combat monopolies, absolute decentralization conflicts with the current societal status quo, meaning additional measures are necessary to effectively counteract monopolistic control \cite{feng2021data}.

    Mitigating the risks of centralization and monopoly in the SB-Metaverse requires a combination of decentralized and community-driven approaches. From a theoretical standpoint, decentralization remains a cornerstone principle. Blockchain technology enables self-sovereign identity and ownership of digital assets, empowering users with control over their data and reducing reliance on central authorities or corporations that might otherwise exploit personal data. Additionally, decentralized finance (DeFi) protocols can reduce economic dependency on traditional financial institutions, providing alternative ways for smaller enterprises to participate in the metaverse economy \cite{ozili2022decentralized}.

    Practical solutions involve leveraging hybrid blockchain architectures that combine decentralized and centralized models, striking a balance between efficiency and user empowerment. For instance, Ethereum Layer-2 solutions can reduce transaction costs, making it more accessible for smaller enterprises to develop and operate within the SB-Metaverse ecosystem \cite{alipanahloo2024maximal}. Community-driven governance models, such as Decentralized Autonomous Organizations (DAOs), can promote democratic decision-making, where users participate in rule-making, voting on policies, and resource allocation, ensuring fairer distribution of power and ownership \cite{santana2022blockchain}. These models reduce the monopolistic influence of large corporations by democratizing control over key aspects of the SB-Metaverse.

	\subsubsection{User Addiction}
	
	As a virtual world which is completely different from the real world, socially beneficial metaverse can be harmful if overused. In situations where the real world is not as desirable, users may escape from reality by immersing themselves into the SB-Metaverse world. In addition, the highly realistic virtual environment enables people to try things that are impossible in real life, after gaining a sense of freshness and excitement, they could be caught in it and unable to withdraw themselves. Studies have shown that when people are addicted to virtual worlds, it can lead to psychological problems and mental disorders, such as depression and loneliness \cite{jeong2015game}. Teenagers are more likely to indulge in them without proper guidance \cite{soh2018parents}. In the current Internet era, Internet addiction is generally relieved by limiting the use of Internet time. However, in SB-Metaverse with high immersion, this method may not be adequate to solve the problem, so user addiction is an important problem in SB-Metaverse that must be solved in the future.

    To mitigate the risk of addiction, a combination of proactive monitoring and interactive management strategies must be implemented. Theoretically, integrating AI-driven systems that monitor usage patterns in real time can provide a foundation for addressing addiction. Machine learning models can detect abnormal behavior, such as prolonged or repetitive use, and trigger interventions. These interventions might include adaptive time constraints or prompts for breaks to encourage users to balance their virtual and real-life activities \cite{gulu2023exploring}. Practically, reinforcement learning algorithms could be employed to dynamically adjust the immersive experience based on individual usage patterns, creating a more balanced interaction. The system could gradually reduce the level of immersion after prolonged use, such as dimming virtual environments or introducing non-immersive elements, to encourage users to take breaks.

    Personalized feedback systems could also be developed to allow users to set their own boundaries for virtual interactions, while still offering flexibility based on their needs. Users could be given the ability to receive personalized prompts or notifications when they exceed a predefined usage threshold. This feedback could be integrated into the SB-Metaverse experience, alerting users to take breaks or engage in offline activities, thus promoting healthier interaction patterns. Collaborative and community-based efforts are essential for mitigating addiction. Peer support systems within the SB-Metaverse, such as moderation groups or support communities, could help users manage their time in the virtual world more effectively. These communities can provide guidance, share best practices for healthy usage, and even intervene when addictive behavior is detected. These solutions can collectively foster a healthy and sustainable user experience, ensuring that users can enjoy the benefits of the metaverse without falling into the trap of excessive immersion.

	\subsubsection{User Safety}
	
	Since the user controls his/her behaviors in SB-Metaverse through the terminal sensing device in the real world, the user becomes visually immersed in SB-Metaverse world, but the body is still in the real world. It is very possible that the surrounding environment of the real world cannot be observed, so there are potential safety hazards. Pokemon Go is an AR mobile game based on Google Maps. It can display Pokemon in the camera screen of the screen through the mobile phone in a real geographical location, and users can capture Pokemon \cite{baranowski2016pokemon}. But this combination of virtual and real-world games can lead to dangerous or risky behavior in the real world. According to reports, playing Pokemon Go while driving has caused many accidents and huge losses in Tippecanoe County, Indiana in the United States \cite{faccio2020death} \cite{drevin2019students}. Thus, we note that, while users are roaming in the metaverse world, how to ensure their safety in the real world remains a serious challenge.

    Advanced mechanisms can be implemented in VR/AR devices to ensure user safety \cite{wang2022survey}. Real-time environmental monitoring, powered by computer vision technologies, can be integrated to continuously scan and map the user's physical surroundings, providing situational awareness within the immersive virtual experience. These systems can detect potential hazards, such as obstacles or traffic, and trigger alerts to prompt users to adjust their actions. Additionally, geofencing technologies can define safe boundaries, warning users when they approach dangerous areas, such as busy streets or restricted zones. This proactive approach ensures that users are notified of potential hazards in real-time, helping them navigate the metaverse without compromising their physical safety \cite{zallio2022designing}. By integrating these technologies, the SB-Metaverse can offer a more secure and balanced experience, allowing users to explore virtual worlds while minimizing risks in the real world.

	\subsection{Non-social Challenges}
	
	Although we have previously described the underlying technologies of SB-Metaverse and their roles, meeting the technical requirements remains a significant challenge. The current hardware access devices are mainly VR and AR devices, but the virtual reality technology is still difficult to produce cost-effectively and adapt it to portable terminal devices. As a result, today, only a small number of people can enjoy the services of SB-Metaverse. Therefore, SB-Metaverse still has to overcome some technical barriers in the future to make SB-Metaverse widely accessible to a broad range of users.

    \subsubsection{Technological Barriers}
    The widespread adoption of the SB-Metaverse is hindered by significant technological barriers, primarily the lack of cost-effective, portable VR/AR hardware and the challenge of achieving low-latency, high-resolution interactions. These limitations impact both the accessibility and quality of the metaverse experience, particularly for users with lower-end devices or those in areas with limited technological infrastructure. To overcome these challenges, it is essential to focus on the development of lightweight, energy-efficient VR/AR hardware that balances performance and portability. Such advancements are critical for reducing user entry barriers, enabling broader participation, and improving immersion for all users.

    In parallel, the integration of edge computing with 5G technology provides a promising solution to address the latency issues inherent in real-time interactions within the metaverse \cite{aung2023edge}. Edge computing brings computation and data storage closer to the user, reducing the reliance on distant data centers and minimizing delays. When combined with the high-speed, low-latency capabilities of 5G networks, edge computing can significantly enhance the responsiveness and interactivity of virtual environments. This integration ensures that high-resolution, real-time interactions are supported with minimal lag, improving the overall user experience \cite{hoa2023dynamic}.

    To evaluate the effectiveness of these solutions, it is necessary to assess the latency reduction achieved through edge-cloud simulations. Additionally, the performance of VR/AR devices should be measured in terms of power consumption and computational efficiency, with benchmarks comparing energy usage and performance under different configurations. The development and deployment of lightweight hardware and edge-5G integration offer both theoretical and practical solutions, which, if successfully implemented, can enhance the accessibility and quality of the SB-Metaverse, paving the way for its broader adoption.

    \subsubsection{Privacy and Security}

    In decentralized, interconnected virtual environments such as the SB-Metaverse, safeguarding user data is a complex challenge due to the frequent exchange of sensitive information among users, systems, and platforms. As users interact within virtual spaces, they generate and transmit personal, behavioral, and transactional data, which makes privacy and security a paramount concern. Privacy refers to the protection of user information, ensuring that personal data is only accessible by authorized parties and not disclosed without consent \cite{huang2023security}. Security, on the other hand, pertains to the protection of data from unauthorized access, tampering, and breaches, ensuring the integrity and confidentiality of the information exchanged across the metaverse.

    Privacy-preserving techniques such as homomorphic encryption and differential privacy should be employed during data processing to safeguard user information. Homomorphic encryption enables computations on encrypted data, allowing for secure processing without exposing sensitive information \cite{marcolla2022survey}. Differential privacy ensures that individual user data remains unidentifiable, even in aggregated analysis \cite{zhao2022survey}. These techniques help mitigate the risks associated with sensitive data exchanges in decentralized environments. Furthermore, blockchain technology plays a crucial role in enhancing privacy and security by providing an immutable, transparent system for record-keeping. Through blockchain, all transactions and interactions within the metaverse can be securely logged, ensuring that the data is tamper-proof and auditable \cite{fu2022survey}. This transparency fosters user trust and promotes accountability within the system, enabling users to verify the integrity of their interactions.

    Evaluating the effectiveness of privacy and security measures requires attention to several aspects. Using tools for formal verification of smart contracts ensures that blockchain-based interactions are free from vulnerabilities. Additionally, penetration testing can simulate potential security breaches, assessing the system's resilience. Privacy audits can be conducted to verify the application of data anonymization techniques such as differential privacy and to ensure that user identities are protected from unauthorized access.

    In terms of privacy and security, a strong theoretical approach involves creating comprehensive frameworks that guide the implementation of cryptographic and privacy-enhancing technologies. On the practical side, this translates into the deployment of encryption methods and blockchain systems that safeguard user data and ensure transparency. When applied effectively, these measures not only protect privacy but also foster accountability, ensuring that the SB-Metaverse operates securely while maintaining trust among users.

    \subsubsection{Interoperability}

    Interoperability within the SB-Metaverse is essential for seamless communication and the efficient sharing of assets across diverse virtual environments. Currently, the lack of standardization among metaverse platforms creates fragmentation, making it challenging for users to transfer assets, data, and experiences between ecosystems. Interoperability refers to the ability of different systems or platforms to work together, share data, and allow assets to be used across various environments without compatibility issues \cite{li2023metaopera}. In the SB-Metaverse, achieving interoperability involves addressing both technical compatibility and data synchronization across platforms.

    To resolve these challenges, cross-chain protocols and APIs should be implemented to enable the secure transfer of assets and data across different blockchain systems and metaverse platforms. Cross-chain technologies can facilitate the seamless exchange of non-fungible tokens (NFTs) and other digital assets, regardless of the underlying blockchain \cite{ren2024hcnct}. Additionally, developing universal standards like OpenXR for VR and AR applications can ensure that metaverse platforms adhere to common protocols, making it easier to share and exchange assets between diverse platforms. By adopting these standards, developers can create more consistent user experiences and foster better communication between platforms, leading to greater connectivity and a unified metaverse ecosystem.

    Developing open standards is a key theoretical solution for ensuring interoperability across metaverse platforms. These standards should define data formats, communication protocols, and asset structures, creating a foundation for future platforms that can seamlessly interact. On the practical side, cross-chain technologies and platform-specific APIs are vital for facilitating the exchange of assets and data between different environments. The effectiveness of these solutions can be evaluated by testing the smooth transfer of assets across platforms, as well as assessing the user experience in terms of latency, usability, and overall functionality \cite{sami2024metaverse}.

    By implementing these solutions, the SB-Metaverse can move towards a more interconnected and cohesive ecosystem, where users can easily navigate between platforms and engage with a wide range of digital assets and experiences.

	\section{Conclusion}
	In this paper, we introduced the definition and described the developments of the metaverse. We proposed the SB-Metaverse and described its framework and the associated layers. Then, we described the basic technologies of SB-Metaverse. Finally, we presented the applications of SB-Metaverse and discussed the challenges faced by it. In addition to the issues covered in this paper, there are still many unsolved social issues in SB-Metaverse. The core motivation of this paper is to guide more people to pay attention to and study SB-Metaverse, and to provide scientific guidance for the future development of SB-Metaverse.

	
	
	
%
	
	
	

	
%
%
	
\bibliographystyle{model3-num-names} 
\bibliography{reference1}

\begin{thebibliography}{176}
\providecommand{\natexlab}[1]{#1}
\providecommand{\url}[1]{\texttt{#1}}
\providecommand{\href}[2]{#2}
\providecommand{\path}[1]{#1}
\providecommand{\eprint}[1]{\href{http://arxiv.org/abs/#1}{\path{#1}}}
\providecommand{\DOIprefix}{doi:}
\providecommand{\ArXivprefix}{arXiv:}
\providecommand{\URLprefix}{URL: }
\providecommand{\Pubmedprefix}{pmid:}
\providecommand{\doi}[1]{\href{http://dx.doi.org/#1}{\path{#1}}}
\providecommand{\Pubmed}[1]{\href{pmid:#1}{\path{#1}}}
\providecommand{\BIBand}{and}
\providecommand{\bibinfo}[2]{#2}
\ifx\xfnm\undefined \def\xfnm[#1]{\unskip,\space#1}\fi
\bibitem[{Chung et~al.(2014)Chung, Gulcehre, Cho and Bengio}]{chung2014empirical}
\bibinfo{author}{Chung\xfnm[ J.]}, \bibinfo{author}{Gulcehre\xfnm[ C.]}, \bibinfo{author}{Cho\xfnm[ K.]}, \bibinfo{author}{Bengio\xfnm[ Y.]}.
\newblock \bibinfo{title}{Empirical evaluation of gated recurrent neural networks on sequence modeling}.
\newblock \bibinfo{journal}{arXiv preprint arXiv:14123555} \bibinfo{year}{2014};.
\bibitem[{Xiang et~al.(2023)Xiang, Zhang and Bilal}]{xiang2023cloud}
\bibinfo{author}{Xiang\xfnm[ H.]}, \bibinfo{author}{Zhang\xfnm[ X.]}, \bibinfo{author}{Bilal\xfnm[ M.]}.
\newblock \bibinfo{title}{A cloud-edge service offloading method for the metaverse in smart manufacturing}.
\newblock \bibinfo{journal}{Software: Practice and Experience} \bibinfo{year}{2023};.
\bibitem[{Duan et~al.(2021)Duan, Li, Fan, Lin, Wu and Cai}]{duan2021metaverse}
\bibinfo{author}{Duan\xfnm[ H.]}, \bibinfo{author}{Li\xfnm[ J.]}, \bibinfo{author}{Fan\xfnm[ S.]}, \bibinfo{author}{Lin\xfnm[ Z.]}, \bibinfo{author}{Wu\xfnm[ X.]}, \bibinfo{author}{Cai\xfnm[ W.]}.
\newblock \bibinfo{title}{Metaverse for social good: A university campus prototype}.
\newblock In: \bibinfo{booktitle}{Proceedings of the 29th ACM International Conference on Multimedia}. \bibinfo{year}{2021}, p. \bibinfo{pages}{153--161}.
\bibitem[{Lee et~al.(2021)Lee, Braud, Zhou, Wang, Xu, Lin et~al.}]{lee2021all}
\bibinfo{author}{Lee\xfnm[ L.H.]}, \bibinfo{author}{Braud\xfnm[ T.]}, \bibinfo{author}{Zhou\xfnm[ P.]}, \bibinfo{author}{Wang\xfnm[ L.]}, \bibinfo{author}{Xu\xfnm[ D.]}, \bibinfo{author}{Lin\xfnm[ Z.]}, et~al.
\newblock \bibinfo{title}{All one needs to know about metaverse: A complete survey on technological singularity, virtual ecosystem, and research agenda}.
\newblock \bibinfo{journal}{arXiv preprint arXiv:211005352} \bibinfo{year}{2021};.
\bibitem[{Kraus et~al.(2022)Kraus, Kanbach, Krysta, Steinhoff and Tomini}]{kraus2022facebook}
\bibinfo{author}{Kraus\xfnm[ S.]}, \bibinfo{author}{Kanbach\xfnm[ D.]}, \bibinfo{author}{Krysta\xfnm[ P.]}, \bibinfo{author}{Steinhoff\xfnm[ M.]}, \bibinfo{author}{Tomini\xfnm[ N.]}.
\newblock \bibinfo{title}{Facebook and the creation of the metaverse: Radical business model innovation or incremental transformation?}
\newblock \bibinfo{journal}{International Journal of Entrepreneurial Behavior \& Research} \bibinfo{year}{2022};.
\bibitem[{Kim(2021)}]{kim2021advertising}
\bibinfo{author}{Kim\xfnm[ J.]}.
\newblock \bibinfo{title}{Advertising in the metaverse: Research agenda}.
\newblock \bibinfo{year}{2021}.
\bibitem[{Liaqat et~al.(2019)Liaqat, Ali, Qadir, Bashir, Bilal and Majeed}]{LiaqatBilal}
\bibinfo{author}{Liaqat\xfnm[ H.B.]}, \bibinfo{author}{Ali\xfnm[ A.]}, \bibinfo{author}{Qadir\xfnm[ J.]}, \bibinfo{author}{Bashir\xfnm[ A.K.]}, \bibinfo{author}{Bilal\xfnm[ M.]}, \bibinfo{author}{Majeed\xfnm[ F.]}.
\newblock \bibinfo{title}{Socially-aware congestion control in ad-hoc networks: Current status and the way forward}.
\newblock \bibinfo{journal}{Future Generation Computer Systems} \bibinfo{year}{2019};\bibinfo{volume}{97}:\bibinfo{pages}{634--660}.
\bibitem[{Stephenson(2003)}]{stephenson2003snow}
\bibinfo{author}{Stephenson\xfnm[ N.]}.
\newblock \bibinfo{title}{Snow Crash: A Novel}.
\newblock \bibinfo{publisher}{Spectra}; \bibinfo{year}{2003}.
\bibitem[{Collins(2008)}]{collins2008looking}
\bibinfo{author}{Collins\xfnm[ C.]}.
\newblock \bibinfo{title}{Looking to the future: Higher education in the metaverse}.
\newblock \bibinfo{journal}{Educause Review} \bibinfo{year}{2008};\bibinfo{volume}{43}(\bibinfo{number}{5}):\bibinfo{pages}{51--63}.
\bibitem[{Jaynes et~al.(2003)Jaynes, Seales, Calvert, Fei and Griffioen}]{jaynes2003metaverse}
\bibinfo{author}{Jaynes\xfnm[ C.]}, \bibinfo{author}{Seales\xfnm[ W.B.]}, \bibinfo{author}{Calvert\xfnm[ K.]}, \bibinfo{author}{Fei\xfnm[ Z.]}, \bibinfo{author}{Griffioen\xfnm[ J.]}.
\newblock \bibinfo{title}{The metaverse: a networked collection of inexpensive, self-configuring, immersive environments}.
\newblock In: \bibinfo{booktitle}{Proceedings of the workshop on Virtual environments 2003}. \bibinfo{year}{2003}, p. \bibinfo{pages}{115--124}.
\bibitem[{Gibson(2019)}]{gibson2019neuromancer}
\bibinfo{author}{Gibson\xfnm[ W.]}.
\newblock \bibinfo{title}{Neuromancer (1984)}.
\newblock In: \bibinfo{booktitle}{Crime and Media}. \bibinfo{publisher}{Routledge}; \bibinfo{year}{2019}, p. \bibinfo{pages}{86--94}.
\bibitem[{Stephenson(1994)}]{stephenson1994snow}
\bibinfo{author}{Stephenson\xfnm[ N.]}.
\newblock \bibinfo{title}{Snow crash}.
\newblock \bibinfo{publisher}{Penguin UK}; \bibinfo{year}{1994}.
\bibitem[{Vinge(2015)}]{vinge2015true}
\bibinfo{author}{Vinge\xfnm[ V.]}.
\newblock \bibinfo{title}{True names and the opening of the cyberspace frontier}.
\newblock \bibinfo{publisher}{Tor Books}; \bibinfo{year}{2015}.
\bibitem[{Dambekodi et~al.(2020)Dambekodi, Frazier, Ammanabrolu and Riedl}]{dambekodi2020playing}
\bibinfo{author}{Dambekodi\xfnm[ S.]}, \bibinfo{author}{Frazier\xfnm[ S.]}, \bibinfo{author}{Ammanabrolu\xfnm[ P.]}, \bibinfo{author}{Riedl\xfnm[ M.O.]}.
\newblock \bibinfo{title}{Playing text-based games with common sense}.
\newblock \bibinfo{journal}{arXiv preprint arXiv:201202757} \bibinfo{year}{2020};.
\bibitem[{Shah and Romine(1995)}]{shah1995playing}
\bibinfo{author}{Shah\xfnm[ R.]}, \bibinfo{author}{Romine\xfnm[ J.]}.
\newblock \bibinfo{title}{Playing MUDS on the Internet}.
\newblock \bibinfo{publisher}{John Wiley \& Sons, Inc.}; \bibinfo{year}{1995}.
\bibitem[{Zen(2003)}]{zen2003impacts}
\bibinfo{author}{Zen\xfnm[ L.]}.
\newblock \bibinfo{title}{The impacts of medievia and medthievia}.
\newblock \bibinfo{year}{2003}.
\bibitem[{Bartle(2003)}]{bartle2003virtual}
\bibinfo{author}{Bartle\xfnm[ R.]}.
\newblock \bibinfo{title}{What are virtual worlds}.
\newblock \bibinfo{journal}{Retrieved March} \bibinfo{year}{2003};\bibinfo{volume}{20}:\bibinfo{pages}{2008}.
\bibitem[{Mauldin(1994)}]{mauldin1994chatterbots}
\bibinfo{author}{Mauldin\xfnm[ M.L.]}.
\newblock \bibinfo{title}{Chatterbots, tinymuds, and the turing test: Entering the loebner prize competition}.
\newblock In: \bibinfo{booktitle}{AAAI}; vol.~\bibinfo{volume}{94}. \bibinfo{year}{1994}, p. \bibinfo{pages}{16--21}.
\bibitem[{Livingstone et~al.(2008)Livingstone, Kemp and Edgar}]{livingstone2008multi}
\bibinfo{author}{Livingstone\xfnm[ D.]}, \bibinfo{author}{Kemp\xfnm[ J.]}, \bibinfo{author}{Edgar\xfnm[ E.]}.
\newblock \bibinfo{title}{From multi-user virtual environment to 3d virtual learning environment}.
\newblock \bibinfo{journal}{ALT-J} \bibinfo{year}{2008};\bibinfo{volume}{16}(\bibinfo{number}{3}):\bibinfo{pages}{139--150}.
\bibitem[{Tatum(2000)}]{tatum2000active}
\bibinfo{author}{Tatum\xfnm[ M.]}.
\newblock \bibinfo{title}{Active worlds}.
\newblock \bibinfo{journal}{ACM SIGGRAPH Computer Graphics} \bibinfo{year}{2000};\bibinfo{volume}{34}(\bibinfo{number}{2}):\bibinfo{pages}{56--57}.
\bibitem[{Kim(2002)}]{kim2002korean}
\bibinfo{author}{Kim\xfnm[ S.Y.]}.
\newblock \bibinfo{title}{Korean college students’ reflections of english language learning via cmc and ffc}.
\newblock \bibinfo{journal}{Multimedia-Assisted Language Learning} \bibinfo{year}{2002};\bibinfo{volume}{5}(\bibinfo{number}{2}):\bibinfo{pages}{9--28}.
\bibitem[{Chan and Chang(2004)}]{chan2004strifeshadow}
\bibinfo{author}{Chan\xfnm[ H.T.]}, \bibinfo{author}{Chang\xfnm[ R.K.]}.
\newblock \bibinfo{title}{Strifeshadow fantasy: a massive multi-player online game}.
\newblock In: \bibinfo{booktitle}{First IEEE Consumer Communications and Networking Conference, 2004. CCNC 2004.} \bibinfo{organization}{IEEE}; \bibinfo{year}{2004}, p. \bibinfo{pages}{557--562}.
\bibitem[{Rymaszewski et~al.(2007)Rymaszewski, Au, Wallace, Winters, Ondrejka and Batstone-Cunningham}]{rymaszewski2007second}
\bibinfo{author}{Rymaszewski\xfnm[ M.]}, \bibinfo{author}{Au\xfnm[ W.J.]}, \bibinfo{author}{Wallace\xfnm[ M.]}, \bibinfo{author}{Winters\xfnm[ C.]}, \bibinfo{author}{Ondrejka\xfnm[ C.]}, \bibinfo{author}{Batstone-Cunningham\xfnm[ B.]}.
\newblock \bibinfo{title}{Second life: The official guide}.
\newblock \bibinfo{publisher}{John Wiley \& Sons}; \bibinfo{year}{2007}.
\bibitem[{Kaplan and Haenlein(2009)}]{kaplan2009fairyland}
\bibinfo{author}{Kaplan\xfnm[ A.M.]}, \bibinfo{author}{Haenlein\xfnm[ M.]}.
\newblock \bibinfo{title}{The fairyland of second life: Virtual social worlds and how to use them}.
\newblock \bibinfo{journal}{Business horizons} \bibinfo{year}{2009};\bibinfo{volume}{52}(\bibinfo{number}{6}):\bibinfo{pages}{563--572}.
\bibitem[{Boulos et~al.(2007)Boulos, Hetherington and Wheeler}]{boulos2007second}
\bibinfo{author}{Boulos\xfnm[ M.N.K.]}, \bibinfo{author}{Hetherington\xfnm[ L.]}, \bibinfo{author}{Wheeler\xfnm[ S.]}.
\newblock \bibinfo{title}{Second life: an overview of the potential of 3-d virtual worlds in medical and health education}.
\newblock \bibinfo{journal}{Health Information \& Libraries Journal} \bibinfo{year}{2007};\bibinfo{volume}{24}(\bibinfo{number}{4}):\bibinfo{pages}{233--245}.
\bibitem[{Jagneaux(2018)}]{jagneaux2018ultimate}
\bibinfo{author}{Jagneaux\xfnm[ D.]}.
\newblock \bibinfo{title}{The Ultimate Roblox Book: An Unofficial Guide}.
\newblock \bibinfo{publisher}{Adams Media}; \bibinfo{year}{2018}.
\bibitem[{Duncan(2011)}]{duncan2011minecraft}
\bibinfo{author}{Duncan\xfnm[ S.C.]}.
\newblock \bibinfo{title}{Minecraft, beyond construction and survival} \bibinfo{year}{2011};.
\bibitem[{Nakamoto et~al.(2008)}]{nakamoto2008bitcoin}
\bibinfo{author}{Nakamoto\xfnm[ S.]}, et~al.
\newblock \bibinfo{title}{Bitcoin}.
\newblock \bibinfo{journal}{A peer-to-peer electronic cash system} \bibinfo{year}{2008};.
\bibitem[{Chen and Bellavitis(2020)}]{chen2020blockchain}
\bibinfo{author}{Chen\xfnm[ Y.]}, \bibinfo{author}{Bellavitis\xfnm[ C.]}.
\newblock \bibinfo{title}{Blockchain disruption and decentralized finance: The rise of decentralized business models}.
\newblock \bibinfo{journal}{Journal of Business Venturing Insights} \bibinfo{year}{2020};\bibinfo{volume}{13}:\bibinfo{pages}{e00151}.
\bibitem[{Chaudhari et~al.(2019)Chaudhari, Laddha and Potdar}]{chaudhari2019decentraland}
\bibinfo{author}{Chaudhari\xfnm[ A.]}, \bibinfo{author}{Laddha\xfnm[ D.]}, \bibinfo{author}{Potdar\xfnm[ M.]}.
\newblock \bibinfo{title}{Decentraland--a blockchain based model for smart property experience}.
\newblock \bibinfo{journal}{International Engineering Journal For Research \& Development} \bibinfo{year}{2019};\bibinfo{volume}{4}(\bibinfo{number}{5}):\bibinfo{pages}{5--5}.
\bibitem[{Kang(2021)}]{kang2021metaverse}
\bibinfo{author}{Kang\xfnm[ Y.m.]}.
\newblock \bibinfo{title}{Metaverse framework and building block}.
\newblock \bibinfo{journal}{Journal of the Korea Institute of Information and Communication Engineering} \bibinfo{year}{2021};\bibinfo{volume}{25}(\bibinfo{number}{9}):\bibinfo{pages}{1263--1266}.
\bibitem[{Ning et~al.(2021)Ning, Wang, Lin, Wang, Dhelim, Farha et~al.}]{ning2021survey}
\bibinfo{author}{Ning\xfnm[ H.]}, \bibinfo{author}{Wang\xfnm[ H.]}, \bibinfo{author}{Lin\xfnm[ Y.]}, \bibinfo{author}{Wang\xfnm[ W.]}, \bibinfo{author}{Dhelim\xfnm[ S.]}, \bibinfo{author}{Farha\xfnm[ F.]}, et~al.
\newblock \bibinfo{title}{A survey on metaverse: the state-of-the-art, technologies, applications, and challenges}.
\newblock \bibinfo{journal}{arXiv preprint arXiv:211109673} \bibinfo{year}{2021};.
\bibitem[{Fuller et~al.(2020)Fuller, Fan, Day and Barlow}]{fuller2020digital}
\bibinfo{author}{Fuller\xfnm[ A.]}, \bibinfo{author}{Fan\xfnm[ Z.]}, \bibinfo{author}{Day\xfnm[ C.]}, \bibinfo{author}{Barlow\xfnm[ C.]}.
\newblock \bibinfo{title}{Digital twin: Enabling technologies, challenges and open research}.
\newblock \bibinfo{journal}{IEEE access} \bibinfo{year}{2020};\bibinfo{volume}{8}:\bibinfo{pages}{108952--108971}.
\bibitem[{Tao et~al.(2018)Tao, Zhang, Liu and Nee}]{tao2018digital}
\bibinfo{author}{Tao\xfnm[ F.]}, \bibinfo{author}{Zhang\xfnm[ H.]}, \bibinfo{author}{Liu\xfnm[ A.]}, \bibinfo{author}{Nee\xfnm[ A.Y.]}.
\newblock \bibinfo{title}{Digital twin in industry: State-of-the-art}.
\newblock \bibinfo{journal}{IEEE Transactions on Industrial Informatics} \bibinfo{year}{2018};\bibinfo{volume}{15}(\bibinfo{number}{4}):\bibinfo{pages}{2405--2415}.
\bibitem[{Divya et~al.(2024)Divya, Prasanth, Devi~Sowndarya and Pham}]{divya2024digital}
\bibinfo{author}{Divya\xfnm[ V.]}, \bibinfo{author}{Prasanth\xfnm[ A.]}, \bibinfo{author}{Devi~Sowndarya\xfnm[ K.]}, \bibinfo{author}{Pham\xfnm[ C.T.]}.
\newblock \bibinfo{title}{Digital twins for hyperautomation for next generation}.
\newblock \bibinfo{journal}{Hyperautomation for Next-Generation Industries} \bibinfo{year}{2024};:\bibinfo{pages}{127--151}.
\bibitem[{Jeyalakshmi et~al.(2024)Jeyalakshmi, Prasanth, Yogeshwari and Elngar}]{jeyalakshmi2024digital}
\bibinfo{author}{Jeyalakshmi\xfnm[ S.]}, \bibinfo{author}{Prasanth\xfnm[ A.]}, \bibinfo{author}{Yogeshwari\xfnm[ M.]}, \bibinfo{author}{Elngar\xfnm[ A.A.]}.
\newblock \bibinfo{title}{Digital twins in flexible industrial production and smart manufacturing: Case study on intelligent logistics and supply chain management}.
\newblock \bibinfo{journal}{Digital Twins in Industrial Production and Smart Manufacturing: An Understanding of Principles, Enhancers, and Obstacles} \bibinfo{year}{2024};:\bibinfo{pages}{161--172}.
\bibitem[{Balasubramaniam et~al.(2024)Balasubramaniam, Sumina, Kumar and Prasanth}]{balasubramaniam2024machine}
\bibinfo{author}{Balasubramaniam\xfnm[ S.]}, \bibinfo{author}{Sumina\xfnm[ S.]}, \bibinfo{author}{Kumar\xfnm[ K.S.]}, \bibinfo{author}{Prasanth\xfnm[ A.]}.
\newblock \bibinfo{title}{Machine learning based models for implementing digital twins in healthcare industry}.
\newblock In: \bibinfo{booktitle}{Metaverse Technologies in Healthcare}. \bibinfo{publisher}{Elsevier}; \bibinfo{year}{2024}, p. \bibinfo{pages}{135--162}.
\bibitem[{Lawrence(2022)}]{lawrence2022epic}
\bibinfo{author}{Lawrence\xfnm[ K.]}.
\newblock \bibinfo{title}{Epic games acquires bandcamp: Will consolidation compromise the artist-friendly streaming platform?}
\newblock In: \bibinfo{booktitle}{SAGE Business Cases}. \bibinfo{publisher}{SAGE Publications: SAGE Business Cases Originals}; \bibinfo{year}{2022},.
\bibitem[{Fang et~al.(2021)Fang, Cai and Wang}]{fang2021metahuman}
\bibinfo{author}{Fang\xfnm[ Z.]}, \bibinfo{author}{Cai\xfnm[ L.]}, \bibinfo{author}{Wang\xfnm[ G.]}.
\newblock \bibinfo{title}{Metahuman creator the starting point of the metaverse}.
\newblock In: \bibinfo{booktitle}{2021 International Symposium on Computer Technology and Information Science (ISCTIS)}. \bibinfo{organization}{IEEE}; \bibinfo{year}{2021}, p. \bibinfo{pages}{154--157}.
\bibitem[{de~Souza et~al.(2021)de~Souza, Maciel and dos Santos~Nunes}]{de2021inspeccao}
\bibinfo{author}{de~Souza\xfnm[ R.L.]}, \bibinfo{author}{Maciel\xfnm[ C.]}, \bibinfo{author}{dos Santos~Nunes\xfnm[ E.P.]}.
\newblock \bibinfo{title}{Inspe{\c{c}}{\~a}o semi{\'o}tica no sistema do metahuman creator: avatares em foco}.
\newblock In: \bibinfo{booktitle}{Anais da XXI Escola Regional de Inform{\'a}tica de Mato Grosso}. \bibinfo{organization}{SBC}; \bibinfo{year}{2021}, p. \bibinfo{pages}{77--83}.
\bibitem[{Kwon(2020)}]{kwon2020study}
\bibinfo{author}{Kwon\xfnm[ D.H.]}.
\newblock \bibinfo{title}{A study on the meanings of half-life: Alyx and the success factors of vr games}.
\newblock \bibinfo{journal}{The Journal of the Korea Contents Association} \bibinfo{year}{2020};\bibinfo{volume}{20}(\bibinfo{number}{9}):\bibinfo{pages}{271--284}.
\bibitem[{Hu et~al.(2019)Hu, Tian, Yang, Taleb, Xiang and Hao}]{hu2019ready}
\bibinfo{author}{Hu\xfnm[ L.]}, \bibinfo{author}{Tian\xfnm[ Y.]}, \bibinfo{author}{Yang\xfnm[ J.]}, \bibinfo{author}{Taleb\xfnm[ T.]}, \bibinfo{author}{Xiang\xfnm[ L.]}, \bibinfo{author}{Hao\xfnm[ Y.]}.
\newblock \bibinfo{title}{Ready player one: Uav-clustering-based multi-task offloading for vehicular vr/ar gaming}.
\newblock \bibinfo{journal}{IEEE Network} \bibinfo{year}{2019};\bibinfo{volume}{33}(\bibinfo{number}{3}):\bibinfo{pages}{42--48}.
\bibitem[{Benford et~al.(1998)Benford, Greenhalgh, Reynard, Brown and Koleva}]{benford1998understanding}
\bibinfo{author}{Benford\xfnm[ S.]}, \bibinfo{author}{Greenhalgh\xfnm[ C.]}, \bibinfo{author}{Reynard\xfnm[ G.]}, \bibinfo{author}{Brown\xfnm[ C.]}, \bibinfo{author}{Koleva\xfnm[ B.]}.
\newblock \bibinfo{title}{Understanding and constructing shared spaces with mixed-reality boundaries}.
\newblock \bibinfo{journal}{ACM Transactions on computer-human interaction (TOCHI)} \bibinfo{year}{1998};\bibinfo{volume}{5}(\bibinfo{number}{3}):\bibinfo{pages}{185--223}.
\bibitem[{Wyrwoll(2014)}]{wyrwoll2014user}
\bibinfo{author}{Wyrwoll\xfnm[ C.]}.
\newblock \bibinfo{title}{User-generated content}.
\newblock In: \bibinfo{booktitle}{Social Media}. \bibinfo{publisher}{Springer}; \bibinfo{year}{2014}, p. \bibinfo{pages}{11--45}.
\bibitem[{Power and Teigland(2013)}]{power2013postcards}
\bibinfo{author}{Power\xfnm[ D.]}, \bibinfo{author}{Teigland\xfnm[ R.]}.
\newblock \bibinfo{title}{Postcards from the metaverse: An introduction to the immersive internet}.
\newblock In: \bibinfo{booktitle}{The Immersive Internet}. \bibinfo{publisher}{Springer}; \bibinfo{year}{2013}, p. \bibinfo{pages}{1--12}.
\bibitem[{Liu et~al.(2018)Liu, Chen, Tang, Xu and Piao}]{liu2018energy}
\bibinfo{author}{Liu\xfnm[ C.H.]}, \bibinfo{author}{Chen\xfnm[ Z.]}, \bibinfo{author}{Tang\xfnm[ J.]}, \bibinfo{author}{Xu\xfnm[ J.]}, \bibinfo{author}{Piao\xfnm[ C.]}.
\newblock \bibinfo{title}{Energy-efficient uav control for effective and fair communication coverage: A deep reinforcement learning approach}.
\newblock \bibinfo{journal}{IEEE Journal on Selected Areas in Communications} \bibinfo{year}{2018};\bibinfo{volume}{36}(\bibinfo{number}{9}):\bibinfo{pages}{2059--2070}.
\bibitem[{Montazerolghaem(2021)}]{montazerolghaem2021software}
\bibinfo{author}{Montazerolghaem\xfnm[ A.]}.
\newblock \bibinfo{title}{Software-defined internet of multimedia things: Energy-efficient and load-balanced resource management}.
\newblock \bibinfo{journal}{IEEE Internet of Things Journal} \bibinfo{year}{2021};.
\bibitem[{Ivkovic et~al.(2015)Ivkovic, Stavness, Gutwin and Sutcliffe}]{ivkovic2015quantifying}
\bibinfo{author}{Ivkovic\xfnm[ Z.]}, \bibinfo{author}{Stavness\xfnm[ I.]}, \bibinfo{author}{Gutwin\xfnm[ C.]}, \bibinfo{author}{Sutcliffe\xfnm[ S.]}.
\newblock \bibinfo{title}{Quantifying and mitigating the negative effects of local latencies on aiming in 3d shooter games}.
\newblock In: \bibinfo{booktitle}{Proceedings of the 33rd Annual ACM Conference on Human Factors in Computing Systems}. \bibinfo{year}{2015}, p. \bibinfo{pages}{135--144}.
\bibitem[{Lincoln et~al.(2016)Lincoln, Blate, Singh, Whitted, State, Lastra et~al.}]{lincoln2016motion}
\bibinfo{author}{Lincoln\xfnm[ P.]}, \bibinfo{author}{Blate\xfnm[ A.]}, \bibinfo{author}{Singh\xfnm[ M.]}, \bibinfo{author}{Whitted\xfnm[ T.]}, \bibinfo{author}{State\xfnm[ A.]}, \bibinfo{author}{Lastra\xfnm[ A.]}, et~al.
\newblock \bibinfo{title}{From motion to photons in 80 microseconds: Towards minimal latency for virtual and augmented reality}.
\newblock \bibinfo{journal}{IEEE transactions on visualization and computer graphics} \bibinfo{year}{2016};\bibinfo{volume}{22}(\bibinfo{number}{4}):\bibinfo{pages}{1367--1376}.
\bibitem[{Shafi et~al.(2017)Shafi, Molisch, Smith, Haustein, Zhu, De~Silva et~al.}]{shafi20175g}
\bibinfo{author}{Shafi\xfnm[ M.]}, \bibinfo{author}{Molisch\xfnm[ A.F.]}, \bibinfo{author}{Smith\xfnm[ P.J.]}, \bibinfo{author}{Haustein\xfnm[ T.]}, \bibinfo{author}{Zhu\xfnm[ P.]}, \bibinfo{author}{De~Silva\xfnm[ P.]}, et~al.
\newblock \bibinfo{title}{5g: A tutorial overview of standards, trials, challenges, deployment, and practice}.
\newblock \bibinfo{journal}{IEEE journal on selected areas in communications} \bibinfo{year}{2017};\bibinfo{volume}{35}(\bibinfo{number}{6}):\bibinfo{pages}{1201--1221}.
\bibitem[{Al-Falahy and Alani(2017)}]{al2017technologies}
\bibinfo{author}{Al-Falahy\xfnm[ N.]}, \bibinfo{author}{Alani\xfnm[ O.Y.]}.
\newblock \bibinfo{title}{Technologies for 5g networks: Challenges and opportunities}.
\newblock \bibinfo{journal}{IT Professional} \bibinfo{year}{2017};\bibinfo{volume}{19}(\bibinfo{number}{1}):\bibinfo{pages}{12--20}.
\bibitem[{Zhang et~al.(2015)Zhang, Cheng, Gamage, Zhang, Mark and Shen}]{zhang2015cloud}
\bibinfo{author}{Zhang\xfnm[ N.]}, \bibinfo{author}{Cheng\xfnm[ N.]}, \bibinfo{author}{Gamage\xfnm[ A.T.]}, \bibinfo{author}{Zhang\xfnm[ K.]}, \bibinfo{author}{Mark\xfnm[ J.W.]}, \bibinfo{author}{Shen\xfnm[ X.]}.
\newblock \bibinfo{title}{Cloud assisted hetnets toward 5g wireless networks}.
\newblock \bibinfo{journal}{IEEE communications magazine} \bibinfo{year}{2015};\bibinfo{volume}{53}(\bibinfo{number}{6}):\bibinfo{pages}{59--65}.
\bibitem[{Agiwal et~al.(2016)Agiwal, Roy and Saxena}]{agiwal2016next}
\bibinfo{author}{Agiwal\xfnm[ M.]}, \bibinfo{author}{Roy\xfnm[ A.]}, \bibinfo{author}{Saxena\xfnm[ N.]}.
\newblock \bibinfo{title}{Next generation 5g wireless networks: A comprehensive survey}.
\newblock \bibinfo{journal}{IEEE Communications Surveys \& Tutorials} \bibinfo{year}{2016};\bibinfo{volume}{18}(\bibinfo{number}{3}):\bibinfo{pages}{1617--1655}.
\bibitem[{Lee et~al.(2014)Lee, Chuah, Loo and Vinel}]{lee2014recent}
\bibinfo{author}{Lee\xfnm[ Y.L.]}, \bibinfo{author}{Chuah\xfnm[ T.C.]}, \bibinfo{author}{Loo\xfnm[ J.]}, \bibinfo{author}{Vinel\xfnm[ A.]}.
\newblock \bibinfo{title}{Recent advances in radio resource management for heterogeneous lte/lte-a networks}.
\newblock \bibinfo{journal}{IEEE Communications Surveys \& Tutorials} \bibinfo{year}{2014};\bibinfo{volume}{16}(\bibinfo{number}{4}):\bibinfo{pages}{2142--2180}.
\bibitem[{Talwar et~al.(2014)Talwar, Choudhury, Dimou, Aryafar, Bangerter and Stewart}]{talwar2014enabling}
\bibinfo{author}{Talwar\xfnm[ S.]}, \bibinfo{author}{Choudhury\xfnm[ D.]}, \bibinfo{author}{Dimou\xfnm[ K.]}, \bibinfo{author}{Aryafar\xfnm[ E.]}, \bibinfo{author}{Bangerter\xfnm[ B.]}, \bibinfo{author}{Stewart\xfnm[ K.]}.
\newblock \bibinfo{title}{Enabling technologies and architectures for 5g wireless}.
\newblock In: \bibinfo{booktitle}{2014 IEEE MTT-S International Microwave Symposium (IMS2014)}. \bibinfo{organization}{IEEE}; \bibinfo{year}{2014}, p. \bibinfo{pages}{1--4}.
\bibitem[{Dhawankar et~al.(2021)Dhawankar, Kumar, Crespi, Busawon, Qureshi, Javed et~al.}]{dhawankar2021next}
\bibinfo{author}{Dhawankar\xfnm[ P.]}, \bibinfo{author}{Kumar\xfnm[ A.]}, \bibinfo{author}{Crespi\xfnm[ N.]}, \bibinfo{author}{Busawon\xfnm[ K.]}, \bibinfo{author}{Qureshi\xfnm[ K.N.]}, \bibinfo{author}{Javed\xfnm[ I.T.]}, et~al.
\newblock \bibinfo{title}{Next-generation indoor wireless systems: Compatibility and migration case study}.
\newblock \bibinfo{journal}{IEEE Access} \bibinfo{year}{2021};.
\bibitem[{Federici et~al.(2016)Federici, Ma and Moeller}]{federici2016review}
\bibinfo{author}{Federici\xfnm[ J.F.]}, \bibinfo{author}{Ma\xfnm[ J.]}, \bibinfo{author}{Moeller\xfnm[ L.]}.
\newblock \bibinfo{title}{Review of weather impact on outdoor terahertz wireless communication links}.
\newblock \bibinfo{journal}{Nano Communication Networks} \bibinfo{year}{2016};\bibinfo{volume}{10}:\bibinfo{pages}{13--26}.
\bibitem[{Marfievici et~al.(2013)Marfievici, Murphy, Picco, Ossi and Cagnacci}]{marfievici2013environmental}
\bibinfo{author}{Marfievici\xfnm[ R.]}, \bibinfo{author}{Murphy\xfnm[ A.L.]}, \bibinfo{author}{Picco\xfnm[ G.P.]}, \bibinfo{author}{Ossi\xfnm[ F.]}, \bibinfo{author}{Cagnacci\xfnm[ F.]}.
\newblock \bibinfo{title}{How environmental factors impact outdoor wireless sensor networks: a case study}.
\newblock In: \bibinfo{booktitle}{2013 IEEE 10th international conference on mobile ad-hoc and sensor systems}. \bibinfo{organization}{IEEE}; \bibinfo{year}{2013}, p. \bibinfo{pages}{565--573}.
\bibitem[{Wang et~al.(2020)Wang, Zhao, Lv, Ma, Zhang and Lin}]{wang2020optimizing}
\bibinfo{author}{Wang\xfnm[ Q.]}, \bibinfo{author}{Zhao\xfnm[ X.]}, \bibinfo{author}{Lv\xfnm[ Z.]}, \bibinfo{author}{Ma\xfnm[ X.]}, \bibinfo{author}{Zhang\xfnm[ R.]}, \bibinfo{author}{Lin\xfnm[ Y.]}.
\newblock \bibinfo{title}{Optimizing the ultra-dense 5g base stations in urban outdoor areas: Coupling gis and heuristic optimization}.
\newblock \bibinfo{journal}{Sustainable Cities and Society} \bibinfo{year}{2020};\bibinfo{volume}{63}:\bibinfo{pages}{102445}.
\bibitem[{Kumari et~al.(2020)Kumari, Kumar and Prasad}]{kumari2020optimization}
\bibinfo{author}{Kumari\xfnm[ M.S.]}, \bibinfo{author}{Kumar\xfnm[ N.]}, \bibinfo{author}{Prasad\xfnm[ R.]}.
\newblock \bibinfo{title}{Optimization of street canyon outdoor channel deployment geometry for mmwave 5g communication}.
\newblock \bibinfo{journal}{AEU-International Journal of Electronics and Communications} \bibinfo{year}{2020};\bibinfo{volume}{125}:\bibinfo{pages}{153368}.
\bibitem[{Moore(1965)}]{moore1965moore}
\bibinfo{author}{Moore\xfnm[ G.]}.
\newblock \bibinfo{title}{Moore’s law}.
\newblock \bibinfo{journal}{Electronics Magazine} \bibinfo{year}{1965};\bibinfo{volume}{38}(\bibinfo{number}{8}):\bibinfo{pages}{114}.
\bibitem[{Torrejon et~al.(2017)Torrejon, Riou, Araujo, Tsunegi, Khalsa, Querlioz et~al.}]{torrejon2017neuromorphic}
\bibinfo{author}{Torrejon\xfnm[ J.]}, \bibinfo{author}{Riou\xfnm[ M.]}, \bibinfo{author}{Araujo\xfnm[ F.A.]}, \bibinfo{author}{Tsunegi\xfnm[ S.]}, \bibinfo{author}{Khalsa\xfnm[ G.]}, \bibinfo{author}{Querlioz\xfnm[ D.]}, et~al.
\newblock \bibinfo{title}{Neuromorphic computing with nanoscale spintronic oscillators}.
\newblock \bibinfo{journal}{Nature} \bibinfo{year}{2017};\bibinfo{volume}{547}(\bibinfo{number}{7664}):\bibinfo{pages}{428--431}.
\bibitem[{Vandoorne et~al.(2014)Vandoorne, Mechet, Van~Vaerenbergh, Fiers, Morthier, Verstraeten et~al.}]{vandoorne2014experimental}
\bibinfo{author}{Vandoorne\xfnm[ K.]}, \bibinfo{author}{Mechet\xfnm[ P.]}, \bibinfo{author}{Van~Vaerenbergh\xfnm[ T.]}, \bibinfo{author}{Fiers\xfnm[ M.]}, \bibinfo{author}{Morthier\xfnm[ G.]}, \bibinfo{author}{Verstraeten\xfnm[ D.]}, et~al.
\newblock \bibinfo{title}{Experimental demonstration of reservoir computing on a silicon photonics chip}.
\newblock \bibinfo{journal}{Nature communications} \bibinfo{year}{2014};\bibinfo{volume}{5}(\bibinfo{number}{1}):\bibinfo{pages}{1--6}.
\bibitem[{Zhong et~al.(2021)Zhong, Tang, Li, Gao, Qian and Wu}]{zhong2021dynamic}
\bibinfo{author}{Zhong\xfnm[ Y.]}, \bibinfo{author}{Tang\xfnm[ J.]}, \bibinfo{author}{Li\xfnm[ X.]}, \bibinfo{author}{Gao\xfnm[ B.]}, \bibinfo{author}{Qian\xfnm[ H.]}, \bibinfo{author}{Wu\xfnm[ H.]}.
\newblock \bibinfo{title}{Dynamic memristor-based reservoir computing for high-efficiency temporal signal processing}.
\newblock \bibinfo{journal}{Nature Communications} \bibinfo{year}{2021};\bibinfo{volume}{12}(\bibinfo{number}{1}):\bibinfo{pages}{1--9}.
\bibitem[{Dale(2021)}]{dale2021gpt}
\bibinfo{author}{Dale\xfnm[ R.]}.
\newblock \bibinfo{title}{Gpt-3: What’s it good for?}
\newblock \bibinfo{journal}{Natural Language Engineering} \bibinfo{year}{2021};\bibinfo{volume}{27}(\bibinfo{number}{1}):\bibinfo{pages}{113--118}.
\bibitem[{Lin et~al.(2021)Lin, Men, Yang, Zhou, Ding, Zhang et~al.}]{lin2021m6}
\bibinfo{author}{Lin\xfnm[ J.]}, \bibinfo{author}{Men\xfnm[ R.]}, \bibinfo{author}{Yang\xfnm[ A.]}, \bibinfo{author}{Zhou\xfnm[ C.]}, \bibinfo{author}{Ding\xfnm[ M.]}, \bibinfo{author}{Zhang\xfnm[ Y.]}, et~al.
\newblock \bibinfo{title}{M6: A chinese multimodal pretrainer}.
\newblock \bibinfo{journal}{arXiv preprint arXiv:210300823} \bibinfo{year}{2021};.
\bibitem[{Cao et~al.(2020)Cao, Liu, Cheng, Zheng, Li, Wu et~al.}]{cao2020polardb}
\bibinfo{author}{Cao\xfnm[ W.]}, \bibinfo{author}{Liu\xfnm[ Y.]}, \bibinfo{author}{Cheng\xfnm[ Z.]}, \bibinfo{author}{Zheng\xfnm[ N.]}, \bibinfo{author}{Li\xfnm[ W.]}, \bibinfo{author}{Wu\xfnm[ W.]}, et~al.
\newblock \bibinfo{title}{$\{$POLARDB$\}$ meets computational storage: Efficiently support analytical workloads in cloud-native relational database}.
\newblock In: \bibinfo{booktitle}{18th $\{$USENIX$\}$ Conference on File and Storage Technologies ($\{$FAST$\}$ 20)}. \bibinfo{year}{2020}, p. \bibinfo{pages}{29--41}.
\bibitem[{Russell and Norvig(2002)}]{russell2002artificial}
\bibinfo{author}{Russell\xfnm[ S.]}, \bibinfo{author}{Norvig\xfnm[ P.]}.
\newblock \bibinfo{title}{Artificial intelligence: a modern approach} \bibinfo{year}{2002};.
\bibitem[{Thiebes et~al.(2021)Thiebes, Lins and Sunyaev}]{thiebes2021trustworthy}
\bibinfo{author}{Thiebes\xfnm[ S.]}, \bibinfo{author}{Lins\xfnm[ S.]}, \bibinfo{author}{Sunyaev\xfnm[ A.]}.
\newblock \bibinfo{title}{Trustworthy artificial intelligence}.
\newblock \bibinfo{journal}{Electronic Markets} \bibinfo{year}{2021};\bibinfo{volume}{31}(\bibinfo{number}{2}):\bibinfo{pages}{447--464}.
\bibitem[{Rao and Frtunikj(2018)}]{rao2018deep}
\bibinfo{author}{Rao\xfnm[ Q.]}, \bibinfo{author}{Frtunikj\xfnm[ J.]}.
\newblock \bibinfo{title}{Deep learning for self-driving cars: Chances and challenges}.
\newblock In: \bibinfo{booktitle}{Proceedings of the 1st International Workshop on Software Engineering for AI in Autonomous Systems}. \bibinfo{year}{2018}, p. \bibinfo{pages}{35--38}.
\bibitem[{Li et~al.(2020)Li, Mu, Li and Peng}]{li2020review}
\bibinfo{author}{Li\xfnm[ L.]}, \bibinfo{author}{Mu\xfnm[ X.]}, \bibinfo{author}{Li\xfnm[ S.]}, \bibinfo{author}{Peng\xfnm[ H.]}.
\newblock \bibinfo{title}{A review of face recognition technology}.
\newblock \bibinfo{journal}{IEEE Access} \bibinfo{year}{2020};\bibinfo{volume}{8}:\bibinfo{pages}{139110--139120}.
\bibitem[{Park and Han(2018)}]{park2018methodologic}
\bibinfo{author}{Park\xfnm[ S.H.]}, \bibinfo{author}{Han\xfnm[ K.]}.
\newblock \bibinfo{title}{Methodologic guide for evaluating clinical performance and effect of artificial intelligence technology for medical diagnosis and prediction}.
\newblock \bibinfo{journal}{Radiology} \bibinfo{year}{2018};\bibinfo{volume}{286}(\bibinfo{number}{3}):\bibinfo{pages}{800--809}.
\bibitem[{Shaker et~al.(2016)Shaker, Togelius and Nelson}]{shaker2016procedural}
\bibinfo{author}{Shaker\xfnm[ N.]}, \bibinfo{author}{Togelius\xfnm[ J.]}, \bibinfo{author}{Nelson\xfnm[ M.J.]}.
\newblock \bibinfo{title}{Procedural content generation in games}.
\newblock \bibinfo{publisher}{Springer}; \bibinfo{year}{2016}.
\bibitem[{Xu et~al.(2016)Xu, Wu, Lin, Tang and Hua}]{xu2016automatic}
\bibinfo{author}{Xu\xfnm[ H.]}, \bibinfo{author}{Wu\xfnm[ Z.]}, \bibinfo{author}{Lin\xfnm[ T.]}, \bibinfo{author}{Tang\xfnm[ N.]}, \bibinfo{author}{Hua\xfnm[ L.]}.
\newblock \bibinfo{title}{Automatic content generation in tetris game based on emotion modeling}.
\newblock In: \bibinfo{booktitle}{2016 Nicograph International (NicoInt)}. \bibinfo{organization}{IEEE}; \bibinfo{year}{2016}, p. \bibinfo{pages}{139--139}.
\bibitem[{Games(2016)}]{games2016no}
\bibinfo{author}{Games\xfnm[ H.]}.
\newblock \bibinfo{title}{No man’s sky}.
\newblock \bibinfo{journal}{Hello Games} \bibinfo{year}{2016};.
\bibitem[{Drago(2019)}]{drago2019no}
\bibinfo{author}{Drago\xfnm[ S.]}.
\newblock \bibinfo{title}{No man's sky: Utilizing maritime law to address the need for space debris removal technology}.
\newblock \bibinfo{journal}{Santa Clara L Rev} \bibinfo{year}{2019};\bibinfo{volume}{59}:\bibinfo{pages}{389}.
\bibitem[{Kobbacy(2012)}]{kobbacy2012application}
\bibinfo{author}{Kobbacy\xfnm[ K.A.]}.
\newblock \bibinfo{title}{Application of artificial intelligence in maintenance modelling and management}.
\newblock \bibinfo{journal}{IFAC Proceedings Volumes} \bibinfo{year}{2012};\bibinfo{volume}{45}(\bibinfo{number}{31}):\bibinfo{pages}{54--59}.
\bibitem[{Marginean et~al.(2019)Marginean, Bader, Chandra, Harman, Jia, Mao et~al.}]{marginean2019sapfix}
\bibinfo{author}{Marginean\xfnm[ A.]}, \bibinfo{author}{Bader\xfnm[ J.]}, \bibinfo{author}{Chandra\xfnm[ S.]}, \bibinfo{author}{Harman\xfnm[ M.]}, \bibinfo{author}{Jia\xfnm[ Y.]}, \bibinfo{author}{Mao\xfnm[ K.]}, et~al.
\newblock \bibinfo{title}{Sapfix: Automated end-to-end repair at scale}.
\newblock In: \bibinfo{booktitle}{2019 IEEE/ACM 41st International Conference on Software Engineering: Software Engineering in Practice (ICSE-SEIP)}. \bibinfo{organization}{IEEE}; \bibinfo{year}{2019}, p. \bibinfo{pages}{269--278}.
\bibitem[{Mao et~al.(2016)Mao, Harman and Jia}]{mao2016sapienz}
\bibinfo{author}{Mao\xfnm[ K.]}, \bibinfo{author}{Harman\xfnm[ M.]}, \bibinfo{author}{Jia\xfnm[ Y.]}.
\newblock \bibinfo{title}{Sapienz: Multi-objective automated testing for android applications}.
\newblock In: \bibinfo{booktitle}{Proceedings of the 25th International Symposium on Software Testing and Analysis}. \bibinfo{year}{2016}, p. \bibinfo{pages}{94--105}.
\bibitem[{Diaconescu(2013)}]{diaconescu2013automated}
\bibinfo{author}{Diaconescu\xfnm[ A.I.]}.
\newblock \bibinfo{title}{Automated code review for fault injection} \bibinfo{year}{2013};.
\bibitem[{Bassuday and Ahmed(2019)}]{bassuday2019fault}
\bibinfo{author}{Bassuday\xfnm[ K.]}, \bibinfo{author}{Ahmed\xfnm[ M.]}.
\newblock \bibinfo{title}{Fault prediction in android systems through ai} \bibinfo{year}{2019};.
\bibitem[{Fletcher(2018)}]{fletcher2018witness}
\bibinfo{author}{Fletcher\xfnm[ D.]}.
\newblock \bibinfo{title}{The witness}.
\newblock In: \bibinfo{booktitle}{Codify}. \bibinfo{publisher}{Routledge}; \bibinfo{year}{2018}, p. \bibinfo{pages}{254--266}.
\bibitem[{Bonner(2016)}]{bonner2016puzzle}
\bibinfo{author}{Bonner\xfnm[ M.]}.
\newblock \bibinfo{title}{Puzzle about the island--multi-perspective studies on knowledge in the witness}.
\newblock In: \bibinfo{booktitle}{Proceedings of the Philosophy of Computer Games Conference, Malta}. \bibinfo{year}{2016},.
\bibitem[{Hayes(2008)}]{hayes2008cloud}
\bibinfo{author}{Hayes\xfnm[ B.]}.
\newblock \bibinfo{title}{Cloud computing}.
\newblock \bibinfo{year}{2008}.
\bibitem[{Sadiku et~al.(2014)Sadiku, Musa and Momoh}]{sadiku2014cloud}
\bibinfo{author}{Sadiku\xfnm[ M.N.]}, \bibinfo{author}{Musa\xfnm[ S.M.]}, \bibinfo{author}{Momoh\xfnm[ O.D.]}.
\newblock \bibinfo{title}{Cloud computing: opportunities and challenges}.
\newblock \bibinfo{journal}{IEEE potentials} \bibinfo{year}{2014};\bibinfo{volume}{33}(\bibinfo{number}{1}):\bibinfo{pages}{34--36}.
\bibitem[{Antonescu et~al.(2012)Antonescu, Robinson and Braun}]{antonescu2012dynamic}
\bibinfo{author}{Antonescu\xfnm[ A.F.]}, \bibinfo{author}{Robinson\xfnm[ P.]}, \bibinfo{author}{Braun\xfnm[ T.]}.
\newblock \bibinfo{title}{Dynamic topology orchestration for distributed cloud-based applications}.
\newblock In: \bibinfo{booktitle}{2012 Second Symposium on Network Cloud Computing and Applications}. \bibinfo{organization}{IEEE}; \bibinfo{year}{2012}, p. \bibinfo{pages}{116--123}.
\bibitem[{Deng et~al.(2010)Deng, Huang, Han and Deng}]{deng2010fault}
\bibinfo{author}{Deng\xfnm[ J.]}, \bibinfo{author}{Huang\xfnm[ S.C.H.]}, \bibinfo{author}{Han\xfnm[ Y.S.]}, \bibinfo{author}{Deng\xfnm[ J.H.]}.
\newblock \bibinfo{title}{Fault-tolerant and reliable computation in cloud computing}.
\newblock In: \bibinfo{booktitle}{2010 IEEE Globecom Workshops}. \bibinfo{organization}{IEEE}; \bibinfo{year}{2010}, p. \bibinfo{pages}{1601--1605}.
\bibitem[{Liu et~al.(2017)Liu, Wang, Liu, Peng and Wu}]{liu2017achieving}
\bibinfo{author}{Liu\xfnm[ Q.]}, \bibinfo{author}{Wang\xfnm[ G.]}, \bibinfo{author}{Liu\xfnm[ X.]}, \bibinfo{author}{Peng\xfnm[ T.]}, \bibinfo{author}{Wu\xfnm[ J.]}.
\newblock \bibinfo{title}{Achieving reliable and secure services in cloud computing environments}.
\newblock \bibinfo{journal}{Computers \& Electrical Engineering} \bibinfo{year}{2017};\bibinfo{volume}{59}:\bibinfo{pages}{153--164}.
\bibitem[{Hameed et~al.(2016)Hameed, Khoshkbarforoushha, Ranjan, Jayaraman, Kolodziej, Balaji et~al.}]{hameed2016survey}
\bibinfo{author}{Hameed\xfnm[ A.]}, \bibinfo{author}{Khoshkbarforoushha\xfnm[ A.]}, \bibinfo{author}{Ranjan\xfnm[ R.]}, \bibinfo{author}{Jayaraman\xfnm[ P.P.]}, \bibinfo{author}{Kolodziej\xfnm[ J.]}, \bibinfo{author}{Balaji\xfnm[ P.]}, et~al.
\newblock \bibinfo{title}{A survey and taxonomy on energy efficient resource allocation techniques for cloud computing systems}.
\newblock \bibinfo{journal}{Computing} \bibinfo{year}{2016};\bibinfo{volume}{98}(\bibinfo{number}{7}):\bibinfo{pages}{751--774}.
\bibitem[{Bui et~al.(2017)Bui, Yoon, Huh, Jun and Lee}]{bui2017energy}
\bibinfo{author}{Bui\xfnm[ D.M.]}, \bibinfo{author}{Yoon\xfnm[ Y.]}, \bibinfo{author}{Huh\xfnm[ E.N.]}, \bibinfo{author}{Jun\xfnm[ S.]}, \bibinfo{author}{Lee\xfnm[ S.]}.
\newblock \bibinfo{title}{Energy efficiency for cloud computing system based on predictive optimization}.
\newblock \bibinfo{journal}{Journal of Parallel and Distributed Computing} \bibinfo{year}{2017};\bibinfo{volume}{102}:\bibinfo{pages}{103--114}.
\bibitem[{Yang et~al.(2010)Yang, Zhou, Liang, He and Sun}]{yang2010sevice}
\bibinfo{author}{Yang\xfnm[ Y.]}, \bibinfo{author}{Zhou\xfnm[ Y.]}, \bibinfo{author}{Liang\xfnm[ L.]}, \bibinfo{author}{He\xfnm[ D.]}, \bibinfo{author}{Sun\xfnm[ Z.]}.
\newblock \bibinfo{title}{A sevice-oriented broker for bulk data transfer in cloud computing}.
\newblock In: \bibinfo{booktitle}{2010 Ninth International Conference on Grid and Cloud Computing}. \bibinfo{organization}{IEEE}; \bibinfo{year}{2010}, p. \bibinfo{pages}{264--269}.
\bibitem[{Banzai et~al.(2010)Banzai, Koizumi, Kanbayashi, Imada, Hanawa and Sato}]{banzai2010d}
\bibinfo{author}{Banzai\xfnm[ T.]}, \bibinfo{author}{Koizumi\xfnm[ H.]}, \bibinfo{author}{Kanbayashi\xfnm[ R.]}, \bibinfo{author}{Imada\xfnm[ T.]}, \bibinfo{author}{Hanawa\xfnm[ T.]}, \bibinfo{author}{Sato\xfnm[ M.]}.
\newblock \bibinfo{title}{D-cloud: Design of a software testing environment for reliable distributed systems using cloud computing technology}.
\newblock In: \bibinfo{booktitle}{2010 10th IEEE/ACM International Conference on Cluster, Cloud and Grid Computing}. \bibinfo{organization}{IEEE}; \bibinfo{year}{2010}, p. \bibinfo{pages}{631--636}.
\bibitem[{Ardagna et~al.(2014)Ardagna, Casale, Ciavotta, P{\'e}rez and Wang}]{ardagna2014quality}
\bibinfo{author}{Ardagna\xfnm[ D.]}, \bibinfo{author}{Casale\xfnm[ G.]}, \bibinfo{author}{Ciavotta\xfnm[ M.]}, \bibinfo{author}{P{\'e}rez\xfnm[ J.F.]}, \bibinfo{author}{Wang\xfnm[ W.]}.
\newblock \bibinfo{title}{Quality-of-service in cloud computing: modeling techniques and their applications}.
\newblock \bibinfo{journal}{Journal of Internet Services and Applications} \bibinfo{year}{2014};\bibinfo{volume}{5}(\bibinfo{number}{1}):\bibinfo{pages}{1--17}.
\bibitem[{Moqbel et~al.(2014)Moqbel, Bartelt and Al-Suqri}]{moqbel2014study}
\bibinfo{author}{Moqbel\xfnm[ M.]}, \bibinfo{author}{Bartelt\xfnm[ V.L.]}, \bibinfo{author}{Al-Suqri\xfnm[ M.]}.
\newblock \bibinfo{title}{A study of personal cloud computing: compatibility, social influence, and moderating role of perceived familiarity} \bibinfo{year}{2014};.
\bibitem[{Paul and Ghose(2018)}]{paul2018novel}
\bibinfo{author}{Paul\xfnm[ P.K.]}, \bibinfo{author}{Ghose\xfnm[ M.K.]}.
\newblock \bibinfo{title}{A novel educational proposal and strategies toward promoting cloud computing, big data, and human--computer interaction in engineering colleges and universities}.
\newblock In: \bibinfo{booktitle}{Advances in Smart Grid and Renewable Energy}. \bibinfo{publisher}{Springer}; \bibinfo{year}{2018}, p. \bibinfo{pages}{93--102}.
\bibitem[{Shi et~al.(2016)Shi, Cao, Zhang, Li and Xu}]{shi2016edge}
\bibinfo{author}{Shi\xfnm[ W.]}, \bibinfo{author}{Cao\xfnm[ J.]}, \bibinfo{author}{Zhang\xfnm[ Q.]}, \bibinfo{author}{Li\xfnm[ Y.]}, \bibinfo{author}{Xu\xfnm[ L.]}.
\newblock \bibinfo{title}{Edge computing: Vision and challenges}.
\newblock \bibinfo{journal}{IEEE internet of things journal} \bibinfo{year}{2016};\bibinfo{volume}{3}(\bibinfo{number}{5}):\bibinfo{pages}{637--646}.
\bibitem[{Shi and Dustdar(2016)}]{shi2016promise}
\bibinfo{author}{Shi\xfnm[ W.]}, \bibinfo{author}{Dustdar\xfnm[ S.]}.
\newblock \bibinfo{title}{The promise of edge computing}.
\newblock \bibinfo{journal}{Computer} \bibinfo{year}{2016};\bibinfo{volume}{49}(\bibinfo{number}{5}):\bibinfo{pages}{78--81}.
\bibitem[{Peng et~al.(2018)Peng, Leung, Xu, Zheng, Wang and Huang}]{peng2018survey}
\bibinfo{author}{Peng\xfnm[ K.]}, \bibinfo{author}{Leung\xfnm[ V.]}, \bibinfo{author}{Xu\xfnm[ X.]}, \bibinfo{author}{Zheng\xfnm[ L.]}, \bibinfo{author}{Wang\xfnm[ J.]}, \bibinfo{author}{Huang\xfnm[ Q.]}.
\newblock \bibinfo{title}{A survey on mobile edge computing: Focusing on service adoption and provision}.
\newblock \bibinfo{journal}{Wireless Communications and Mobile Computing} \bibinfo{year}{2018};\bibinfo{volume}{2018}.
\bibitem[{Dolui and Datta(2017)}]{dolui2017comparison}
\bibinfo{author}{Dolui\xfnm[ K.]}, \bibinfo{author}{Datta\xfnm[ S.K.]}.
\newblock \bibinfo{title}{Comparison of edge computing implementations: Fog computing, cloudlet and mobile edge computing}.
\newblock In: \bibinfo{booktitle}{2017 Global Internet of Things Summit (GIoTS)}. \bibinfo{organization}{IEEE}; \bibinfo{year}{2017}, p. \bibinfo{pages}{1--6}.
\bibitem[{Lin et~al.(2020)Lin, Zeadally, Chen, Labiod and Wang}]{lin2020survey}
\bibinfo{author}{Lin\xfnm[ H.]}, \bibinfo{author}{Zeadally\xfnm[ S.]}, \bibinfo{author}{Chen\xfnm[ Z.]}, \bibinfo{author}{Labiod\xfnm[ H.]}, \bibinfo{author}{Wang\xfnm[ L.]}.
\newblock \bibinfo{title}{A survey on computation offloading modeling for edge computing}.
\newblock \bibinfo{journal}{Journal of Network and Computer Applications} \bibinfo{year}{2020};\bibinfo{volume}{169}:\bibinfo{pages}{102781}.
\bibitem[{Ha et~al.(2014)Ha, Chen, Hu, Richter, Pillai and Satyanarayanan}]{ha2014towards}
\bibinfo{author}{Ha\xfnm[ K.]}, \bibinfo{author}{Chen\xfnm[ Z.]}, \bibinfo{author}{Hu\xfnm[ W.]}, \bibinfo{author}{Richter\xfnm[ W.]}, \bibinfo{author}{Pillai\xfnm[ P.]}, \bibinfo{author}{Satyanarayanan\xfnm[ M.]}.
\newblock \bibinfo{title}{Towards wearable cognitive assistance}.
\newblock In: \bibinfo{booktitle}{Proceedings of the 12th annual international conference on Mobile systems, applications, and services}. \bibinfo{year}{2014}, p. \bibinfo{pages}{68--81}.
\bibitem[{Xiao et~al.(2020)Xiao, Liu, Li and Li}]{xiao2020system}
\bibinfo{author}{Xiao\xfnm[ S.]}, \bibinfo{author}{Liu\xfnm[ C.]}, \bibinfo{author}{Li\xfnm[ K.]}, \bibinfo{author}{Li\xfnm[ K.]}.
\newblock \bibinfo{title}{System delay optimization for mobile edge computing}.
\newblock \bibinfo{journal}{Future Generation Computer Systems} \bibinfo{year}{2020};\bibinfo{volume}{109}:\bibinfo{pages}{17--28}.
\bibitem[{Niu et~al.(2019)Niu, Shao, Xin, Zhou, Guo, Chen et~al.}]{niu2019workload}
\bibinfo{author}{Niu\xfnm[ X.]}, \bibinfo{author}{Shao\xfnm[ S.]}, \bibinfo{author}{Xin\xfnm[ C.]}, \bibinfo{author}{Zhou\xfnm[ J.]}, \bibinfo{author}{Guo\xfnm[ S.]}, \bibinfo{author}{Chen\xfnm[ X.]}, et~al.
\newblock \bibinfo{title}{Workload allocation mechanism for minimum service delay in edge computing-based power internet of things}.
\newblock \bibinfo{journal}{IEEE Access} \bibinfo{year}{2019};\bibinfo{volume}{7}:\bibinfo{pages}{83771--83784}.
\bibitem[{Zhao et~al.(2017)Zhao, Zhou, Guo and Niu}]{zhao2017tasks}
\bibinfo{author}{Zhao\xfnm[ T.]}, \bibinfo{author}{Zhou\xfnm[ S.]}, \bibinfo{author}{Guo\xfnm[ X.]}, \bibinfo{author}{Niu\xfnm[ Z.]}.
\newblock \bibinfo{title}{Tasks scheduling and resource allocation in heterogeneous cloud for delay-bounded mobile edge computing}.
\newblock In: \bibinfo{booktitle}{2017 IEEE international conference on communications (ICC)}. \bibinfo{organization}{IEEE}; \bibinfo{year}{2017}, p. \bibinfo{pages}{1--7}.
\bibitem[{Mao et~al.(2019)Mao, Leng, Maharjan and Zhang}]{mao2019energy}
\bibinfo{author}{Mao\xfnm[ S.]}, \bibinfo{author}{Leng\xfnm[ S.]}, \bibinfo{author}{Maharjan\xfnm[ S.]}, \bibinfo{author}{Zhang\xfnm[ Y.]}.
\newblock \bibinfo{title}{Energy efficiency and delay tradeoff for wireless powered mobile-edge computing systems with multi-access schemes}.
\newblock \bibinfo{journal}{IEEE Transactions on Wireless Communications} \bibinfo{year}{2019};\bibinfo{volume}{19}(\bibinfo{number}{3}):\bibinfo{pages}{1855--1867}.
\bibitem[{Leenes(2007)}]{leenes2007privacy}
\bibinfo{author}{Leenes\xfnm[ R.]}.
\newblock \bibinfo{title}{Privacy in the metaverse}.
\newblock In: \bibinfo{booktitle}{IFIP International Summer School on the Future of Identity in the Information Society}. \bibinfo{organization}{Springer}; \bibinfo{year}{2007}, p. \bibinfo{pages}{95--112}.
\bibitem[{Zhang et~al.(2018)Zhang, Chen, Zhao, Cheng and Hu}]{zhang2018data}
\bibinfo{author}{Zhang\xfnm[ J.]}, \bibinfo{author}{Chen\xfnm[ B.]}, \bibinfo{author}{Zhao\xfnm[ Y.]}, \bibinfo{author}{Cheng\xfnm[ X.]}, \bibinfo{author}{Hu\xfnm[ F.]}.
\newblock \bibinfo{title}{Data security and privacy-preserving in edge computing paradigm: Survey and open issues}.
\newblock \bibinfo{journal}{IEEE access} \bibinfo{year}{2018};\bibinfo{volume}{6}:\bibinfo{pages}{18209--18237}.
\bibitem[{Li et~al.(2018)Li, Liu, Wu, Kumari and Rodrigues}]{li2018privacy}
\bibinfo{author}{Li\xfnm[ X.]}, \bibinfo{author}{Liu\xfnm[ S.]}, \bibinfo{author}{Wu\xfnm[ F.]}, \bibinfo{author}{Kumari\xfnm[ S.]}, \bibinfo{author}{Rodrigues\xfnm[ J.J.]}.
\newblock \bibinfo{title}{Privacy preserving data aggregation scheme for mobile edge computing assisted iot applications}.
\newblock \bibinfo{journal}{IEEE Internet of Things Journal} \bibinfo{year}{2018};\bibinfo{volume}{6}(\bibinfo{number}{3}):\bibinfo{pages}{4755--4763}.
\bibitem[{Gheisari et~al.(2020)Gheisari, Wang and Chen}]{gheisari2020edge}
\bibinfo{author}{Gheisari\xfnm[ M.]}, \bibinfo{author}{Wang\xfnm[ G.]}, \bibinfo{author}{Chen\xfnm[ S.]}.
\newblock \bibinfo{title}{An edge computing-enhanced internet of things framework for privacy-preserving in smart city}.
\newblock \bibinfo{journal}{Computers \& Electrical Engineering} \bibinfo{year}{2020};\bibinfo{volume}{81}:\bibinfo{pages}{106504}.
\bibitem[{Ranaweera et~al.(2021)Ranaweera, Jurcut and Liyanage}]{ranaweera2021survey}
\bibinfo{author}{Ranaweera\xfnm[ P.]}, \bibinfo{author}{Jurcut\xfnm[ A.D.]}, \bibinfo{author}{Liyanage\xfnm[ M.]}.
\newblock \bibinfo{title}{Survey on multi-access edge computing security and privacy}.
\newblock \bibinfo{journal}{IEEE Communications Surveys \& Tutorials} \bibinfo{year}{2021};\bibinfo{volume}{23}(\bibinfo{number}{2}):\bibinfo{pages}{1078--1124}.
\bibitem[{Mehrabi et~al.(2019)Mehrabi, You, Latzko, Salah, Reisslein and Fitzek}]{mehrabi2019device}
\bibinfo{author}{Mehrabi\xfnm[ M.]}, \bibinfo{author}{You\xfnm[ D.]}, \bibinfo{author}{Latzko\xfnm[ V.]}, \bibinfo{author}{Salah\xfnm[ H.]}, \bibinfo{author}{Reisslein\xfnm[ M.]}, \bibinfo{author}{Fitzek\xfnm[ F.H.]}.
\newblock \bibinfo{title}{Device-enhanced mec: Multi-access edge computing (mec) aided by end device computation and caching: A survey}.
\newblock \bibinfo{journal}{IEEE Access} \bibinfo{year}{2019};\bibinfo{volume}{7}:\bibinfo{pages}{166079--166108}.
\bibitem[{Hu et~al.(2015)Hu, Patel, Sabella, Sprecher and Young}]{hu2015mobile}
\bibinfo{author}{Hu\xfnm[ Y.C.]}, \bibinfo{author}{Patel\xfnm[ M.]}, \bibinfo{author}{Sabella\xfnm[ D.]}, \bibinfo{author}{Sprecher\xfnm[ N.]}, \bibinfo{author}{Young\xfnm[ V.]}.
\newblock \bibinfo{title}{Mobile edge computing—a key technology towards 5g}.
\newblock \bibinfo{journal}{ETSI white paper} \bibinfo{year}{2015};\bibinfo{volume}{11}(\bibinfo{number}{11}):\bibinfo{pages}{1--16}.
\bibitem[{Zheng et~al.(2018)Zheng, Xie, Dai, Chen and Wang}]{zheng2018blockchain}
\bibinfo{author}{Zheng\xfnm[ Z.]}, \bibinfo{author}{Xie\xfnm[ S.]}, \bibinfo{author}{Dai\xfnm[ H.N.]}, \bibinfo{author}{Chen\xfnm[ X.]}, \bibinfo{author}{Wang\xfnm[ H.]}.
\newblock \bibinfo{title}{Blockchain challenges and opportunities: A survey}.
\newblock \bibinfo{journal}{International Journal of Web and Grid Services} \bibinfo{year}{2018};\bibinfo{volume}{14}(\bibinfo{number}{4}):\bibinfo{pages}{352--375}.
\bibitem[{Nguyen et~al.(2021)Nguyen, Hoang, Nguyen and Dutkiewicz}]{nguyen2021metachain}
\bibinfo{author}{Nguyen\xfnm[ C.T.]}, \bibinfo{author}{Hoang\xfnm[ D.T.]}, \bibinfo{author}{Nguyen\xfnm[ D.N.]}, \bibinfo{author}{Dutkiewicz\xfnm[ E.]}.
\newblock \bibinfo{title}{Metachain: A novel blockchain-based framework for metaverse applications}.
\newblock \bibinfo{journal}{arXiv preprint arXiv:220100759} \bibinfo{year}{2021};.
\bibitem[{Jacobovitz(2016)}]{jacobovitz2016blockchain}
\bibinfo{author}{Jacobovitz\xfnm[ O.]}.
\newblock \bibinfo{title}{Blockchain for identity management}.
\newblock \bibinfo{journal}{The Lynne and William Frankel Center for Computer Science Department of Computer Science Ben-Gurion University, Beer Sheva} \bibinfo{year}{2016};.
\bibitem[{Mazzoni et~al.(2021)Mazzoni, Corradi and Di~Nicola}]{mazzoni2021performance}
\bibinfo{author}{Mazzoni\xfnm[ M.]}, \bibinfo{author}{Corradi\xfnm[ A.]}, \bibinfo{author}{Di~Nicola\xfnm[ V.]}.
\newblock \bibinfo{title}{Performance evaluation of permissioned blockchains for financial applications: The consensys quorum case study}.
\newblock \bibinfo{journal}{Blockchain: Research and Applications} \bibinfo{year}{2021};:\bibinfo{pages}{100026}.
\bibitem[{Pollitt(2005)}]{pollitt2005decentralization}
\bibinfo{author}{Pollitt\xfnm[ C.]}.
\newblock \bibinfo{title}{Decentralization}.
\newblock In: \bibinfo{booktitle}{The Oxford handbook of public management}. \bibinfo{year}{2005},.
\bibitem[{Zhang et~al.(2019)Zhang, Xue and Liu}]{zhang2019security}
\bibinfo{author}{Zhang\xfnm[ R.]}, \bibinfo{author}{Xue\xfnm[ R.]}, \bibinfo{author}{Liu\xfnm[ L.]}.
\newblock \bibinfo{title}{Security and privacy on blockchain}.
\newblock \bibinfo{journal}{ACM Computing Surveys (CSUR)} \bibinfo{year}{2019};\bibinfo{volume}{52}(\bibinfo{number}{3}):\bibinfo{pages}{1--34}.
\bibitem[{Zhu et~al.(2018)Zhu, Qin, Zhou, Song, Liu and Chu}]{zhu2018digital}
\bibinfo{author}{Zhu\xfnm[ Y.]}, \bibinfo{author}{Qin\xfnm[ Y.]}, \bibinfo{author}{Zhou\xfnm[ Z.]}, \bibinfo{author}{Song\xfnm[ X.]}, \bibinfo{author}{Liu\xfnm[ G.]}, \bibinfo{author}{Chu\xfnm[ W.C.C.]}.
\newblock \bibinfo{title}{Digital asset management with distributed permission over blockchain and attribute-based access control}.
\newblock In: \bibinfo{booktitle}{2018 IEEE International Conference on Services Computing (SCC)}. \bibinfo{organization}{IEEE}; \bibinfo{year}{2018}, p. \bibinfo{pages}{193--200}.
\bibitem[{Harish et~al.(2021)Harish, Liu, Zhong and Huang}]{harish2021log}
\bibinfo{author}{Harish\xfnm[ A.R.]}, \bibinfo{author}{Liu\xfnm[ X.]}, \bibinfo{author}{Zhong\xfnm[ R.Y.]}, \bibinfo{author}{Huang\xfnm[ G.Q.]}.
\newblock \bibinfo{title}{Log-flock: A blockchain-enabled platform for digital asset valuation and risk assessment in e-commerce logistics financing}.
\newblock \bibinfo{journal}{Computers \& Industrial Engineering} \bibinfo{year}{2021};\bibinfo{volume}{151}:\bibinfo{pages}{107001}.
\bibitem[{Sherman and Craig(2003)}]{sherman2003understanding}
\bibinfo{author}{Sherman\xfnm[ W.R.]}, \bibinfo{author}{Craig\xfnm[ A.B.]}.
\newblock \bibinfo{title}{Understanding virtual reality}.
\newblock \bibinfo{journal}{San Francisco, CA: Morgan Kauffman} \bibinfo{year}{2003};.
\bibitem[{Forsyth and Ponce(2011)}]{forsyth2011computer}
\bibinfo{author}{Forsyth\xfnm[ D.]}, \bibinfo{author}{Ponce\xfnm[ J.]}.
\newblock \bibinfo{title}{Computer vision: A modern approach.}
\newblock \bibinfo{publisher}{Prentice hall}; \bibinfo{year}{2011}.
\bibitem[{Hickok and Poeppel(2007)}]{hickok2007cortical}
\bibinfo{author}{Hickok\xfnm[ G.]}, \bibinfo{author}{Poeppel\xfnm[ D.]}.
\newblock \bibinfo{title}{The cortical organization of speech processing}.
\newblock \bibinfo{journal}{Nature reviews neuroscience} \bibinfo{year}{2007};\bibinfo{volume}{8}(\bibinfo{number}{5}):\bibinfo{pages}{393--402}.
\bibitem[{Spencer~Jr et~al.(2004)Spencer~Jr, Ruiz-Sandoval and Kurata}]{spencer2004smart}
\bibinfo{author}{Spencer~Jr\xfnm[ B.]}, \bibinfo{author}{Ruiz-Sandoval\xfnm[ M.E.]}, \bibinfo{author}{Kurata\xfnm[ N.]}.
\newblock \bibinfo{title}{Smart sensing technology: opportunities and challenges}.
\newblock \bibinfo{journal}{Structural Control and Health Monitoring} \bibinfo{year}{2004};\bibinfo{volume}{11}(\bibinfo{number}{4}):\bibinfo{pages}{349--368}.
\bibitem[{Ag{\'o}cs et~al.(2006)Ag{\'o}cs, Balogh, Forg{\'a}cs, Bettio, Gobbetti, Zanetti et~al.}]{agocs2006large}
\bibinfo{author}{Ag{\'o}cs\xfnm[ T.]}, \bibinfo{author}{Balogh\xfnm[ T.]}, \bibinfo{author}{Forg{\'a}cs\xfnm[ T.]}, \bibinfo{author}{Bettio\xfnm[ F.]}, \bibinfo{author}{Gobbetti\xfnm[ E.]}, \bibinfo{author}{Zanetti\xfnm[ G.]}, et~al.
\newblock \bibinfo{title}{A large scale interactive holographic display}.
\newblock In: \bibinfo{booktitle}{IEEE Virtual Reality Conference (VR 2006)}. \bibinfo{organization}{IEEE}; \bibinfo{year}{2006}, p. \bibinfo{pages}{311--311}.
\bibitem[{Warburton et~al.(2007)Warburton, Bredin, Horita, Zbogar, Scott, Esch et~al.}]{warburton2007health}
\bibinfo{author}{Warburton\xfnm[ D.E.]}, \bibinfo{author}{Bredin\xfnm[ S.S.]}, \bibinfo{author}{Horita\xfnm[ L.T.]}, \bibinfo{author}{Zbogar\xfnm[ D.]}, \bibinfo{author}{Scott\xfnm[ J.M.]}, \bibinfo{author}{Esch\xfnm[ B.T.]}, et~al.
\newblock \bibinfo{title}{The health benefits of interactive video game exercise}.
\newblock \bibinfo{journal}{Applied Physiology, Nutrition, and Metabolism} \bibinfo{year}{2007};\bibinfo{volume}{32}(\bibinfo{number}{4}):\bibinfo{pages}{655--663}.
\bibitem[{Flintham et~al.(2003)Flintham, Benford, Anastasi, Hemmings, Crabtree, Greenhalgh et~al.}]{flintham2003line}
\bibinfo{author}{Flintham\xfnm[ M.]}, \bibinfo{author}{Benford\xfnm[ S.]}, \bibinfo{author}{Anastasi\xfnm[ R.]}, \bibinfo{author}{Hemmings\xfnm[ T.]}, \bibinfo{author}{Crabtree\xfnm[ A.]}, \bibinfo{author}{Greenhalgh\xfnm[ C.]}, et~al.
\newblock \bibinfo{title}{Where on-line meets on the streets: experiences with mobile mixed reality games}.
\newblock In: \bibinfo{booktitle}{Proceedings of the SIGCHI conference on Human factors in computing systems}. \bibinfo{year}{2003}, p. \bibinfo{pages}{569--576}.
\bibitem[{Oh et~al.(2016)Oh, Chockalingam, Lee et~al.}]{oh2016control}
\bibinfo{author}{Oh\xfnm[ J.]}, \bibinfo{author}{Chockalingam\xfnm[ V.]}, \bibinfo{author}{Lee\xfnm[ H.]}, et~al.
\newblock \bibinfo{title}{Control of memory, active perception, and action in minecraft}.
\newblock In: \bibinfo{booktitle}{International Conference on Machine Learning}. \bibinfo{organization}{PMLR}; \bibinfo{year}{2016}, p. \bibinfo{pages}{2790--2799}.
\bibitem[{Long(2019)}]{long2019roblox}
\bibinfo{author}{Long\xfnm[ R.U.]}.
\newblock \bibinfo{title}{Roblox and effect on education}.
\newblock Ph.D. thesis; Master’s Thesis, Drury University, Springfield, MO, USA; \bibinfo{year}{2019}.
\bibitem[{Goanta(2020)}]{goanta2020selling}
\bibinfo{author}{Goanta\xfnm[ C.]}.
\newblock \bibinfo{title}{Selling land in decentraland: The regime of non-fungible tokens on the ethereum blockchain under the digital content directive}.
\newblock In: \bibinfo{booktitle}{Disruptive Technology, Legal Innovation, and the Future of Real Estate}. \bibinfo{publisher}{Springer}; \bibinfo{year}{2020}, p. \bibinfo{pages}{139--154}.
\bibitem[{De~Jesus et~al.(2022)De~Jesus, Austria, Marcelo, Ocampo, Tibudan and Tus}]{de2022play}
\bibinfo{author}{De~Jesus\xfnm[ S.B.]}, \bibinfo{author}{Austria\xfnm[ D.]}, \bibinfo{author}{Marcelo\xfnm[ D.R.]}, \bibinfo{author}{Ocampo\xfnm[ C.]}, \bibinfo{author}{Tibudan\xfnm[ A.J.]}, \bibinfo{author}{Tus\xfnm[ J.]}.
\newblock \bibinfo{title}{Play-to-earn: A qualitative analysis of the experiences and challenges faced by axie infinity online gamers amidst the covid-19 pandemic}.
\newblock \bibinfo{journal}{International Journal of Psychology and Counseling} \bibinfo{year}{2022};\bibinfo{volume}{1}(\bibinfo{number}{12}):\bibinfo{pages}{291--424}.
\bibitem[{Dowling(2022)}]{dowling2022fertile}
\bibinfo{author}{Dowling\xfnm[ M.]}.
\newblock \bibinfo{title}{Fertile land: Pricing non-fungible tokens}.
\newblock \bibinfo{journal}{Finance Research Letters} \bibinfo{year}{2022};\bibinfo{volume}{44}:\bibinfo{pages}{102096}.
\bibitem[{Jeon and Kim(2016)}]{jeon2016effect}
\bibinfo{author}{Jeon\xfnm[ I.]}, \bibinfo{author}{Kim\xfnm[ J.]}.
\newblock \bibinfo{title}{Effect of game based learning utilized sandbox game on creative problem-solving ability and learning flow}.
\newblock \bibinfo{journal}{Journal of the Korean Association of Information Education} \bibinfo{year}{2016};\bibinfo{volume}{20}(\bibinfo{number}{3}):\bibinfo{pages}{313--322}.
\bibitem[{Moore(2017)}]{moore2017expression}
\bibinfo{author}{Moore\xfnm[ S.R.]}.
\newblock \bibinfo{title}{Expression of Identity and Finding Meaning in The Sims 4}.
\newblock \bibinfo{publisher}{Illinois State University}; \bibinfo{year}{2017}.
\bibitem[{Kanematsu et~al.(2014)Kanematsu, Kobayashi, Barry, Fukumura, Dharmawansa and Ogawa}]{kanematsu2014virtual}
\bibinfo{author}{Kanematsu\xfnm[ H.]}, \bibinfo{author}{Kobayashi\xfnm[ T.]}, \bibinfo{author}{Barry\xfnm[ D.M.]}, \bibinfo{author}{Fukumura\xfnm[ Y.]}, \bibinfo{author}{Dharmawansa\xfnm[ A.]}, \bibinfo{author}{Ogawa\xfnm[ N.]}.
\newblock \bibinfo{title}{Virtual stem class for nuclear safety education in metaverse}.
\newblock \bibinfo{journal}{Procedia Computer Science} \bibinfo{year}{2014};\bibinfo{volume}{35}:\bibinfo{pages}{1255--1261}.
\bibitem[{Bhirangi et~al.(2021)Bhirangi, Hellebrekers, Majidi and Gupta}]{bhirangi2021reskin}
\bibinfo{author}{Bhirangi\xfnm[ R.]}, \bibinfo{author}{Hellebrekers\xfnm[ T.]}, \bibinfo{author}{Majidi\xfnm[ C.]}, \bibinfo{author}{Gupta\xfnm[ A.]}.
\newblock \bibinfo{title}{Reskin: versatile, replaceable, lasting tactile skins}.
\newblock \bibinfo{journal}{arXiv preprint arXiv:211100071} \bibinfo{year}{2021};.
\bibitem[{Javaid and Haleem(2020)}]{javaid2020virtual}
\bibinfo{author}{Javaid\xfnm[ M.]}, \bibinfo{author}{Haleem\xfnm[ A.]}.
\newblock \bibinfo{title}{Virtual reality applications toward medical field}.
\newblock \bibinfo{journal}{Clinical Epidemiology and Global Health} \bibinfo{year}{2020};\bibinfo{volume}{8}(\bibinfo{number}{2}):\bibinfo{pages}{600--605}.
\bibitem[{Lanfranco et~al.(2004)Lanfranco, Castellanos, Desai and Meyers}]{lanfranco2004robotic}
\bibinfo{author}{Lanfranco\xfnm[ A.R.]}, \bibinfo{author}{Castellanos\xfnm[ A.E.]}, \bibinfo{author}{Desai\xfnm[ J.P.]}, \bibinfo{author}{Meyers\xfnm[ W.C.]}.
\newblock \bibinfo{title}{Robotic surgery: a current perspective}.
\newblock \bibinfo{journal}{Annals of surgery} \bibinfo{year}{2004};\bibinfo{volume}{239}(\bibinfo{number}{1}):\bibinfo{pages}{14}.
\bibitem[{Klinger et~al.(2005)Klinger, Bouchard, L{\'e}geron, Roy, Lauer, Chemin et~al.}]{klinger2005virtual}
\bibinfo{author}{Klinger\xfnm[ E.]}, \bibinfo{author}{Bouchard\xfnm[ S.]}, \bibinfo{author}{L{\'e}geron\xfnm[ P.]}, \bibinfo{author}{Roy\xfnm[ S.]}, \bibinfo{author}{Lauer\xfnm[ F.]}, \bibinfo{author}{Chemin\xfnm[ I.]}, et~al.
\newblock \bibinfo{title}{Virtual reality therapy versus cognitive behavior therapy for social phobia: A preliminary controlled study}.
\newblock \bibinfo{journal}{Cyberpsychology \& behavior} \bibinfo{year}{2005};\bibinfo{volume}{8}(\bibinfo{number}{1}):\bibinfo{pages}{76--88}.
\bibitem[{Vincelli et~al.(2002)Vincelli, Choi, Molinari, Wiederhold, Bouchard and Riva}]{vincelli2002virtual}
\bibinfo{author}{Vincelli\xfnm[ F.]}, \bibinfo{author}{Choi\xfnm[ H.]}, \bibinfo{author}{Molinari\xfnm[ E.]}, \bibinfo{author}{Wiederhold\xfnm[ B.]}, \bibinfo{author}{Bouchard\xfnm[ S.]}, \bibinfo{author}{Riva\xfnm[ G.]}.
\newblock \bibinfo{title}{Virtual reality assisted cognitive behavioral therapy for the treatment of panic disorders with agoraphobia}.
\newblock In: \bibinfo{booktitle}{Medicine Meets Virtual Reality 02/10}. \bibinfo{publisher}{IOS Press}; \bibinfo{year}{2002}, p. \bibinfo{pages}{552--559}.
\bibitem[{Pavlou et~al.(2012)Pavlou, Kanegaonkar, Swapp, Bamiou, Slater and Luxon}]{pavlou2012effect}
\bibinfo{author}{Pavlou\xfnm[ M.]}, \bibinfo{author}{Kanegaonkar\xfnm[ R.]}, \bibinfo{author}{Swapp\xfnm[ D.]}, \bibinfo{author}{Bamiou\xfnm[ D.]}, \bibinfo{author}{Slater\xfnm[ M.]}, \bibinfo{author}{Luxon\xfnm[ L.]}.
\newblock \bibinfo{title}{The effect of virtual reality on visual vertigo symptoms in patients with peripheral vestibular dysfunction: a pilot study}.
\newblock \bibinfo{journal}{Journal of vestibular research} \bibinfo{year}{2012};\bibinfo{volume}{22}(\bibinfo{number}{5-6}):\bibinfo{pages}{273--281}.
\bibitem[{Freeman et~al.(2017)Freeman, Reeve, Robinson, Ehlers, Clark, Spanlang et~al.}]{freeman2017virtual}
\bibinfo{author}{Freeman\xfnm[ D.]}, \bibinfo{author}{Reeve\xfnm[ S.]}, \bibinfo{author}{Robinson\xfnm[ A.]}, \bibinfo{author}{Ehlers\xfnm[ A.]}, \bibinfo{author}{Clark\xfnm[ D.]}, \bibinfo{author}{Spanlang\xfnm[ B.]}, et~al.
\newblock \bibinfo{title}{Virtual reality in the assessment, understanding, and treatment of mental health disorders}.
\newblock \bibinfo{journal}{Psychological medicine} \bibinfo{year}{2017};\bibinfo{volume}{47}(\bibinfo{number}{14}):\bibinfo{pages}{2393--2400}.
\bibitem[{Kumar et~al.(2021)Kumar, Lim, Pandey and Christopher~Westland}]{kumar202120}
\bibinfo{author}{Kumar\xfnm[ S.]}, \bibinfo{author}{Lim\xfnm[ W.M.]}, \bibinfo{author}{Pandey\xfnm[ N.]}, \bibinfo{author}{Christopher~Westland\xfnm[ J.]}.
\newblock \bibinfo{title}{20 years of electronic commerce research}.
\newblock \bibinfo{journal}{Electronic Commerce Research} \bibinfo{year}{2021};\bibinfo{volume}{21}(\bibinfo{number}{1}):\bibinfo{pages}{1--40}.
\bibitem[{Ren et~al.(2021)Ren, Tang, Meng, Ding, Shao, Torr et~al.}]{ren2021cloth}
\bibinfo{author}{Ren\xfnm[ B.]}, \bibinfo{author}{Tang\xfnm[ H.]}, \bibinfo{author}{Meng\xfnm[ F.]}, \bibinfo{author}{Ding\xfnm[ R.]}, \bibinfo{author}{Shao\xfnm[ L.]}, \bibinfo{author}{Torr\xfnm[ P.H.]}, et~al.
\newblock \bibinfo{title}{Cloth interactive transformer for virtual try-on}.
\newblock \bibinfo{journal}{arXiv preprint arXiv:210405519} \bibinfo{year}{2021};.
\bibitem[{Anderson(2019)}]{anderson2019getting}
\bibinfo{author}{Anderson\xfnm[ K.E.]}.
\newblock \bibinfo{title}{Getting acquainted with social networks and apps: figuring out fortnite in (hopefully) less than a fortnight}.
\newblock \bibinfo{journal}{Library Hi Tech News} \bibinfo{year}{2019};.
\bibitem[{Zuckerberg and King(2021)}]{zuckerberg2021facebook}
\bibinfo{author}{Zuckerberg\xfnm[ M.]}, \bibinfo{author}{King\xfnm[ G.]}.
\newblock \bibinfo{title}{Facebook launches" horizon workrooms." here's how it works} \bibinfo{year}{2021};.
\bibitem[{Slater et~al.(2000)Slater, Howell, Steed, Pertaub and Garau}]{slater2000acting}
\bibinfo{author}{Slater\xfnm[ M.]}, \bibinfo{author}{Howell\xfnm[ J.]}, \bibinfo{author}{Steed\xfnm[ A.]}, \bibinfo{author}{Pertaub\xfnm[ D.P.]}, \bibinfo{author}{Garau\xfnm[ M.]}.
\newblock \bibinfo{title}{Acting in virtual reality}.
\newblock In: \bibinfo{booktitle}{Proceedings of the third international conference on Collaborative virtual environments}. \bibinfo{year}{2000}, p. \bibinfo{pages}{103--110}.
\bibitem[{Greenhalgh and Benford(1995)}]{greenhalgh1995massive}
\bibinfo{author}{Greenhalgh\xfnm[ C.]}, \bibinfo{author}{Benford\xfnm[ S.]}.
\newblock \bibinfo{title}{Massive: a collaborative virtual environment for teleconferencing}.
\newblock \bibinfo{journal}{ACM Transactions on Computer-Human Interaction (TOCHI)} \bibinfo{year}{1995};\bibinfo{volume}{2}(\bibinfo{number}{3}):\bibinfo{pages}{239--261}.
\bibitem[{Allal-Ch{\'e}rif(2022)}]{allal2022intelligent}
\bibinfo{author}{Allal-Ch{\'e}rif\xfnm[ O.]}.
\newblock \bibinfo{title}{Intelligent cathedrals: Using augmented reality, virtual reality, and artificial intelligence to provide an intense cultural, historical, and religious visitor experience}.
\newblock \bibinfo{journal}{Technological Forecasting and Social Change} \bibinfo{year}{2022};\bibinfo{volume}{178}:\bibinfo{pages}{121604}.
\bibitem[{Woodruff et~al.(2018)Woodruff, Fox, Rousso-Schindler and Warshaw}]{woodruff2018qualitative}
\bibinfo{author}{Woodruff\xfnm[ A.]}, \bibinfo{author}{Fox\xfnm[ S.E.]}, \bibinfo{author}{Rousso-Schindler\xfnm[ S.]}, \bibinfo{author}{Warshaw\xfnm[ J.]}.
\newblock \bibinfo{title}{A qualitative exploration of perceptions of algorithmic fairness}.
\newblock In: \bibinfo{booktitle}{Proceedings of the 2018 chi conference on human factors in computing systems}. \bibinfo{year}{2018}, p. \bibinfo{pages}{1--14}.
\bibitem[{Mehrabi et~al.(2021)Mehrabi, Morstatter, Saxena, Lerman and Galstyan}]{mehrabi2021survey}
\bibinfo{author}{Mehrabi\xfnm[ N.]}, \bibinfo{author}{Morstatter\xfnm[ F.]}, \bibinfo{author}{Saxena\xfnm[ N.]}, \bibinfo{author}{Lerman\xfnm[ K.]}, \bibinfo{author}{Galstyan\xfnm[ A.]}.
\newblock \bibinfo{title}{A survey on bias and fairness in machine learning}.
\newblock \bibinfo{journal}{ACM computing surveys (CSUR)} \bibinfo{year}{2021};\bibinfo{volume}{54}(\bibinfo{number}{6}):\bibinfo{pages}{1--35}.
\bibitem[{Wan et~al.(2023)Wan, Zha, Liu and Zou}]{wan2023processing}
\bibinfo{author}{Wan\xfnm[ M.]}, \bibinfo{author}{Zha\xfnm[ D.]}, \bibinfo{author}{Liu\xfnm[ N.]}, \bibinfo{author}{Zou\xfnm[ N.]}.
\newblock \bibinfo{title}{In-processing modeling techniques for machine learning fairness: A survey}.
\newblock \bibinfo{journal}{ACM Transactions on Knowledge Discovery from Data} \bibinfo{year}{2023};\bibinfo{volume}{17}(\bibinfo{number}{3}):\bibinfo{pages}{1--27}.
\bibitem[{Hong et~al.(2021)Hong, Zhu, Yu, Wang, Dodge and Zhou}]{hong2021federated}
\bibinfo{author}{Hong\xfnm[ J.]}, \bibinfo{author}{Zhu\xfnm[ Z.]}, \bibinfo{author}{Yu\xfnm[ S.]}, \bibinfo{author}{Wang\xfnm[ Z.]}, \bibinfo{author}{Dodge\xfnm[ H.H.]}, \bibinfo{author}{Zhou\xfnm[ J.]}.
\newblock \bibinfo{title}{Federated adversarial debiasing for fair and transferable representations}.
\newblock In: \bibinfo{booktitle}{Proceedings of the 27th ACM SIGKDD Conference on Knowledge Discovery \& Data Mining}. \bibinfo{year}{2021}, p. \bibinfo{pages}{617--627}.
\bibitem[{Ge et~al.(2022)Ge, Zhao, Yu, Paul, Hu, Hsieh et~al.}]{ge2022toward}
\bibinfo{author}{Ge\xfnm[ Y.]}, \bibinfo{author}{Zhao\xfnm[ X.]}, \bibinfo{author}{Yu\xfnm[ L.]}, \bibinfo{author}{Paul\xfnm[ S.]}, \bibinfo{author}{Hu\xfnm[ D.]}, \bibinfo{author}{Hsieh\xfnm[ C.C.]}, et~al.
\newblock \bibinfo{title}{Toward pareto efficient fairness-utility trade-off in recommendation through reinforcement learning}.
\newblock In: \bibinfo{booktitle}{Proceedings of the fifteenth ACM international conference on web search and data mining}. \bibinfo{year}{2022}, p. \bibinfo{pages}{316--324}.
\bibitem[{Qin et~al.(2022)Qin, Ding, Li, Guan, Wang, Ren et~al.}]{qin2022web3}
\bibinfo{author}{Qin\xfnm[ R.]}, \bibinfo{author}{Ding\xfnm[ W.]}, \bibinfo{author}{Li\xfnm[ J.]}, \bibinfo{author}{Guan\xfnm[ S.]}, \bibinfo{author}{Wang\xfnm[ G.]}, \bibinfo{author}{Ren\xfnm[ Y.]}, et~al.
\newblock \bibinfo{title}{Web3-based decentralized autonomous organizations and operations: Architectures, models, and mechanisms}.
\newblock \bibinfo{journal}{IEEE Transactions on Systems, Man, and Cybernetics: Systems} \bibinfo{year}{2022};\bibinfo{volume}{53}(\bibinfo{number}{4}):\bibinfo{pages}{2073--2082}.
\bibitem[{Feng and Cai(2021)}]{feng2021data}
\bibinfo{author}{Feng\xfnm[ Q.]}, \bibinfo{author}{Cai\xfnm[ R.]}.
\newblock \bibinfo{title}{“data hegemony”: Reflections for the application and development direction of metaverse technology in urban design based on digital}.
\newblock \bibinfo{journal}{Journal of World Architecture} \bibinfo{year}{2021};\bibinfo{volume}{5}(\bibinfo{number}{6}):\bibinfo{pages}{52--61}.
\bibitem[{Ozili(2022)}]{ozili2022decentralized}
\bibinfo{author}{Ozili\xfnm[ P.K.]}.
\newblock \bibinfo{title}{Decentralized finance research and developments around the world}.
\newblock \bibinfo{journal}{Journal of Banking and Financial Technology} \bibinfo{year}{2022};\bibinfo{volume}{6}(\bibinfo{number}{2}):\bibinfo{pages}{117--133}.
\bibitem[{Alipanahloo et~al.(2024)Alipanahloo, Hafid and Zhang}]{alipanahloo2024maximal}
\bibinfo{author}{Alipanahloo\xfnm[ Z.]}, \bibinfo{author}{Hafid\xfnm[ A.S.]}, \bibinfo{author}{Zhang\xfnm[ K.]}.
\newblock \bibinfo{title}{Maximal extractable value mitigation approaches in ethereum and layer-2 chains: A comprehensive survey}.
\newblock \bibinfo{journal}{arXiv preprint arXiv:240719572} \bibinfo{year}{2024};.
\bibitem[{Santana and Albareda(2022)}]{santana2022blockchain}
\bibinfo{author}{Santana\xfnm[ C.]}, \bibinfo{author}{Albareda\xfnm[ L.]}.
\newblock \bibinfo{title}{Blockchain and the emergence of decentralized autonomous organizations (daos): An integrative model and research agenda}.
\newblock \bibinfo{journal}{Technological Forecasting and Social Change} \bibinfo{year}{2022};\bibinfo{volume}{182}:\bibinfo{pages}{121806}.
\bibitem[{Jeong et~al.(2015)Jeong, Kim and Lee}]{jeong2015game}
\bibinfo{author}{Jeong\xfnm[ E.J.]}, \bibinfo{author}{Kim\xfnm[ D.J.]}, \bibinfo{author}{Lee\xfnm[ D.M.]}.
\newblock \bibinfo{title}{Game addiction from psychosocial health perspective}.
\newblock In: \bibinfo{booktitle}{Proceedings of the 17th International Conference on Electronic Commerce 2015}. \bibinfo{year}{2015}, p. \bibinfo{pages}{1--9}.
\bibitem[{Soh et~al.(2018)Soh, Chew, Koay and Ang}]{soh2018parents}
\bibinfo{author}{Soh\xfnm[ P.C.H.]}, \bibinfo{author}{Chew\xfnm[ K.W.]}, \bibinfo{author}{Koay\xfnm[ K.Y.]}, \bibinfo{author}{Ang\xfnm[ P.H.]}.
\newblock \bibinfo{title}{Parents vs peers’ influence on teenagers’ internet addiction and risky online activities}.
\newblock \bibinfo{journal}{Telematics and Informatics} \bibinfo{year}{2018};\bibinfo{volume}{35}(\bibinfo{number}{1}):\bibinfo{pages}{225--236}.
\bibitem[{G{\"u}l{\"u} et~al.(2023)G{\"u}l{\"u}, Yagin, Gocer, Yapici, Ayyildiz, Clemente et~al.}]{gulu2023exploring}
\bibinfo{author}{G{\"u}l{\"u}\xfnm[ M.]}, \bibinfo{author}{Yagin\xfnm[ F.H.]}, \bibinfo{author}{Gocer\xfnm[ I.]}, \bibinfo{author}{Yapici\xfnm[ H.]}, \bibinfo{author}{Ayyildiz\xfnm[ E.]}, \bibinfo{author}{Clemente\xfnm[ F.M.]}, et~al.
\newblock \bibinfo{title}{Exploring obesity, physical activity, and digital game addiction levels among adolescents: A study on machine learning-based prediction of digital game addiction}.
\newblock \bibinfo{journal}{Frontiers in Psychology} \bibinfo{year}{2023};\bibinfo{volume}{14}:\bibinfo{pages}{1097145}.
\bibitem[{Baranowski(2016)}]{baranowski2016pokemon}
\bibinfo{author}{Baranowski\xfnm[ T.]}.
\newblock \bibinfo{title}{Pok{\'e}mon go, go, go, gone?}
\newblock \bibinfo{year}{2016}.
\bibitem[{Faccio and McConnell(2020)}]{faccio2020death}
\bibinfo{author}{Faccio\xfnm[ M.]}, \bibinfo{author}{McConnell\xfnm[ J.J.]}.
\newblock \bibinfo{title}{Death by pok{\'e}mon go: The economic and human cost of using apps while driving}.
\newblock \bibinfo{journal}{Journal of Risk and Insurance} \bibinfo{year}{2020};\bibinfo{volume}{87}(\bibinfo{number}{3}):\bibinfo{pages}{815--849}.
\bibitem[{Drevin and Kim(2019)}]{drevin2019students}
\bibinfo{author}{Drevin\xfnm[ L.]}, \bibinfo{author}{Kim\xfnm[ J.D.]}.
\newblock \bibinfo{title}{It students’awareness of the negative effects of technology}.
\newblock \bibinfo{journal}{Institute of Science and Technology Education College of Graduate Studies University of South Africa PO Box 329, UNISA, 0003} \bibinfo{year}{2019};:\bibinfo{pages}{55}.
\bibitem[{Wang et~al.(2022)Wang, Su, Zhang, Xing, Liu, Luan et~al.}]{wang2022survey}
\bibinfo{author}{Wang\xfnm[ Y.]}, \bibinfo{author}{Su\xfnm[ Z.]}, \bibinfo{author}{Zhang\xfnm[ N.]}, \bibinfo{author}{Xing\xfnm[ R.]}, \bibinfo{author}{Liu\xfnm[ D.]}, \bibinfo{author}{Luan\xfnm[ T.H.]}, et~al.
\newblock \bibinfo{title}{A survey on metaverse: Fundamentals, security, and privacy}.
\newblock \bibinfo{journal}{IEEE communications surveys \& tutorials} \bibinfo{year}{2022};\bibinfo{volume}{25}(\bibinfo{number}{1}):\bibinfo{pages}{319--352}.
\bibitem[{Zallio and Clarkson(2022)}]{zallio2022designing}
\bibinfo{author}{Zallio\xfnm[ M.]}, \bibinfo{author}{Clarkson\xfnm[ P.J.]}.
\newblock \bibinfo{title}{Designing the metaverse: A study on inclusion, diversity, equity, accessibility and safety for digital immersive environments}.
\newblock \bibinfo{journal}{Telematics and Informatics} \bibinfo{year}{2022};\bibinfo{volume}{75}:\bibinfo{pages}{101909}.
\bibitem[{Aung et~al.(2023)Aung, Dhelim, Chen, Ning, Atzori and Kechadi}]{aung2023edge}
\bibinfo{author}{Aung\xfnm[ N.]}, \bibinfo{author}{Dhelim\xfnm[ S.]}, \bibinfo{author}{Chen\xfnm[ L.]}, \bibinfo{author}{Ning\xfnm[ H.]}, \bibinfo{author}{Atzori\xfnm[ L.]}, \bibinfo{author}{Kechadi\xfnm[ T.]}.
\newblock \bibinfo{title}{Edge-enabled metaverse: The convergence of metaverse and mobile edge computing}.
\newblock \bibinfo{journal}{Tsinghua Science and Technology} \bibinfo{year}{2023};\bibinfo{volume}{29}(\bibinfo{number}{3}):\bibinfo{pages}{795--805}.
\bibitem[{Hoa et~al.(2023)Hoa, Son, Luong, Niyato et~al.}]{hoa2023dynamic}
\bibinfo{author}{Hoa\xfnm[ N.T.]}, \bibinfo{author}{Son\xfnm[ B.D.]}, \bibinfo{author}{Luong\xfnm[ N.C.]}, \bibinfo{author}{Niyato\xfnm[ D.]}, et~al.
\newblock \bibinfo{title}{Dynamic offloading for edge computing-assisted metaverse systems}.
\newblock \bibinfo{journal}{IEEE Communications Letters} \bibinfo{year}{2023};\bibinfo{volume}{27}(\bibinfo{number}{7}):\bibinfo{pages}{1749--1753}.
\bibitem[{Huang et~al.(2023)Huang, Li and Cai}]{huang2023security}
\bibinfo{author}{Huang\xfnm[ Y.]}, \bibinfo{author}{Li\xfnm[ Y.J.]}, \bibinfo{author}{Cai\xfnm[ Z.]}.
\newblock \bibinfo{title}{Security and privacy in metaverse: A comprehensive survey}.
\newblock \bibinfo{journal}{Big Data Mining and Analytics} \bibinfo{year}{2023};\bibinfo{volume}{6}(\bibinfo{number}{2}):\bibinfo{pages}{234--247}.
\bibitem[{Marcolla et~al.(2022)Marcolla, Sucasas, Manzano, Bassoli, Fitzek and Aaraj}]{marcolla2022survey}
\bibinfo{author}{Marcolla\xfnm[ C.]}, \bibinfo{author}{Sucasas\xfnm[ V.]}, \bibinfo{author}{Manzano\xfnm[ M.]}, \bibinfo{author}{Bassoli\xfnm[ R.]}, \bibinfo{author}{Fitzek\xfnm[ F.H.]}, \bibinfo{author}{Aaraj\xfnm[ N.]}.
\newblock \bibinfo{title}{Survey on fully homomorphic encryption, theory, and applications}.
\newblock \bibinfo{journal}{Proceedings of the IEEE} \bibinfo{year}{2022};\bibinfo{volume}{110}(\bibinfo{number}{10}):\bibinfo{pages}{1572--1609}.
\bibitem[{Zhao and Chen(2022)}]{zhao2022survey}
\bibinfo{author}{Zhao\xfnm[ Y.]}, \bibinfo{author}{Chen\xfnm[ J.]}.
\newblock \bibinfo{title}{A survey on differential privacy for unstructured data content}.
\newblock \bibinfo{journal}{ACM Computing Surveys (CSUR)} \bibinfo{year}{2022};\bibinfo{volume}{54}(\bibinfo{number}{10s}):\bibinfo{pages}{1--28}.
\bibitem[{Fu et~al.(2022)Fu, Li, Yu, Luan, Zhao and Liu}]{fu2022survey}
\bibinfo{author}{Fu\xfnm[ Y.]}, \bibinfo{author}{Li\xfnm[ C.]}, \bibinfo{author}{Yu\xfnm[ F.R.]}, \bibinfo{author}{Luan\xfnm[ T.H.]}, \bibinfo{author}{Zhao\xfnm[ P.]}, \bibinfo{author}{Liu\xfnm[ S.]}.
\newblock \bibinfo{title}{A survey of blockchain and intelligent networking for the metaverse}.
\newblock \bibinfo{journal}{IEEE Internet of Things Journal} \bibinfo{year}{2022};\bibinfo{volume}{10}(\bibinfo{number}{4}):\bibinfo{pages}{3587--3610}.
\bibitem[{Li et~al.(2023)Li, Yang, Yang, Lan, Zhou, Luo et~al.}]{li2023metaopera}
\bibinfo{author}{Li\xfnm[ T.]}, \bibinfo{author}{Yang\xfnm[ C.]}, \bibinfo{author}{Yang\xfnm[ Q.]}, \bibinfo{author}{Lan\xfnm[ S.]}, \bibinfo{author}{Zhou\xfnm[ S.]}, \bibinfo{author}{Luo\xfnm[ X.]}, et~al.
\newblock \bibinfo{title}{Metaopera: A cross-metaverse interoperability protocol}.
\newblock \bibinfo{journal}{IEEE Wireless Communications} \bibinfo{year}{2023};\bibinfo{volume}{30}(\bibinfo{number}{5}):\bibinfo{pages}{136--143}.
\bibitem[{Ren et~al.(2024)Ren, Lv, Xiong and Wang}]{ren2024hcnct}
\bibinfo{author}{Ren\xfnm[ Y.]}, \bibinfo{author}{Lv\xfnm[ Z.]}, \bibinfo{author}{Xiong\xfnm[ N.N.]}, \bibinfo{author}{Wang\xfnm[ J.]}.
\newblock \bibinfo{title}{Hcnct: A cross-chain interaction scheme for the blockchain-based metaverse}.
\newblock \bibinfo{journal}{ACM Transactions on Multimedia Computing, Communications and Applications} \bibinfo{year}{2024};\bibinfo{volume}{20}(\bibinfo{number}{7}):\bibinfo{pages}{1--23}.
\bibitem[{Sami et~al.(2024)Sami, Hammoud, Arafeh, Wazzeh, Arisdakessian, Chahoud et~al.}]{sami2024metaverse}
\bibinfo{author}{Sami\xfnm[ H.]}, \bibinfo{author}{Hammoud\xfnm[ A.]}, \bibinfo{author}{Arafeh\xfnm[ M.]}, \bibinfo{author}{Wazzeh\xfnm[ M.]}, \bibinfo{author}{Arisdakessian\xfnm[ S.]}, \bibinfo{author}{Chahoud\xfnm[ M.]}, et~al.
\newblock \bibinfo{title}{The metaverse: Survey, trends, novel pipeline ecosystem \& future directions}.
\newblock \bibinfo{journal}{IEEE Communications Surveys \& Tutorials} \bibinfo{year}{2024};.

\end{thebibliography}

\end{document}